%% file: Manuscript.tex
\documentclass{article}

\usepackage{amsmath, amsfonts, amssymb}
\usepackage{bbold,siunitx,textgreek}
\usepackage{subcaption}
\usepackage{tikz}
\usepackage{placeins}
\usepackage{mathtools}
\usepackage{etoolbox}
\usepackage{ulem}
\usepackage[]{inputenc}
\usepackage{authblk}
\usepackage{hyperref}
\usepackage[bottom=8em,top=8em,right=8em,left=8em]{geometry}

\definecolor{linkcolor}{rgb}{0.5,0.1,0.1}
\definecolor{urlcolor} {rgb}{0.1,0.1,0.5}
\definecolor{citecolor}{rgb}{0.1,0.5,0.1}
\hypersetup{colorlinks=true,linkcolor=linkcolor,urlcolor=urlcolor,citecolor=citecolor,
            pdfborder={0 0 0}}

\newcommand{\secref}[1]{Section~\ref{#1}}
\newcommand{\suppref}[1]{Supplementary Section \textcolor{red!50!black}{S#1}}
\newcommand{\figref}[1]{Fig.~\ref{#1}}
\renewcommand{\eqref}[1]{Eq.~(\ref{#1})}
\newcommand{\eq}{\,{=}\,}
\newcommand{\deq}{\,{:=}\,}

\newcommand{\identity}{\mathbb{1}}

\newcommand{\innert}[2]{\left(#1,#2\right)_2}
\renewcommand{\vec}[1]{\boldsymbol{#1}}
\renewcommand{\matrix}[1]{\underline{\underline{\mathbf{#1}}}}
\newcommand{\transpose}[1]{#1^{\phantom{}^\intercal}}

\makeatletter
\DeclareRobustCommand{\volume}{V}
\newcommand{\volumedash}{%
  \makebox[0pt][l]{%
    \ooalign{\hfil\hphantom{$\m@th V$}\hfil\cr\kern0.08em--\hfil\cr}%
  }%
}
\makeatother

\title{Metamaterial Eigenmodes beyond Homogenization}

\author[1]{Antonio G\"{u}nzler}
\author[1]{Cedric Schumacher}
\author[1,*]{Matthias Saba}
\affil[1]{Adolphe Merkle Institute, University of Fribourg, Chemin des Verdiers 4, 1700 Fribourg, Switzerland}
\affil[*]{ \href{https://www.ami.swiss/physics/en/groups/plasmonic-networx/}{www.ami.swiss/physics/en/groups/plasmonic-networx/}\newline\href{mailto:matthias.saba@unifr.ch}{matthias.saba@unifr.ch}}

\begin{document}

\maketitle

\begin{abstract}
Metamaterial homogenization theories usually start with crude approximations that are valid in certain limits in zero order, such as small frequencies, wave vectors and material fill fractions.
In some cases they remain surprisingly robust exceeding their initial assumptions, such as the well-established Maxwell-Garnett theory for elliptical inclusions that can produce reliable results for fill fractions far above its theoretical limitations.
We here present a rigorous solution of Maxwell's equations in binary periodic materials employing a combined Greens-Galerkin procedure to obtain a low-dimensional eigenproblem for the evanescent Floquet eigenmodes of the material.
In its general form, our method provides an accurate solution of the multi-valued complex Floquet bandstructure, which currently cannot be obtained with established solvers.
It is thus shown to be valid in regimes where homogenization theories naturally break down.
For small frequencies and wave numbers in lowest order, our method simplifies to the Maxwell-Garnett result for 2D cylinder and 3D sphere packings.
It therefore provides the missing explanation why Maxwell-Garnett works well up to extremely high fill fractions of approximately $50\%$ depending on the base materials, provided the inclusions are arranged on an isotropic lattice.
\end{abstract}

\section{\label{sec:intro}Introduction}

In the original sense, metamaterials are bespoke plasmonic, periodic structures designed to manipulate the propagation of light.
Their success is rooted in their ability to generate otherwise unavailable electromagnetic material properties leading to real-life applications such as the perfect lens.
The popularity of metamaterials is, however, just as much caused by their accessibility through effective material parameters that formally resemble those in natural materials, but with the potential to generate values that are otherwise unattainable \cite{2006_Metamaterials,zouhdi2009metamaterials,Chang2020}
Perhaps the most notorious such optical characteristic is negative refraction, which has been demonstrated in the microwave regime\cite{Smith2000,Smith2003,Belov2003} and subsequently for smaller wavelengths in the infrared\cite{Shalaev2005}. However, the range of ``unnatural'' optical responses extends far beyond negative index materials, from science fiction cloaking to chiral metamaterials exhibiting polarization-dependent birefringence.  In addition, they enable a broad range of applications in sensing, waveguiding, and imaging \cite{Russell358,JorgensenPhysRevLett.107.143902,NAIR201089,Yabloinhib,JOANNOPOULOS1997165,joannopoulos2008photonic,Kim:12}.

As the structuring at nanometer length-scales required for visible-light metamaterials is still very challenging, it is of the utmost importance to develop a strong theoretical understanding of how the observed optical responses are achieved from  particular geometries.
Thanks to the computing power that is available today, trial and error strategies are more feasible than ever, as brute-force numerical methods such as finite elements or finite difference discretization is readily available to facilitate the study of hypothetical metamaterials without their fabrication.
Optical simulations are powerful tools to predict and understand experimental findings, and can even be employed for machine learning pathways towards bespoke metamaterial functionalities \cite{Pestourie2020}.
From a fundamental perspective, they provide however little to no insight into the physical mechanism behind a particular metamaterial response.
An understanding of the underlying electrodynamic modes can instead provide guidance in the targeted template design for desired optical properties.

For optical metamaterials, theoretical models generally fall into the category of effective medium theories (EMTs), which either work on first principles, accessing limited geometries \cite{Banhegyi1986,MaxwellGarnett,Bruggeman}, or are retrieved by analyzing their scattering behavior or microscopic fields \cite{PhysRevE.71.036617,Smith:06}. 
Even more complex EMTs cannot cover the full physical picture as they, by definition, assume a homogeneous medium with effective material parameters, where the optical response is effectively fitted to a model described by effective permittivity $\varepsilon_\text{eff}$ and permeability $\mu_\text{eff}$ tensors\cite{Chebykin2015,Belyaev2018,Zhang2015}, and sometimes at the cost of increasing complexity, additionally by chirality, that is magnetoelectric cross-coupling, tensors\cite{Sihvola_1996,Demetriadou2012ATM,Ciattoni2015}. 
Yet, even very crude EMTs for elliptical meta-atoms, such as the Maxwell-Garnett (MGA) or Bruggeman approximations, have proven to be surprisingly accurate even for metamaterials violating their underlying assumptions, e.g.\ for high metal fill fractions \cite{Dolan2016}.

Here, we explore a more complete theoretical metamaterial description, inspired by the photonic crystal \cite{Saba2015,SACKEY2018} and mechanical metamaterials \cite{TALLARICO2020115499} communities.
Instead of a full homogenization, which naturally can yield only one mode per polarization direction, this description employs the \textit{complex bandstructure}, which is the set of complex-valued Floquet wave numbers found for a given (real) frequency and crystal inclination.
These modes form a complete solution basis of Maxwell's equations in a semi-infinite metamaterial domain and can be correlated with EMTs, scattering and emission experiments alike.
In this article, we combine a Greens and a Galerkin approach to transform Maxwell's equations into a low-dimensional non-linear eigenproblem that is generally valid for any two-component metamaterial and acts on explicit currents in one material domain (generally the metal domain) only.
We show that for small frequencies and wave vectors ($\omega,|\vec{k}|\,{\ll}\,2\pi/a$ with lattice constant $a$) and without explicit restrictions on the volume fill fraction, our method recovers the MGA formulae for cylinders[spheres] arranged on isotropic 2D[3D] lattices.
We here define a square or hexagonal lattice in 2D, and a cubic lattice (with arbitrary centering) in 3D as isotropic.
Formally, only these lattices have sufficient point symmetry such that no 2D/3D matrix (apart from a multiple of the identity matrix) is invariant under their point group operations.\footnote{It is for that reason that any homogenization model using local effective medium tensors only is isotropic for these lattices.}

\section{Method \label{sec:method}}

\subsection{Floquet modes in linear binary media}
Consider an arbitrary binary periodic metamaterial with primitive unit cell $\Omega = \Omega_1 \cup \Omega_2$ (where $\Omega_1 \cap \Omega_2 \eq \emptyset$), such that the respective materials determine the wavelength-dependent permittivity $\varepsilon(\vec{r},\omega) = \varepsilon_i(\omega)$ for $\vec{r}\in\Omega_i$, $i=1,2$. Most commonly, such a metamaterial  consists of a metal domain and a dielectric background.
Using a monochromatic ansatz for a fixed angular frequency $\omega\ne0$, such that $\delta_t \rightarrow -\imath \omega$, the macroscopic Maxwell equations yield a Helmholtz-type wave equation\footnote{We here assume a constant permeability $\mu \eq 1$, which is a very good approximation for most dielectric matrix materials at optical frequencies.} for the electric field $\vec{E}(\vec{r})$.
Defining the vacuum wave number $k_0 \deq \omega / c$ with the vacuum speed of light $c$, and the material wave number $k(\vec{r}) = k_i \deq \sqrt{\varepsilon_i} k_0$ on $\Omega_i$, with $\Omega_2$ the metal domain, the wave equation takes the form
\begin{subequations}
\begin{align}
    \vec{C}(\vec{r}) &= \underbrace{\left( k_1^2+\Delta-\nabla\otimes\nabla \right)}_{=:\mathcal{H}}
        \vec{E}(\vec{r}) \text{ ,} \label{eq:MWW_differential} \\
    \text{with }\vec{C}(\vec{r}) &:= \left[k_1^2-k^2(\vec{r})\right]\vec{E}(\vec{r})\text{ .} \label{eq:MWW_currentdef}
\end{align}
\label{eq:MaxwellWave}
\end{subequations}
The driving current $\vec{C}$ has evidently compact support on $\Omega_2$, such that the wave equation \eqref{eq:MWW_differential} is homogeneous on $\Omega_1$.
For the permittivity of the metal domain,no specific frequency dependence is required, as in a Drude model, for example.
Instead, the numerical results below are based on interpolated experimental data from \cite{JCRI} for silver and gold.

By the Floquet theorem, the (non-dimensionalized) electric field is of the form
$$\vec{E}(\vec{r})\eq u(\vec{r})\exp\{\imath \boldsymbol{\kappa}\cdot\vec{r}\}\text{ ,}$$
with a Bloch wave vector $\boldsymbol{\kappa}$ and some periodic function $u{:}\,\mathbb{R}^3{\rightarrow}\mathbb{C}^3$ such that
$$u(\vec{r}{+}\vec{T})\eq u(\vec{r}) \;{\forall}\, \vec{T}\eq\matrix{A}.\vec{n},\,\vec{n}{\in}\mathbb{Z}^d\text{ ,}$$
where $d$ is the lattice dimension and $\matrix{A}$ the lattice matrix with primitive lattice vectors as columns.
Remarkably, the Floquet theory applies to a half-space with a periodic shift-operator (not only an infinite periodic structure), such that all field solutions are a superposition of waves of the above form in a semi-infinite metamaterial domain and even a finite metamaterial slab, if the wave vector component in the finite direction is allowed to be complex (that is, the fields are allowed to be evanescent) \cite{Saba2015,PhysRevB.98.155138}.

While there is a large number of solvers available that can calculate the real bandstructure and the bulk modes with complex eigenfrequencies, computing evanescent Floquet modes with complex wave numbers is much less straight-forward.
We show in the following, that \eqref{eq:MaxwellWave} is an ideal starting point to compute evanescent Floquet modes in a computationally affordable and physically meaningful way.

\subsection{Floquet modes from a low-dimensional eigenproblem}
We begin with a plane-wave expansion of the electric field, which is obtained by expressing $\vec{u}$ by its $d$-dimensional Fourier series
\begin{align}
    \vec{E}(\vec{r}) &= \sum_{\vec{G}} \boldsymbol{\mathcal{E}}_{\vec{G}}\,e^{\imath(\boldsymbol{\kappa}+\vec{G})\cdot\vec{r}}
    =:e^{\imath\boldsymbol{\kappa}\cdot\vec{r}}\sum_{\vec{G}} \boldsymbol{\mathcal{E}}_{\vec{G}}\,p_{\vec{G}}(\vec{r})\text{ ,} \label{eq:ExpandE}
\end{align}
where the lattice sum iterates over all reciprocal lattice vectors $\vec{G}\eq \matrix{B}.\vec{n},\,\vec{n}{\in}\mathbb{Z}^d$ ($\matrix{B}\deq2\pi(\transpose{\matrix{A}})^{-1}$).
This is a very convenient set of basis functions to span the vector space of periodic $\vec{u}$ over the unit cell, as the boundary conditions are automatically satisfied and the Helmholtz operator becomes diagonal and analytically invertible.
It is, however, incomplete as the vector space to be spanned contains fields, which are not continuous at the interface between $\Omega_1$ and $\Omega_2$, while the Fourier series only rigorously expresses fields that are analytical over the entire unit cell $\Omega$.
The convergence behavior is therefore only linear in the numerical truncation of the $\vec{G}$ vectors (in this manuscript, we use a convenient parallelepiped cut-off through $|n_i|{\le}N_G$ with the numerical parameter $N_G{\in}\mathbb{N}$).

As we show below, however, a very large number of reciprocal lattice vectors can be afforded numerically, as the computational cost scales linearly with the number of lattice vectors. These lattice vectors only contribute to a lattice sum and not to the dimension of the computationally expensive eigenproblem itself.
Nevertheless, a more efficient expansion of $\vec{E}$ is conceivable, for example using finite elements.
The driving currents in \eqref{eq:MaxwellWave} are efficiently expanded through polynomial basis functions $P_\alpha$ with compact support on $\Omega_2$, for convenience scaled by the Bloch phase\footnote{The polynomial basis functions are chosen as monomials of the type $x^my^nz^l$ with $m{+}n{+}l{\leq}8$. This very low maximum degree is more than sufficient for the exponential convergence behavior in the polynomial degrees \cite{boyd1989chebyshev} in the cases studied. An orthogonal set of polynomials is, however, numerically more stable, should higher orders be required.}
\begin{align}
    \vec{C}(\vec{r}) &= e^{\imath\boldsymbol{\kappa}\cdot\vec{r}}\sum_\alpha \vec{c}_\alpha\,P_\alpha(\vec{r}) 
    \label{eq:ExpandC}\text{ .}
\end{align}
As we shall demonstrate, the driving field $\vec{C}$ can be approximated with sufficient accuracy by a very small set of these polynomial basis functions.

Using these two expansions, we derive the Floquet eigenproblem in \suppref{1}.
In summary, we first analytically invert the wave operater $\mathcal{H}$ via Galerkin testing of \eqref{eq:MWW_differential} with the plane wave basis.
Here lies the advantage of the plane-wave basis, in which $\mathcal{H}$ is diagonal and can be analytically inverted.
Formally, the procedure is equivalent with using the spectral lattice Greens function \cite{Silveirinha2005Lattice}.
In the second step, we substitute the result of the first step into \eqref{eq:MWW_currentdef}, and Galerkin test with the polynomial basis functions.
Using $N$ polynomial basis functions $P_\alpha$, this results in a $3N$-dimensional homogeneous algebraic equation on the $N$-dimensional vector space of polynomial coefficients $c_\alpha$,
\begin{align}
    \sum_\beta \innert{P_\alpha}{P_\beta}\vec{c}_\beta &= \frac{\delta k^2}{\volume(\Omega)}\sum_{\vec{G},\beta}\innert{P_\alpha}{p_{\vec{G}}}\innert{p_{\vec{G}}}{P_\beta}\matrix{\mathcal{H}}_{\boldsymbol{\kappa}+\vec{G}}^{-1}\cdot\vec{c}_\beta \label{eq:NLEVP} \text{ .}
\end{align}
We have here introduced the sesquilinear form
$$(v,w)_2:=\int_{\Omega_2} \mathrm{d}^dr\,v^*(\vec{r})\, w(\vec{r})\quad \text{on } \mathcal{V}_2:=\{\text{ analytical}\,v:\, \Omega_2 \rightarrow \mathbb{C}\,\}$$
and $\delta k^2\deq k_1^2-k_2^2$.
The volume of a domain is denoted $\volume(\cdot)$, while the inverted spectral Helmholtz operator evaluates to
\begin{align*}
    \matrix{\mathcal{H}}_{\boldsymbol{\kappa}+\vec{G}}^{-1} &= \frac{1}{k_1^2-(\boldsymbol{\kappa}+\vec{G})^2}\left[\identity-\frac{(\boldsymbol{\kappa}+\vec{G})\otimes(\boldsymbol{\kappa}+\vec{G})}{k_1^2}\right]\text{ .}
\end{align*}

In order to understand \eqref{eq:NLEVP} as an eigenproblem, we define a coordinate system such that the normal of a chosen crystal inclination $(hkl)$[$(hk)$] for a 3D[2D] lattice is the $z$-direction, which we also refer to as propagation direction.
Equation \ref{eq:NLEVP} now forms a family of eigenproblems with the in-plane components of the Bloch wave vector $\vec{\kappa}_\parallel$ and the vacuum wave number $k_0$ as input parameters and $\kappa_z$ as (non-linear) eigenvalue.
Generally, a complete solution set for the metamaterial half-space for a given $k_0$ is formed by the solutions for all $\vec{\kappa}_\parallel$ in the surface[edge] Brillouin zone for $(hkl)$[$(hk)$]. 
For a scattering problem with a plane-wave incident on a metamaterial slab with $z$ inclination, it evidently suffices to only consider the solutions with $\vec{\kappa}_\parallel\eq k_0\sin\theta\transpose{(\cos\varphi,\sin\varphi)}$, with the angle of incidence defined by the polar angle $\theta$ and the azimuth angle $\varphi$.
The far-field radiation of a classical dipole embedded in the metamaterial can be obtained using the same idea, but with the complication that in oder to satisfy Maxwell's equation on the dipole plane, the fields above and below this plane must be matched.

\subsection{Numerical solution of the eigenproblem}

To numerically solve the non-linear eigenvalue problem (NLEVP) for any set of polynomial basis functions $P_\alpha$, we transform it into a matrix equation,
\begin{align}
    \underbrace{\left[\matrix{Q} \otimes \mathbb{1} - 
    \delta k^2\eta \sum_{\vec{G}}\,\matrix{P}(\vec{G}) \otimes \matrix{\mathcal{H}}_{\boldsymbol{\kappa}+\vec{G}}^{-1} \right]}_{=:\matrix{M}(k_0,\vec{\kappa}_\parallel;\kappa_z)} \vec{c} = 0 \text{ ,}
    \label{eq:NLEVP_num}
\end{align}
where $\otimes$ is the tensor product between the vector spaces of monomial degrees and the 3D Eucledian space so that $\transpose{\vec{c}}\deq(\transpose{\vec{c}_0},\transpose{\vec{c}_1},\dots,\transpose{\vec{c}_N})$ and
\begin{align*}
    P_{\alpha\beta}(\vec{G}) &:= \frac{1}{\volume_2^2}\, \innert{P_\alpha}{p_{\vec{G}}} \innert{p_{\vec{G}}}{P_\beta} \\
    Q_{\alpha\beta} &:= \frac{1}{\volume_2}\,\innert{P_\alpha}{P_\beta} \text{ .}
\end{align*}
A closed-form expression for the inner products is derived in \suppref{2}. 
We obtain the eigenvalues $\kappa_z$ by solving the characteristic equation $\det(\matrix{M})(\kappa_z)\eq0$ using a standard Newton procedure.\footnote{This approach only works for small $N\,{\lessapprox}\,30$ as the determinant quickly becomes numerically unstable due to an increasing condition number of $\matrix{M}$ that can be mitigated by using orthogonal polynomials instead of monomials. Generally more sophisticated algorithms to solve the non-linear eigenproblem are implemented in \cite{jarlebring2018neppack}.}
Since the corresponding eigenvectors $\vec{c}$ are in the nullspace of $\matrix{M}(\kappa_z)$, we only need to perform a pivoted QR decomposition $\matrix{M}^\dagger\eq\matrix{Q}.\matrix{R}.\matrix{P}$, with $\matrix{Q}$ unitary, $\matrix{R}$ upper-right triangular, and $\matrix{P}$ a permutation.
The eigenvectors are the last $N_a$ columns of $\matrix{Q}$ for algebraic multiplicity of $N_a\eq\dim(\matrix{R}){-}\text{rank}(\matrix{R})$, since
$\matrix{Q}^\dagger.\matrix{Q}\eq\identity$ and the last $N_a$ columns of $\matrix{R}^\dagger$ vanish to numerical precision by definition.

\section{Results and Discussion \label{sec:results}}

General solutions of \eqref{eq:NLEVP} require a numerical evaluation of \eqref{eq:NLEVP_num} with cut-offs in both the reciprocal lattice vectors (through $N_G$) and the number $N$ of polynomial basis functions used.
We show in \secref{sec:NLEVP_const} that analytical solutions of \eqref{eq:NLEVP} can be found for cylinder[sphere] packings for small wavenumbers (${\ll}2\pi/a$) and resulting slowly varying driving currents $\vec{C}$ ($N\eq1$).
The dispersion relation of our analytical solution matches the MGA prediction exactly.
A comprehensive analysis of the full numerical solution for aligned cylinders is presented in 
\secref{sec:TM}. We thereby show that the above approximation produces reliable results for an electric field polarized along the cylinder axis over the whole optical spectrum and even into the near-ultraviolet region.
For polarization perpendicular to the cylinder axis, the approximation predicts the mode with lowest imaginary part of the eigenvalue $\kappa_z'$ well for most of the spectrum, apart from the region close to the fundamental dipole Mie resonance, where two modes of the same symmetry classification with comparable $\kappa_z'$ exist.

\subsection{The approximate eigenproblem for cylinders and spheres on a lattice \label{sec:NLEVP_const}}
We here consider a metamaterial in the low wavelength limit, where we assume $k_0,\kappa\,{\ll}\,2\pi/a$.
In lowest non-vanishing order, this immediately simplifies the inverse Helmholtz operator to $\mathcal{H}^{-1}_{\vec{\kappa}{+}\vec{G}}\overset{\vec{G}{\ne}0}{\approx}k_1^{-1}\hat{\vec{G}}\otimes\hat{\vec{G}}$, where $\hat{\vec{G}}\deq\vec{G}/G$ is the direction of the reciprocal lattice vector.
Given that the skin depth of gold and silver at optical frequencies is above \SI{10}{\nano\metre} for optical frequencies and below \cite{novotny2012principles}, we further assume that the slowly varying field fully penetrates objects with less than ${\approx}\SI{20}{\nano\metre}$ diameter.
We therefore consider constant driving currents $\vec{C}(\vec{r})\eq \vec{c}_0\exp\{\imath\vec{\kappa}\cdot\vec{r}\}$ on $\Omega_2$, that is, with $P_0\eq1$.
With these approximations, the eigenproblem \eqref{eq:NLEVP} simplifies to
\begin{align*}
    \vec{c}_0 &= \eta \left(1-\frac{\varepsilon_2}{\varepsilon_1}\right)
    \left[ \frac{k_1^2\identity-\vec{\kappa}\otimes\vec{\kappa}}{k_1^2-\kappa^2}
        + \underbrace{\sum_{\vec{G}\ne0} \left|\frac{\innert{P_0}{p_{\vec{G}}}}{\volume(\Omega_2)}\right|^2
        \hat{\vec{G}}\otimes\hat{\vec{G}}}_{=:\matrix{L}}
    \right].\,\vec{c}_0 \text{ ,}
\end{align*}
where $\eta\deq\volume(\Omega_2)/\volume(\Omega)$ is the volume fill fraction.

The dyadic lattice sum $\matrix{L}$ is a constant matrix, independent of $\kappa$ and $k_0$.
While it generally needs to be computed numerically, we show in \suppref{3} that it evaluates to
\begin{align*}
    \matrix{L} &= \frac{1-\eta}{d\,\eta}\,\identity_d
\end{align*}
for an isotropic lattice and a domain $\Omega_2$ that is invariant under the rotational symmetries of the lattice, where $\identity_d$ is the identity matrix $\identity$ for a 3D lattice, and $\identity{-}\vec{e}_x{\otimes} \vec{e}_x$ for a 2D lattice, with $\vec{e}_x$ the unit normal of the lattice plane.
Introducing $A\deq\eta(1-\varepsilon_2/\varepsilon_1)$
and an effective permittivity $\bar{\varepsilon}\deq\kappa^2/k_0^2$, we arrive at\footnote{We need to of course ignore spurious solutions with $\varepsilon_1{-}\bar{\varepsilon}\eq0$ for this equation.}
\begin{align}
    \left[ A\varepsilon_1 - (\varepsilon_1-\bar{\varepsilon}) \right]\vec{c}_0 &= A \left[k_0^{-2}\,\vec{\kappa}\otimes\vec{\kappa}
    - (\varepsilon_1-\bar{\varepsilon})\,\matrix{L} \right]
    .\,\vec{c}_0 \text{ .}
    \label{eq:NELVP_approx}
\end{align}

\begin{figure}[t]
\centering
\includegraphics[width=\textwidth]{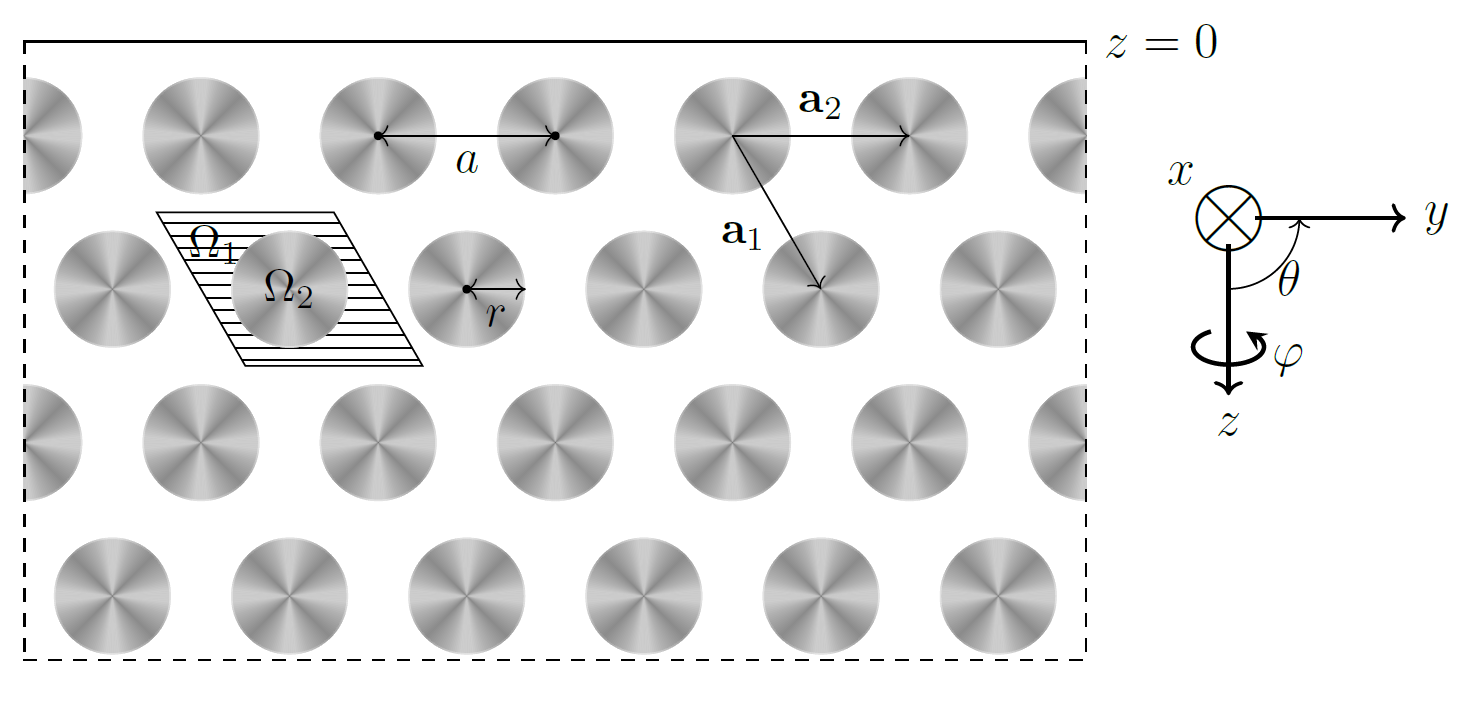}
\caption{Schematic of a 2D-periodic metamaterial consisting of cylindrical metal wires stacked parallel to the boundary surface with $(10)$ inclination on a hexagonal lattice.}
\label{fig:wireMMschem}
\end{figure}

Let us first consider the conceptually simpler, fully isotropic 3D case, for which $\identity_d$ is the 3D identity matrix.
Equation \ref{eq:NELVP_approx} is evidently solved if $\vec{c}_0\eq\vec{\kappa}$ (longitudinal mode) or if $\vec{c}_0\cdot\vec{\kappa}\eq0$ (transverse modes).
The formal definition of longitudinal/transverse can be used in any case.
Note that the longitudinal/transverse terminology, however, provides an exact description for real-valued $\vec{\kappa}$ only.
Generally, the currents are elliptically polarized for $\kappa_z\,{\notin}\,\mathbb{R}$ and $\vec{\kappa}_\parallel\,{\neq}\,0$.

For the longitudinal mode, all coefficients are proportional to $\varepsilon_1{-}\bar{\varepsilon}$, which generally has no solution.\footnote{Only for a lossless metal, a non-dispersive band at the frequency, for which $(\eta+1/2)\varepsilon_2\eq(\eta-1)\varepsilon_1$ is formally found.}
For the transverse mode, \eqref{eq:NELVP_approx} reproduces the well-known MGA result \cite{Banhegyi1986,Markel:16}
\begin{align}
    \bar{\varepsilon}_\text{3D} &= \varepsilon_1\,
    \frac{(2-2\eta)\varepsilon_1 + (1+2\eta)\varepsilon_2}{(2+\eta)\varepsilon_1 + (1-\eta)\varepsilon_2}
    \text{ .}
    \label{eq:MG3D}
\end{align}

We now consider a wire metamaterial \cite{Simovski2012}, consisting of aligned metal wires arranged on an isotropic 2D lattice.
An example of such a metamaterial was fabricated in \cite{Kilchoer2020}.
It consists of gold or silver cylinders with radius $R\eq\SI{10}{nm}$ and lattice constant $a\eq\SI{30}{nm}$, arranged on a hexagonal lattice with $(10)$ inclination, as illustrated in \figref{fig:wireMMschem}.
We define the wire direction as $x$.
Let us first restrict the discussion to a vanishing ordinary (non-Floquet) wave number $\kappa_x\eq0$.
In this situation, the currents are polarized along the cylinder axis or perpendicular to it, so that a comparison with MGA is possible.
Indeed, \eqref{eq:NELVP_approx} is solved if the current modes are longitudinal ($\vec{c}_0\eq\vec{\kappa}$), transverse electric (TE, $\vec{c}_0\eq\vec{e}_x$), or transverse magnetic (TM, $\vec{c}_0\eq\vec{e}_x\,{\times}\,\vec{\kappa}$).
The TE/TM symmetry classification is with respect to the $x\,{\mapsto}\,{-}x$ mirror and only possible for invariant $\vec{\kappa}$, that is, for $\kappa_x\eq0$.
As in the 3D case, the longitudinal mode does generally not exist.
For the TE mode, \eqref{eq:NELVP_approx} yields the MGA result for infinitely long ellipsoids with field polarization along their long axis \cite{Banhegyi1986}
\begin{align}
    \bar{\varepsilon}_\text{TE} &= (1-\eta)\varepsilon_1 + \eta\varepsilon_2 \text{ .}
    \label{eq:MGTE}
\end{align}
For the TM mode, we obtain the MGA result for field polarization perpendicular to the infinitely long ellipsoid axis
\begin{align}
    \bar{\varepsilon}_\text{TM} &= \varepsilon_1\,
    \frac{(1-\eta)\varepsilon_1 + (1+\eta)\varepsilon_2}{(1+\eta)\varepsilon_1 + (1-\eta)\varepsilon_2}
    \text{ .}
    \label{eq:MGTM}
\end{align}

The two general analytical solutions of \eqref{eq:NELVP_approx} for arbitrary\footnote{For this splitting to make sense, we require that the inclination-normal (with complex $\kappa$ component) is either in $x$-direction (wires in substrate-normal direction) or perpendicular to it (substrate-parallel wires). Since any other inclination would cut the cylinders at an oblique angle, this is not too restrictive for experimental metamaterials.} $\vec{\kappa}\eq\vec{\kappa}_\parallel+\kappa_x\vec{e}_x$ require considering the full three-dimensional problem. This, however, splits into a 1D problem and a 2D problem. The former yields a quasi-TM mode with $\vec{c}_0\cdot\vec{e}_x\eq\vec{c}_0\cdot\kappa\eq0$ and the dispersion relation $\bar{\varepsilon}_\text{TM} k_0^2\eq\kappa^2$. For non-zero $\kappa_x$ and $\vec{\kappa}_\parallel$, the 2D problem generally has only one solution. The corresponding (non-normalized) eigenvector is
\begin{align}
    \vec{c}_0 = \kappa_\parallel^2\vec{e}_x
                - \frac{2\,\bar{\varepsilon}_\text{TE}}{\bar{\varepsilon}_\text{TE}+\varepsilon_2}
                    \,\kappa_x \vec{\kappa}_\parallel\text{ .}
    \label{eq:quasiTE}
\end{align}
The eigenvalue obeys the well-known hyperbolic dispersion relation \cite{AlvarezFernandez2021,Shekhar2014}
\begin{align}
    k_0^2 = \frac{\kappa_x^2}{\bar{\varepsilon}_\text{TM}}
            + \frac{\vec{\kappa}_\parallel^2}{\bar{\varepsilon}_\text{TE}} \text{ .}
\end{align}
Note that this quasi-TE solution approaches the TE solution for $\kappa_x\,{\to}\,0$.

\subsection{The breakdown of the constant current approximation for TM modes}\label{sec:TM}

\begin{figure}[t]
    \centering
    \input{gnu/BS-3modes_pd43}
    \caption{
    Bandstructure of the 2D wire metamaterial: Real (solid) and imaginary (dotted) part of $\kappa_z$ of the TE mode (blue) and the two TM modes (red and yellow). The Maxwell-Garnett result for TE case coincides exactly with the calculated mode and is thus omitted, the TM case (black) coincides with either TM mode above and below the resonance frequency when the fields in the wires are approximately constant.}
    \label{fig:bandstructure}
\end{figure}
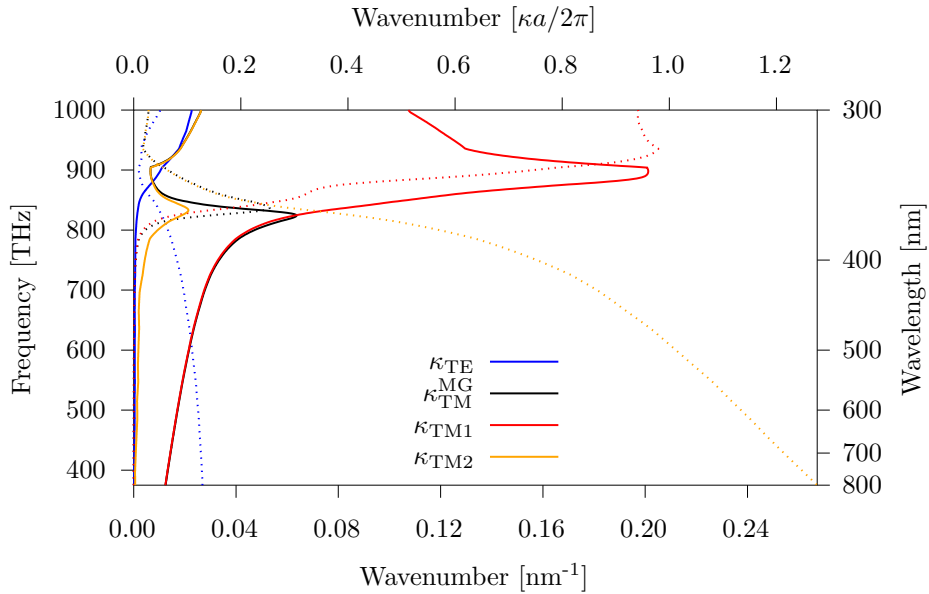 

In this section, we concentrate on the hyperbolic wire metamaterial fabricated in \cite{Kilchoer2020}, to investigate the accuracy of the approximate MGA predictions \eqref{eq:MGTE} and \eqref{eq:MGTM} above. 
This metamaterial consists of silver or gold cylinders (in a vacuum background), arranged on a hexagonal lattice as illustrated in \figref{fig:wireMMschem}.
We here focus on the physically richer silver metamaterial with lattice constant $a\eq\SI{30}{\nano\metre}$ and cylinder radius $R\eq\SI{10}{\nano\metre}$ of \cite{Kilchoer2020}.
More cases with gold wires and different radii are considered in \suppref{4}.
As we have seen above, the solutions of the NLEVP for the 2D wire metamaterial agree exactly with the MGA prediction in the low wavelength limit.
If the latter restriction is relaxed, the NLEVP \eqref{eq:NLEVP} remains three-dimensional, but the lattice sum becomes $k_0$ and $\vec{\kappa}$ dependent and needs to be evaluated numerically.
For the metamaterial in question in the wavelength range of interest ($\SI{300}{\nano\metre}<\lambda<\SI{800}{\nano\metre}$), we find that the analytical results above are reproduced to extremely good precision.
The same holds for the constant current approximation in case of the TE modes, for which \eqref{eq:MGTE} is exact for any practical application.
This is probably expected since the electric field is always tangential to the boundary surface between $\Omega_1$ and $\Omega_2$, so that \eqref{eq:MGTE} is simply the weighted volume average of the two permittivities.
This formula, which is the same as for a binary Bragg reflector, can be immediately deduced from Fatou's theorem \cite{10000133129}, and approximates the dispersion relation well for frequencies below the fundamental bandgap (as long as $\lambda\gtrapprox2\sqrt{\bar{\varepsilon}(\lambda)}a$).
The resulting silver metamaterial TE bandstructure for $\vec{\kappa}_\parallel\eq0$, for which the MGA and the exact solution are optically indistinguishable, is shown in \figref{fig:bandstructure}.

The physics is naturally richer for the TM modes, where the fundamental dipole Mie resonance (the first pole of the scattering coefficients \cite{Lahart2004,Kettunen2015}) of the individual cylinders lies in the frequency range of interest.
Within a small band close to this dipole resonance, the constant current approach is insufficient.
To demonstrate this behavior, we solve the numerical problem \eqref{eq:NLEVP_num} in the spectral range of interest between the frequencies of $\nu_0\eq\SI{375}{\tera\hertz}$ ($\lambda_0\,{\approx}\,\SI{800}{\nano\metre}$) and $\nu_1\eq\SI{1000}{\tera\hertz}$ ($\lambda_1\,{\approx}\,\SI{300}{\nano\metre}$).
As initial $\kappa_z$ for the Newton procedure, we use the MGA approximation \eqref{eq:MGTM} at $\nu_0$ and $\nu_1$, sufficiently far away from the dipole resonance, and iterate along the spectrum in both directions.
Figure \ref{fig:bandstructure} shows the bandstructure for monomials up to degree $4$ in $y$ direction and degree $3$ in $z$ direction, denoted pd\eq(4,3), where the solutions converge sufficiently over the whole spectrum.
The bandstructure is not as simple as MGA suggests and there are indeed two TM modes with reasonably small $\kappa_z'$, of which one (labeled TM1) converges to the MGA band for small frequencies, and the other (TM2) converges to the MGA band for large frequencies.
The bandstructure for pd\eq(2,2) and a detailed convergence analysis is provided in \suppref{5}.

\begin{figure}[t]
    \centering
    
    \begin{minipage}[c]{0.89\textwidth}
    \begin{subfigure}[b]{.32\textwidth}
    \centering
    \caption{TM1, \SI{400}{\tera\hertz}}\label{fig:TM1_400}
    \resizebox{.99\linewidth}{!}{\includegraphics[bb=70 0 410 350,clip,width=\textwidth]{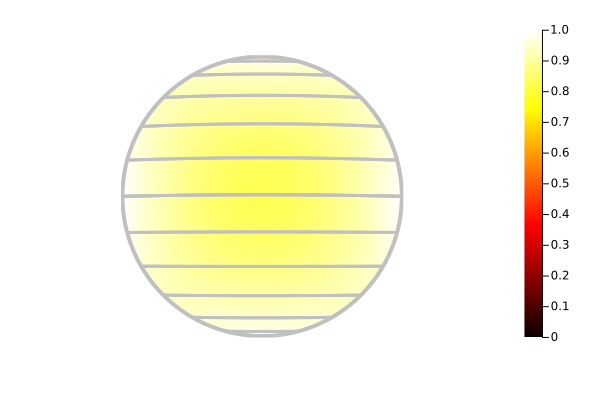}}
    \end{subfigure}
    \hfill    
    \begin{subfigure}[b]{.32\textwidth}
    \centering
    \caption{TM1, \SI{838}{\tera\hertz}}\label{fig:TM1_838}
    \resizebox{.99\linewidth}{!}{\includegraphics[bb=70 0 410 350,clip,width=\textwidth]{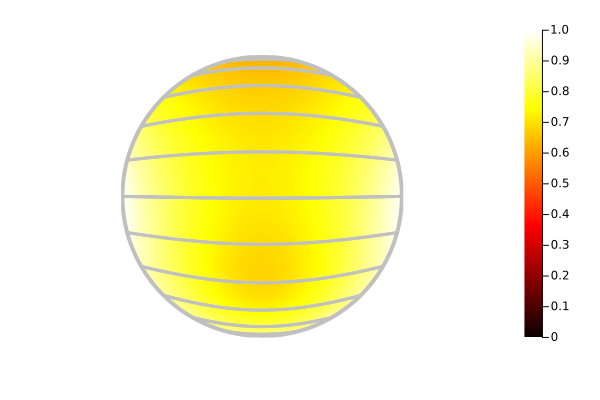}}
    \end{subfigure}
    \hfill
    \begin{subfigure}[b]{.32\textwidth}
    \centering
    \caption{TM1, \SI{900}{\tera\hertz}}\label{fig:TM1_900}
    \resizebox{.99\linewidth}{!}{\includegraphics[bb=70 0 410 350,clip,width=\textwidth]{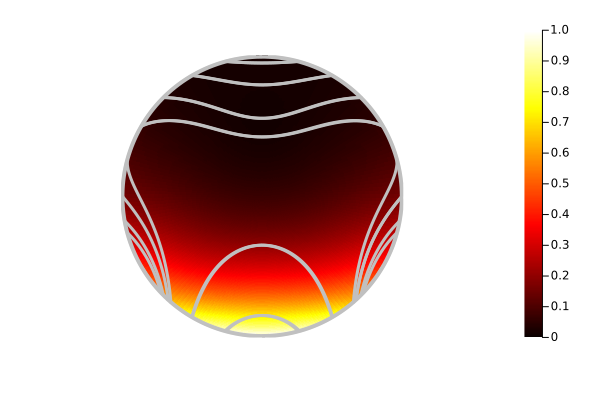}}
    \end{subfigure}
    
    \begin{subfigure}[b]{.32\textwidth}
    \centering
    \caption{TM2, \SI{400}{\tera\hertz}}\label{fig:TM2_400}
    \resizebox{.99\linewidth}{!}{\includegraphics[bb=70 0 410 350,clip,width=\textwidth]{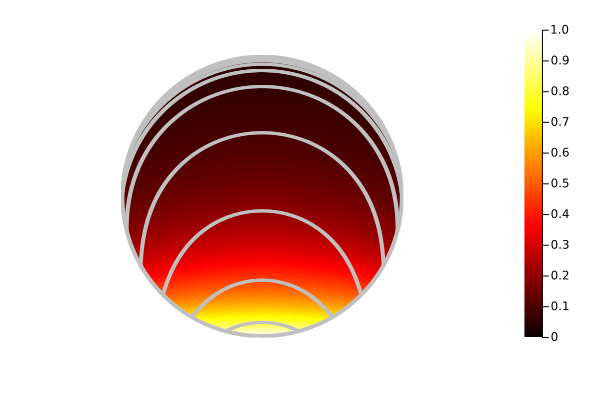}}
    \end{subfigure}
    \hfill
    \begin{subfigure}[b]{.32\textwidth}
    \centering
    \caption{TM2, \SI{838}{\tera\hertz}}\label{fig:TM2_838}
    \resizebox{.99\linewidth}{!}{\includegraphics[bb=70 0 410 350,clip,width=\textwidth]{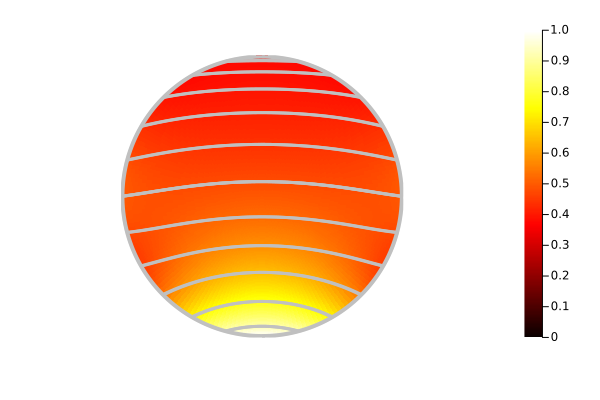}}
    \end{subfigure}
    \hfill
    \begin{subfigure}[b]{.32\textwidth}
    \centering
    \caption{TM2, \SI{900}{\tera\hertz}}\label{fig:TM2_900}
    \resizebox{.99\linewidth}{!}{\includegraphics[bb=70 0 410 350,clip,width=\textwidth]{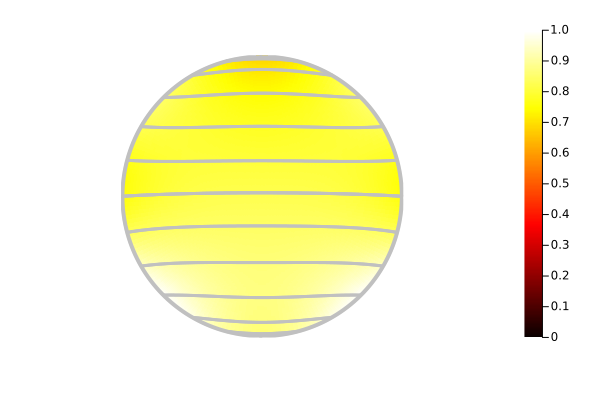}}
    \end{subfigure}
    \end{minipage}
    \begin{minipage}[c]{0.1\textwidth}
        \includegraphics[bb=500 50 570 380,clip,width=\textwidth]{figures/Current_NG1600pd44_TMk2_900cb01.png}
    \end{minipage}

    \caption{
    Heatmaps of normalized current intensities $|\vec{C}|(\vec{r})/\max\{|\vec{C}|(\vec{r})\}$ and field lines of the two TM modes TM1 (a-c) and TM2 (d-f) at \SI{400}{THz} (a,d), \SI{838}{THz} (b,e) and \SI{900}{THz} (c,f).
  }
    \label{fig:current-plots}
\end{figure}

To better illustrate the breakdown of the constant current approximation, we calculate the well converged numerical currents for pd\eq(4,4) from the eigenvectors $\vec{c}$ using \eqref{eq:ExpandC}.
The current intensities and field lines are shown in \figref{fig:current-plots} for the two TM modes at three different frequencies well below (\SI{400}{THz}), above (\SI{900}{THz}) and at the dipole resonance (\SI{838}{THz}).
In the long wavelength limit, the TM1 mode that is well approximated by \eqref{eq:MGTM} exhibits near constant currents as expected (\figref{fig:TM1_400}).
The TM2 mode with relatively large $\kappa_z'$ is on the other hand strongly evanescent within the cylinder with strongly curved field lines (\figref{fig:TM2_400}), which explains why it is not contained in the constant current model.
At the resonance, both modes exhibit varying currents to some degree (\figref{fig:TM1_838} and \ref{fig:TM2_838}).
While they are clearly different, with concentrated currents at the left and the right of the cylinder in case of TM1, neither field is well resolved using a constant current approximation.
The approximation \eqref{eq:MGTM} indeed moves from the TM1 branch to the TM2 branch between ${\sim}800$ and ${\sim}\SI{850}{\tera\hertz}$ (\figref{fig:bandstructure}).
At \SI{900}{\tera\hertz}, the TM1 mode is strongly evanescent within the cylinder (\figref{fig:TM1_900}) in agreement with the large imaginary part of $\kappa_z$ (\figref{fig:bandstructure}).
The field profile is generally relatively complex, explaining why the constant current approximation is not able to resolve this mode.
The TM2 mode, on the other hand, exhibits a weakly varying field at \SI{900}{\tera\hertz} (\figref{fig:TM2_900}).
The dispersion is consequently well approximated by \eqref{eq:MGTM} in this frequency regime.

Generally, the MGA provides good results at long wavelengths if the radius of the cylinders is small compared to the lattice constant, as expected from the assumptions in its derivation.
It is still a good approximation away from resonances for very high fill fractions way beyond its theoretical limitations.
Reasonable results are obtained even at a resonance, as we show in \suppref{4}. For gold, where the resonance is damped through losses, the MGA is valid over the whole visible spectrum up to $R/a\,{\sim}\,0.43$ ($\eta\,{\sim}\,\SI{70}{\percent}$), while for silver, the threshold is $R/a\,{\sim}\,0.26$ ($\eta\,{\sim}\,\SI{25}{\percent}$).
As the MG prediction gradually changes from TM1 to TM2 at the resonance for higher fill fractions, while only one mode suffices at lower fill fractions, the two TM bands must belong to one connected, topologically non-trivial solution manifold in the $R$-$\omega$ parameter space.
We discuss the non-trivial topology, which stems from a band degeneracy with an associated winding number of $2$, in \suppref{6}.


\section{Conclusion}

We have developed a low-dimensional eigenproblem on the vector space of explicit metal currents to solve Maxwell's Floquet problem in binary metamaterials.
Our approach not only provides an intuitive and generalized method to solve for the complex bandstructure, but also justifies the use of the well-known Maxwell-Garnett approximation (MGA) for metal fill fractions far exceeding its original assumptions.
We have indeed shown that the permittivities from the MGA are analytically reproduced by our modal solutions in the long wavelength limit for constant currents with no explicit restriction on the volume fill fraction.
Our results rigorously apply to perfectly ordered isotropic arrangements, but is seems clear that the they can be extended to less ordered cases as long as a certain amount of order remains, regarding the distance between the individual spheres or cylinders, such as in hyperuniform packings \cite{PhysRevE.68.041113}.

At the same time, we revealed the necessity of two TM modes in the spectral vicinity of the single-wire dipole resonance of a 2D-periodic nanowire array, one of which approaches the MGA solution below the resonance and the other one above the resonance.
We have shown that this is due to a topological transition that occurs in the bandstructure of the TM modes at a critical resonance strength (determined by the dielectric background material and volume fill fraction), above which the constant current approach fails to provide an sensible solution.
Although our approach is generally valid for an arbitrary metal domain $\Omega_2$, we have here restricted the discussion to cylinder and sphere packings, for which the inner Galerkin products can be evaluated analytically.
While many other sufficiently symmetric shapes of $\Omega_2$ allow  an analytical evaluation of the sesquilinear products, an implementation for arbitrary geometries would be useful.
An efficient numerical algorithm to compute the  the required integrals from a simplex representation of the boundary of $\Omega_2$ in both 2D and 3D can be found in \cite{Gabard:Integrals}.

{\small
\bibliographystyle{unsrt}
\bibliography{bibliography}
}
\end{document}


\maketitle

\section{NLEVP derivation}\label{sec:NLEVP}

To span the vector space of all holomorphic functions $\mathcal{V}_2\deq\{\,\text{analytical }f:\,\Omega_2\rightarrow\mathbb{C}^3\,\}$, the degree of the $P_\alpha$'s has to be unbounded.
In most realistic cases, however, polynomials of small degree suffice as long as the unit cell size is well below the wavelength length scale and $\Omega_2$ is a reasonably simple domain (for example a sphere, a convex polyhedron or a topologically simple wire-mesh).
As described in the main text, a coordinate system is chosen such that $\boldsymbol{\kappa}$ has real $x$- and $y$-components and the complex $z$ component, which is referred to as the propagation direction, takes the role of an eigenvalue.
Equivalently, it suffices to consider an angle of incidence determined by a polar angle, $\theta$ and an azimuth angle $\varphi$. In a slab-like configuration this means $z$ is perpendicular to the slab interface.

Consider the vector space $\mathcal{V}\deq\{\,\text{analytical }f:\,\Omega \rightarrow \mathbb{C}\mid f(\vec{r}+\vec{a}_i)\eq f(\vec{r})\,\}$ of holomorphic functions on $\Omega$ with the periodicity requirement (we defined $\Omega$ as the parallelepiped spanned by the primitive basis vector $\vec{a}_i$).
This vector space is spanned by the plane-wave basis functions $p_{\vec{G}}\deq\exp\{\imath\vec{G}\cdot\vec{r}\}$, with $\vec{G}$ reciprocal lattice vectors  $\vec{G}\eq \matrix{B}.\vec{n},\,\vec{n}{\in}\mathbb{Z}^d$ ($\matrix{B}\deq2\pi(\transpose{\matrix{A}})^{-1}$).
The corresponding norm for $v \in \mathcal{V}$ is $|v|\deq\sqrt{(v,v)}$, with the sesquilinear form $(v,w)\eq\int_\Omega \mathrm{d}^3r\,v^*(\vec{r})\,w(\vec{r})$, where $(\cdot)^*$ denotes the complex conjugate.
We now transform \maineqref{1a}, that is $\mathcal{H}\vec{E}\eq\vec{C}$, into a weak form via a standard Galerkin method.
First, we expand the periodic part of the electric field according to \maineqref{2}
\begin{align}
    \vec{E}(\vec{r}) &= e^{\imath\boldsymbol{\kappa}\cdot\vec{r}}\sum_{\vec{G}} \vec{\mathcal{E}}_{\vec{G}}\,p_{\vec{G}}(\vec{r})\text{ ,} \label{eq:ExpandE}
\end{align}
and the currents according to \maineqref{3}
\begin{align}
    \vec{C}(\vec{r}) &= e^{\imath\boldsymbol{\kappa}\cdot\vec{r}}\sum_\alpha \vec{c}_\alpha\,P_\alpha(\vec{r}) 
    \label{eq:ExpandC}\text{ ,}
\end{align}
with polynomial functions $P_\alpha(\vec{r})$ with compact support on $\Omega_2$, that is $P_\alpha(\vec{r})\eq0$ for $\vec{r}\,{\in}\,\Omega_1$.
The resulting equation is
\begin{align}
    \sum_{\vec{G}} p_{\vec{G}} \mathcal{H}_{\vec{\kappa}+\vec{G}}.\vec{\mathcal{E}}_{\vec{G}}
    &= \sum_\alpha P_\alpha\,\vec{c}_\alpha\text{ ,}
    \label{eq:strong}
\end{align}
with $\mathcal{H}$ acting on the plane wave components takes the form $$\matrix{\mathcal{H}}_{\boldsymbol{\kappa}+\vec{G}}\eq \left[k_1^2-(\boldsymbol{\kappa}+\vec{G})^2\right]\identity+(\boldsymbol{\kappa}+\vec{G})\otimes(\boldsymbol{\kappa}+\vec{G})\text{ .}$$
Note that this form remains valid for complex $\vec{\kappa}$. Next, we Galerkin test \eqref{eq:strong} with the plane waves $p_{\vec{G}}$.
The orthogonality $(p_{\vec{G}},p_{\vec{G}}')\eq\volume(\Omega)\delta_{\vec{G}\vec{G}'}$, with the $\volume(\cdot)$ the volume of a domain, and a simple algebraic inversion of the Helmholtz matrix
\begin{align*}
    \matrix{\mathcal{H}}_{\boldsymbol{\kappa}+\vec{G}}^{-1} &= \frac{1}{k_1^2-(\boldsymbol{\kappa}+\vec{G})^2}\left[\identity-\frac{(\boldsymbol{\kappa}+\vec{G})\otimes(\boldsymbol{\kappa}+\vec{G})}{k_1^2}\right]
\end{align*}
yields the electric field components $\vec{\mathcal{E}}_{\vec{G}}$ expressed in terms of the current components $\vec{c}_\alpha$:
\begin{align}
    \boldsymbol{\mathcal{E}}_{\vec{G}} &= \sum_\alpha\frac{\inner{p_{\vec{G}}}{P_\alpha}}{\volume(\Omega)}\matrix{\mathcal{H}}_{\boldsymbol{\kappa}+\vec{G}}^{-1}.\vec{c}_\alpha \text{ .}
    \label{eq:EGfromca}
\end{align}
Note that the inner product in this expression is strictly speaking not well-defined as $P_\alpha\notin\mathcal{V}$, which is at the heart of the slow convergence behavior of \eqref{eq:EGfromca}.
On the other hand, the inner product is the only part of \eqref{eq:EGfromca}, which is not a simple analytical expression.
It can be transformed into a well-defined inner product $\innert{p_{\vec{G}}}{P_\alpha}$ on the vector space $\mathcal{V}_2$ of holomorphic functions on $\Omega_2$ and is calculated analytically for a disk below in \secref{sec:innerprods}.
We finally apply the expansions \eqref{eq:ExpandE} and \eqref{eq:ExpandC} to \maineqref{1b}, $\delta k^2 \vec{E}\eq\vec{C}$ on $\Omega_2$, and substitute \eqref{eq:EGfromca} into the resulting expression. Galerkin testing with $P_\alpha$ this time yields the algebraic eigenproblem \maineqref{4}:
\begin{align}
    \sum_\beta \innert{P_\alpha}{P_\beta}\vec{c}_\beta &= \frac{\delta k^2}{\volume(\Omega)}\sum_{\vec{G},\beta}\innert{P_\alpha}{p_{\vec{G}}}\innert{p_{\vec{G}}}{P_\beta}\matrix{\mathcal{H}}_{\boldsymbol{\kappa}+\vec{G}}^{-1}\cdot\vec{c}_\beta \label{eq:NLEVP} \text{ .}
\end{align}
Note that this NLEVP has only dimension $3N$ for the number $N$ of driving current basis functions $P_\alpha$.
The eigenvalues are represented by possible wave vectors components $\kappa_z$, while the associated eigenvectors are the coefficient vectors $\vec{c}_\alpha$ of the expansion of driven field $C$.

\section{Sesquilinear products for polynomial basis functions on a disk}\label{sec:innerprods}

We here limit ourselves to the analytical calculation of the inner products in \eqref{eq:NLEVP} on a disk $\Omega_2\deq\{ \vec{r}\in\mathbb{R}^2 \mid |\vec{r}|{<}R \}$, although similar expressions may be found for ellipses, polygons, or simple 3D objects.
We further restrict the discussion to non-dimensionalized monomials of arbitrary degree as basis functions, such that $P_\alpha\eq y^mz^n/R^{m+n}$.
To evaluate the inner products in the left-hand side of \eqref{eq:NLEVP}, it suffices to integrate a monomial of the above form. 
Using polar coordinates and, it can be shown that
\begin{align}
    \int_{\Omega_2} d^2r\,\frac{y^mz^n}{R^{m+n}} =
    \begin{cases}
        0 & \text{if $m$ or $n$ odd}\\
        V(\Omega_2)\,\frac{2\,(m-1)!!(n-1)!!}{(m+n+2)\,2^{(m+n)/2}[(m+n)/2]!} & \text{else}
    \end{cases}\text{ .}
    \label{eq:inner_LHS}
\end{align}
To evaluate the inner products on the right-hand side of \eqref{eq:NLEVP}, $\innert{P_\alpha}{p_{\vec{G}}}$, we substitute $u\deq GR$ and $u_i\deq G_i R$, and derive a relation for the partial derivatives through $\partial_{u_i}\eq(u_i/u)\partial_u$ and successive application of the product rule 
\begin{align*}
    \partial_{u_i}^m \partial_{u_j}^n = \sum_{\alpha=0}^{\floor*{\frac{m}{2}}} \sum_{\beta=0}^{\floor*{\frac{n}{2}}} c_\alpha^{(m)} c_\beta^{(n)} u_i^{m-2\alpha} u_j^{n-2\beta} \left(\frac{1}{u}\partial_u\right)^{m+n-\alpha-\beta}
\end{align*}
with $c_0^{(m)}\eq1$, $c_\alpha^{(0)}\eq0$ for $\alpha{>}0$, and the recursion $c_{\alpha+1}^{(m+1)}\eq c_{\alpha+1}^{(m)}\,{+}\,(m{-}2\alpha)c_\alpha^{(m)}$. We can further apply the recurrence formulae of the Bessel functions
\begin{align*}
    \frac{1}{u}\partial_u \frac{J_n(u)}{u^n} &= \frac{J'_n(u)-n J_n(u)/u}{u^{n+1}}
    = \frac{J_{n-1}(u)-J_{n+1}(u) - (J_{n-1}(u)+J_{n+1}(u))}{2u^{n+1}} = -\frac{J_{n+1}(u)}{u^{n+1}}
\end{align*}
to  obtain:
\begin{align}
    \int_{\Omega_2} d^2r\,\frac{y^m z^n}{R^{n+m}}\, e^{\imath\vec{G}\cdot\vec{r}} &=
    (-\imath\partial_{u_y})^m(-\imath\partial_{u_z})^n \int_{\Omega_2} d^2r\, e^{\imath\vec{G}\cdot\vec{r}} \\
    &= 2 \volume(\Omega_2) \imath^{m+n} \sum_{\alpha=0}^{\floor*{\frac{m}{2}}} \sum_{\beta=0}^{\floor*{\frac{n}{2}}} (-1)^{\alpha+\beta} c_\alpha^{(m)} c_\beta^{(n)} \frac{u_y^{m-2\alpha} u_z^{n-2\beta}}{u^p} \frac{J_{p+1}(u)}{u}    \text{ ,}
    \label{eq:inner_RHS}
\end{align}
with $p\eq m{+}n{-}\alpha{-}\beta$.
Note that \eqref{eq:inner_LHS} arises as the limit $u\,{\to}\,0$, in which $J_n(u)/u^n\to1/(2^n n!)$, from \eqref{eq:inner_RHS}.

\section{Analytical lattice sums in the constant current approximation}\label{sec:latsum}

We here show that the lattice sum in \mainsecref{3} is given by:
\begin{align}
    \matrix{L} &:= \sum_{\vec{G}\ne0}
        \left|\frac{\innert{1}{p_{\vec{G}}}}{\volume(\Omega_2)}\right|^2
        \hat{\vec{G}}\otimes\hat{\vec{G}}
        \overset{!}{=} \frac{1-\eta}{d\,\eta}\,\identity_d
    \label{eq:latsum}
\end{align}

We start the proof by expanding the indicator function for $\Omega_2$ into plane waves on the unit cell:
\begin{align*}
    \chi(\vec{r})=\sum_{\vec{G}}\chi_{\vec{G}}e^{\imath\vec{G}\cdot\vec{r}} :=
    \begin{cases}
        1 & \text{if }\vec{r}\in\Omega_2 \\
        0 & \text{else}
    \end{cases}\text{ . }
\end{align*}
Testing with the plane wave basis functions yields
\begin{align*}
    \chi_{\vec{G}} = \frac{1}{V}\innert{e^{\imath\vec{G}\cdot\vec{r}}}{1}
\end{align*}
and thus
\begin{align*}
    \chi(\vec{r}) = \frac{1}{V}\sum_{\vec{G}}\innert{e^{\imath\vec{G}\cdot\vec{r}}}{1}e^{\imath\vec{G}\cdot\vec{r}}\text{ .}
\end{align*}
We now express $V_2$ with this indicator function in the plane-wave basis:
\begin{align}
    V_2 &= \int_\Omega d^2r\,\chi = \int_{\Omega_2} d^2r\,\chi \nonumber\\
        &= \frac{1}{V}\sum_{\vec{G}} \left|\innert{1}{p_{\vec{G}}}\right|^2\text{ .}
        \label{eq:V_2}
\end{align}
At this point, we return to \eqref{eq:latsum}.
Let us first consider a current polarized in $y$-direction in a 2D square or hexagonal lattice of cylinders as in the main text.
The relevant component of the outer product in \eqref{eq:latsum} can in this case be rearranged to
\begin{align}
    \frac{G_y^2}{G^2} = \frac{G_y^2+G_z^2}{2G^2} + \frac{G_y^2-G_z^2}{2G^2}
    =  \frac{1}{2} + \frac{G_y^2-G_z^2}{2G^2}\text{ .}
    \label{eq:rearrange}
\end{align}
The last term vanishes under the sum if the lattice has a $\mathcal{C}_4$ or $\mathcal{C}_6$ symmetry,\footnote{Easily seen by rotating the reciprocal vectors, so that for example $C_z$ becomes $C_y$ under $C_4$ and noting that the lattice remains invariant under the rotation.} so that re-substitution into \eqref{eq:latsum} and comparison with \eqref{eq:V_2} yields
\begin{align*}
    L_{yy} = \frac{1-\eta}{2\,\eta} \text{ .}
\end{align*}

The general case can be derived by noting that the rearrangement \eqref{eq:rearrange} is just the projection of the addends in \eqref{eq:latsum} onto the irreducible representations $i$ of the corresponding rotation group \cite{dresselhaus2008group}.
As long as the domain $\Omega_2$ is invariant under the point group symmetries, this is equivalent to a projection of the outer products, which yields:
\begin{align}
    \hat{P}_i\hat{\vec{G}}\otimes\hat{\vec{G}}
    &= \frac{1}{\sum_{S\in \mathcal{P}}}\sum_{S\in \mathcal{P}}\chi_i(S)
    (S\hat{\vec{G}})\otimes(S\hat{\vec{G}})\text{ ,}
    \label{eq:project}
\end{align}
where $\chi_i(S)$ are the characters of the irreducible representation of the point group $\mathcal{P}$ and $\hat{\vec{G}}{\otimes}\hat{\vec{G}}\eq\sum_i\hat{P}_i\hat{\vec{G}}{\otimes}\hat{\vec{G}}$.
For the cyclic abelian point group $\mathcal{C}_4\deq\{\identity,C_4,C_2,C_4^3\}$ for example, we have $4$ irreducible representations with $\chi_i(C_4)\eq\imath^i$.
For $i$ odd, the projection vanishes, while for $i\eq0$ (trivial representation) and $i\eq2$, we obtain
\begin{align*}
    \hat{P}_0\hat{\vec{G}}\otimes\hat{\vec{G}}
    &= \frac{\hat{\vec{G}}\otimes\hat{\vec{G}}+(\vec{e}_h\times\hat{\vec{G}})\otimes(\vec{e}_h\times\hat{\vec{G}})}{2} = \frac{1}{2}\identity_2 \text{ , and}\\
    \hat{P}_2\hat{\vec{G}}\otimes\hat{\vec{G}}
    &= \frac{\hat{\vec{G}}\otimes\hat{\vec{G}}-(\vec{e}_h\times\hat{\vec{G}})\otimes(\vec{e}_h\times\hat{\vec{G}})}{2}
    \text{ ,}
\end{align*}
which is the full tensorial form of the rearrangement \eqref{eq:rearrange} above.
As the lattice is symmetric, any projection onto a non-trivial representation vanishes under the sum (which is a consequence of the wonderful orthogonality theorem for irreducible representations), in analogy to the result for the last term in \eqref{eq:rearrange}.
In 3D for a tetrahedral $\mathcal{T}$ or octahedral $\mathcal{O}$ point symmetry, we analogously obtain
$$\hat{P}_0\hat{\vec{G}}\otimes\hat{\vec{G}} = \frac{1}{3}\identity_3\text{ .}$$
For $\mathcal{T}$ for example, which is the isogonal point group of the space group $P23$ (195), we see this by substituting $\hat{\vec{G}}\deq(x,y,z)$ with $x^2+y^2+z^2\eq1$ and applying the $12$ symmetry elements listed for a general Wyckoff position 12j in \cite{0792365909} to
\begin{align*}
    \frac{1}{12}\sum_{S\in\mathcal{T}}\hat{\vec{G}}\otimes\hat{\vec{G}} &= \frac{1}{12}\sum_{S\in\mathcal{T}}\begin{pmatrix}
        x^2 & xy & xz \\ xy & y^2 & yz \\ xz & yz & z^2
    \end{pmatrix}\\
    &=\frac{1}{12}\begin{pmatrix}
        4(x^2+y^2+z^2) & 0 & 0 \\ 0 & 4(x^2+y^2+z^2) & 0 \\ 0 & 0 & 4(x^2+y^2+z^2)
    \end{pmatrix}
    = \frac{1}{3}\identity_3\text{ .}
\end{align*}
We have now generally shown that
\begin{align}
    \hat{P}_0 \hat{\vec{G}}\otimes\hat{\vec{G}} = \frac{1}{d}\identity_d
    \label{eq:project0}
\end{align}
for isotropic lattices. As long as the domain $\Omega_2$ is invariant under the rotational symmetries of the lattice, which of course includes cylinders and spheres, but generally even chiral objects, we obtain:
\begin{align*}
    \matrix{L} &= \sum_{\vec{G}\ne0}
        \left|\frac{\innert{1}{p_{\vec{G}}}}{\volume(\Omega_2)}\right|^2
        \hat{\vec{G}}\otimes\hat{\vec{G}}
        =\sum_{\vec{G}\ne0}
        \left|\frac{\innert{1}{p_{\vec{G}}}}{\volume(\Omega_2)}\right|^2 \hat{P}_0
        \hat{\vec{G}}\otimes\hat{\vec{G}} \\
        &\overset{\ref{eq:project0}}{=} \frac{1}{d}\identity_d\sum_{\vec{G}\ne0}
        \left|\frac{\innert{1}{p_{\vec{G}}}}{\volume(\Omega_2)}\right|^2
        = \frac{1}{d}\identity_d\Bigg[\underbrace{\sum_{\vec{G}}
        \left|\frac{\innert{1}{p_{\vec{G}}}}{\volume(\Omega_2)}\right|^2}_{\overset{\ref{eq:V_2}}{=}1/\eta} - 1\Bigg] = \frac{1-\eta}{d\,\eta}\identity_d
        \text{ q.e.d.}
\end{align*}

\section{Bandstructures for different wire metamaterials}

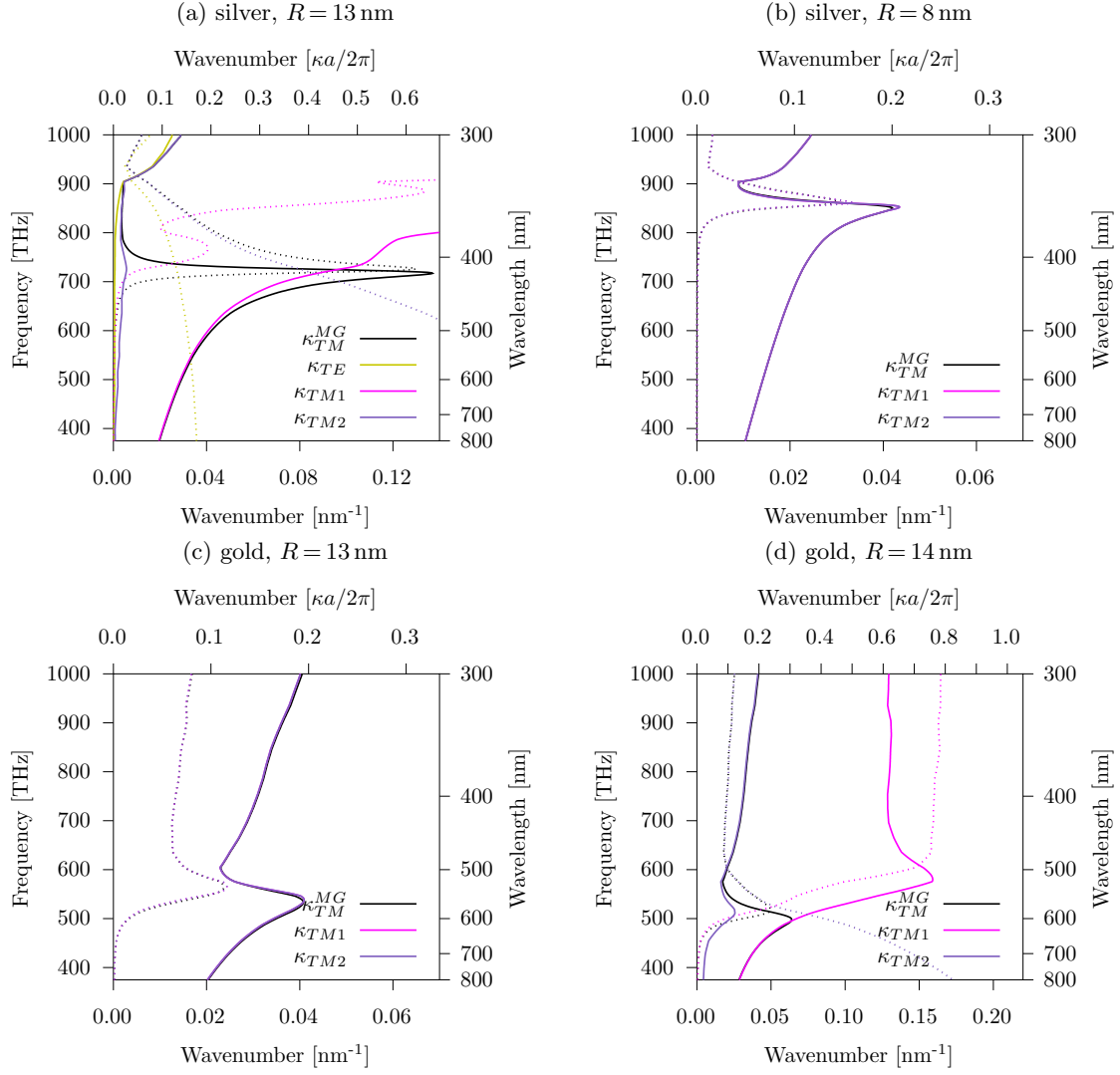
\begin{figure}[ht]
    \centering
    \begin{subfigure}[b]{.48\textwidth}
    \centering
    \caption{silver, $R\eq\SI{13}{\nano\metre}$}
    \resizebox{.99\linewidth}{!}{\input{gnu/BS-R13}}
    \end{subfigure} 
    \begin{subfigure}[b]{.48\textwidth}
    \centering
    \caption{silver, $R\eq\SI{8}{\nano\metre}$}
    \resizebox{.99\linewidth}{!}{\input{gnu/BS-R8}}
    \end{subfigure} 
    
    \begin{subfigure}[b]{.48\textwidth}
    \centering
    \caption{gold, $R\eq\SI{13}{\nano\metre}$}
    \resizebox{.99\linewidth}{!}{\input{gnu/BS-AuR13}}
    \end{subfigure} 
    \begin{subfigure}[b]{.48\textwidth}
    \centering
    \caption{gold, $R\eq\SI{14}{\nano\metre}$}
    \resizebox{.99\linewidth}{!}{\input{gnu/BS-AuR14}}
    \end{subfigure} 
    \caption{Bandstructures of the hexagonal wire metamaterial with lattice constant of $a\eq\SI{30}{\nano\metre}$.
    Four cases with cylinders of a different metal and radius are shown: (a) silver with $R\eq\SI{13}{\nano\metre}$, (b) silver with $R\eq\SI{8}{\nano\metre}$, (c) gold with $R\eq\SI{13}{\nano\metre}$, and (d) gold with $R\eq\SI{14}{\nano\metre}$.
    Each plot shows the real (solid line) and imaginary (dotted line) part of $\kappa_z$ of the two TM modes (magenta and purple), alongside the MGA solution (black). The TE mode (yellow) is shown in (a) and omitted from the remaining cases.
    }
    \label{fig:Au-Ag-radii}
\end{figure}

In the studied 2D nanowire metamaterial the TE mode behaves similar to the chosen metal, with a thinned-out electron density and thus a shift in the plasma frequency, accounted for through the metal fill fraction.
This is expected as the electric field is oscillating parallel to the (infinitely) long cylinder axis and currents can therefore flow freely as in a bulk metal.
For silver nanowires with $R\eq\SI{13}{\nano\metre}$, this results in an effective plasma frequency of \SI{900}{\tera\hertz}, and an evanescent mode with short decay length below that frequency as seen in \figref{fig:Au-Ag-radii} (a).

\FloatBarrier

For the TM modes, where the electric field is oscillating perpendicular to the nanowire axis, a dipole resonance occurs at a certain frequency depending on the chosen metal and the nanowire radius. In the model system with lattice constant $a\eq\SI{30}{\nano\metre}$ and nanowire radius $R\eq\SI{10}{\nano\metre}$, the resonance is strongest at approximately \SI{838}{\tera\hertz}. When the diameter of the nanowires is increased, the dipole resonance is not only red-shifted (to ~\SI{700}{\tera\hertz} for $R\eq\SI{13}{\nano\metre}$) but also intensified, as shown in the bandstructure in \figref{fig:Au-Ag-radii} (a). The increased resonance is expected due to the closer vicinity of opposite cylinder surfaces. Likewise the red-shift of the dipole resonance is a direct consequence of the larger cylinder diameter \cite{HUANG201013}. The topological degeneracy (see \secref{sec:degen}) observed for the TM modes occurs when the resonance exceeds a threshold intensity. For silver nanowires, this threshold is reached at approximately \SI{8.775}{\nano\metre} (see \secref{sec:degen} for details). For radii sufficiently far below the threshold, e.g.~$R\eq\SI{8}{\nano\metre}$ as shown in \figref{fig:Au-Ag-radii} (b), the fundamental TM mode is well described by the MGA formula over the whole frequency range. Hence, the MGA formula indeed breaks down with increasing metal fill fraction, however, at much higher fill fractions than its assumptions suggest and relatively quickly close to the threshold rather than gradually.

The underlying reason of the breakdown seems to be related to an increased dipole resonance, caused by the field hybridization of next-nearest cylinder neighbours, which leads to a stronger electric field variance inside the individual cylinders.
This interpretation can be further illustrated when gold is chosen as the metal constituent. The more strongly pronounced Ohmic losses in gold dampen the resonance strength such that the threshold nanowire radius is between \SI{13}{\nano\metre} and \SI{14}{\nano\metre}, i.e. when the wires are almost touching. Therefore, below and up to  $R\eq\SI{13}{\nano\metre}$ the MGA formula remains valid for the fundamental TM mode in the whole frequency range of interest as seen in \figref{fig:Au-Ag-radii} (c). At $R\eq\SI{14}{\nano\metre}$, however, the same topological transition as for silver has occured, as depicted in \figref{fig:Au-Ag-radii} (d). Note that the resonance is strongly red-shifted compared to silver, consistent with the shift in permittivities between gold and silver \cite{JCRI}. 

\section{Convergence analysis}\label{sec:conv}

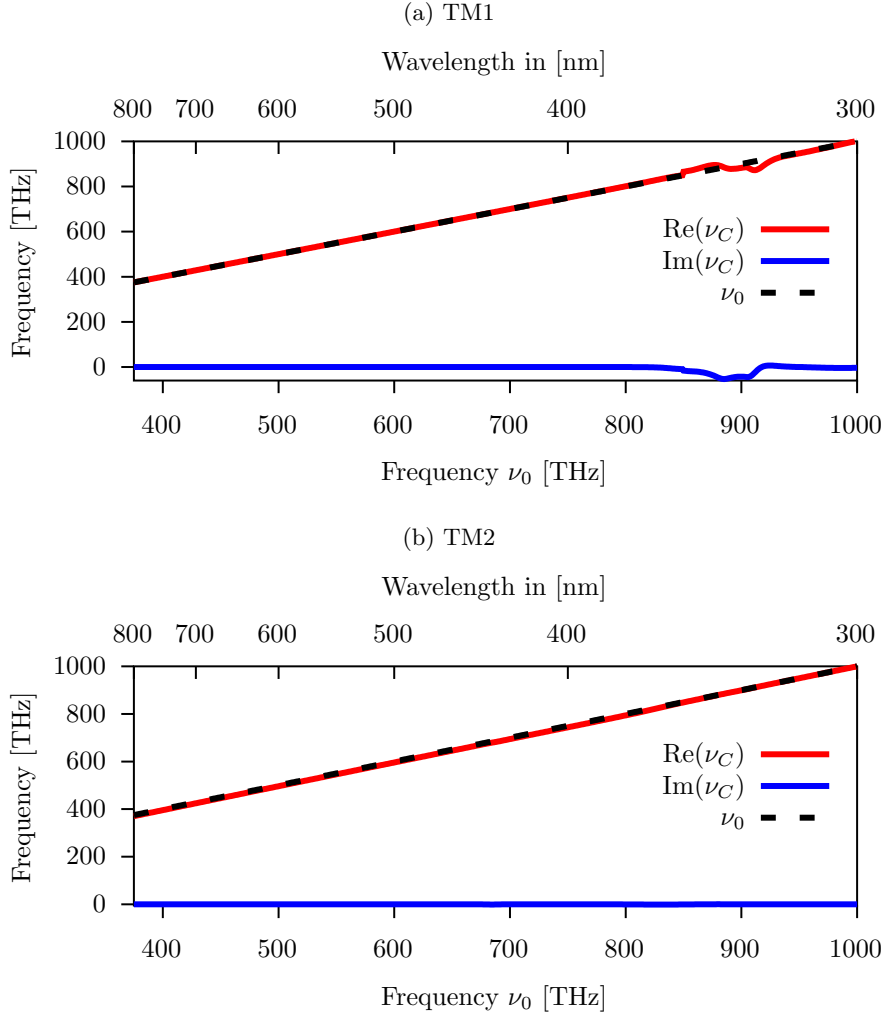
\begin{figure}[ht]
    \centering
    \begin{subfigure}[t]{.99\textwidth}
    \centering
    \subcaption{TM1}\label{fig:COMSOL_BS_Check_TM1}
    \input{gnu/TM1_COMSOL_CHECK_nearest}
    \end{subfigure} 
    \begin{subfigure}[t]{.99\textwidth}\vspace{1em}
    \centering
    \subcaption{TM2}\label{fig:COMSOL_BS_Check_TM2}
    \input{gnu/TM2_COMSOL_CHECK_nearest}
    \end{subfigure} 
    \caption{Complex eigenfrequencies $\nu_C$ calculated with COMSOL using the frequency $\nu_0$ and the wavenumber $\kappa$ from a computation with $\text{pd}\eq(4,3)$ and $N_G\eq800$ as input parameters for the generic silver metamaterial from the main text.}
    \label{fig:COMSOL_BS_Check}
\end{figure}

While the NLEVP \eqref{eq:NLEVP} is exact, the numerical evaluation of the lattice sum requires a cut-off of the reciprocal lattice vectors $\vec{G}\eq \matrix{B}.\vec{n}$.
We choose a convenient square/cubic cut-off $|n_i|{\le}N_G$.
As discussed in the main text, the lattice sum only converges linearly with $N_G$, because of the inconsistency of vector spaces.
To understand the exact behavior, let us consider the 2D case with $P_\alpha\deq y^{\alpha_1}z^{\alpha_2}/R^{\alpha_1{+}\alpha_2}$.
Employing the limiting behavior of the Bessel function $J_n(x)\overset{x\to\infty}{\sim}1/\sqrt{x}$, using a disk cut-off and an integral comparison, the residual vanishes if any of the $\alpha_i$ or $\beta_i$ is odd.
If all exponents are even, we obtain a residual of:
\begin{align*}
    \int_{N_G}^\infty \rm{d}G\,G^{1-l}\overset{l>2}{\propto} N_G^{2-l}
    \text{ .}
\end{align*}
The leading (smallest) degree $l$ of the addend can be read off in \eqref{eq:inner_RHS}.
For pd\eq(0,0), $l\eq3$.
For higher polynomial cut-offs, the leading degree is (for even exponents)
$$l\eq\frac{3}{4}(\alpha_1+\alpha_2+\beta_1+\beta_2)+3\text{ .}$$
This is important to understand that a much lower $N_G$ suffices for the lattice sum addends involving $\alpha_i$ or $\beta_i$ larger than $0$, since the residual of the corresponding matrix entries is proportional to $N_G^{2-l}$.
The convergence dependence on the polynomial degrees can thus be employed to make a numerical implementation with pd$\,{\ne}\,$(0,0) much more efficient.
The limiting matrix entries are, however, for any polynomial cut-off the ones with all $\alpha_i\eq\beta_i\eq0$, which determines the linear convergence behavior of the algorithm, with a residual proportional to $N_G^{-1}$.
For the polynomial degrees, on the other hand, we expect exponential convergence behavior as this is effectively a spectral method that spans a vector space of holomorphic functions \cite{boyd1989chebyshev}.
Consequently, a very small maximum degree suffices for the fundamental (physically relevant) modes.

To study the convergence of our algorithm numerically at a frequency $\nu_0\eq\omega_0/(2\pi)$, we simulate the same geometry with COMSOL Multiphysics.
The material data for this simulation is taken as $\varepsilon_2(\nu_0)$, and the Floquet boundary conditions are determined by our converged $\kappa_z(\nu_0,N_G,\text{pd})$.
We use a fine simulation mesh, so that the complex eigenfrequency from the simulation $\nu_C$ converges to $\nu_0$ at the same rate (albeit not with the same coefficient) as the numerical $\kappa_z(\nu_0,N_G,\text{pd})$ approaches the converged $\kappa_z(\nu_0)$.
\figref{fig:COMSOL_BS_Check} shows the result of this analysis over the whole spectral range for both modes with $N_G\eq800$ and pd\eq(4,3).
For these numerical parameters, the TM2 mode is well converged over the whole spectral range, while the TM1 mode shows minor discrepancies at wavelengths between ${\sim}\SI{850}{\nano\metre}$ and ${\sim}\SI{950}{\nano\metre}$.

\begin{figure}
    \centering
    \begin{subfigure}[b]{.48\textwidth}
    \centering
    \caption{\SI{400}{\tera\hertz}, TM1}\label{fig:conv_TM1_400}
    \resizebox{.99\linewidth}{!}{\input{gnu/Conv_TM1_400THz}}
    \end{subfigure} 
    \begin{subfigure}[b]{.48\textwidth}
    \centering
    \caption{\SI{400}{\tera\hertz}, TM2}\label{fig:conv_TM2_400}
    \resizebox{.99\linewidth}{!}{\input{gnu/Conv_TM2_400THz}}
    \end{subfigure} 
    
    \begin{subfigure}[b]{.48\textwidth}
    \centering
    \caption{\SI{838}{\tera\hertz}, TM1}\label{fig:conv_TM1_838}
    \resizebox{.99\linewidth}{!}{\input{gnu/Conv_TM1_838THz}}
    \end{subfigure} 
    \begin{subfigure}[b]{.48\textwidth}
    \centering
    \caption{\SI{838}{\tera\hertz}, TM2}\label{fig:conv_TM2_838}
    \resizebox{.99\linewidth}{!}{\input{gnu/Conv_TM2_838THz}}
    \end{subfigure} 
    
    \begin{subfigure}[b]{.48\textwidth}
    \centering
    \caption{\SI{900}{\tera\hertz}, TM1}\label{fig:conv_TM1_900}
    \resizebox{.99\linewidth}{!}{\input{gnu/Conv_TM1_900THz}}
    \end{subfigure} 
    \begin{subfigure}[b]{.48\textwidth}
    \centering
    \caption{\SI{900}{\tera\hertz}, TM2}\label{fig:conv_TM2_900}
    \resizebox{.99\linewidth}{!}{\input{gnu/Conv_TM2_900THz}}
    \end{subfigure} 
    \caption{Logarithmic relative error as a function of increasing logarithmic reciprocal lattice cutoff $\log_{10}(N_G)$ of the two TM modes for, \SI{400}{THz} (a,b), \SI{838}{THz} (c,d), \SI{900}{THz} (e,f) measured as relative error according to the COMSOL control. The convergence is not always monotone and for TM1 at \SI{400}{THz} (a) even $pd\eq(2,2)$ produces a more accurate solution.}
    \label{fig:Convergence}
\end{figure}
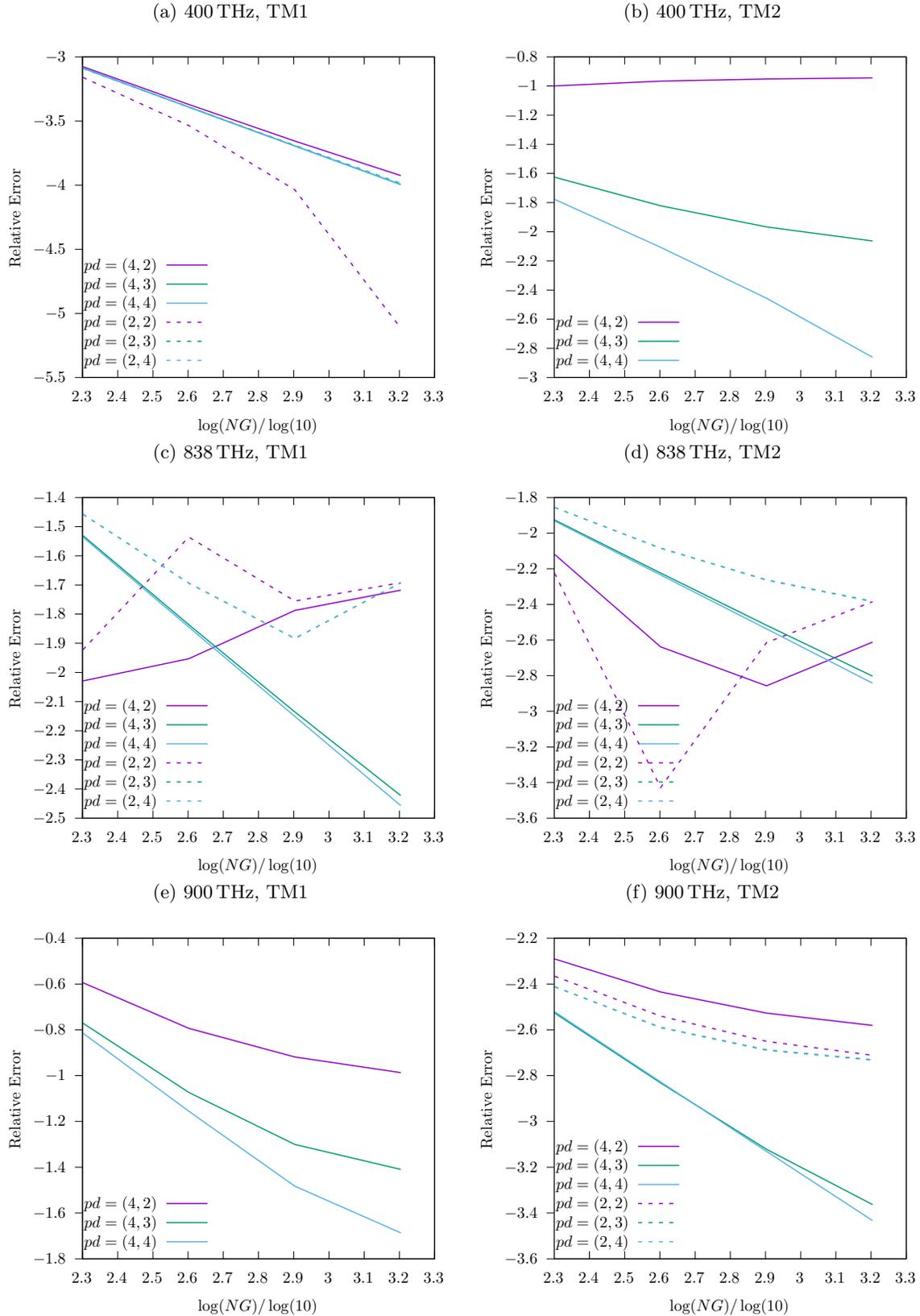

We have computed the relative logarithmic error
\begin{align*}
    \epsilon(N_G,\text{pd}):=\log_{10}\left|1-\frac{\nu_C}{\nu_0}\right|
\end{align*}
for $N_G\,{\in}\,\{200,400,800,1600\}$ and $\text{pd}\,{\in}\,\{(2,2);(2,3);(2,4);(4,2);(4,3);(4,4)$.\footnote{
Note that an even exponent for the $y$ dependence is omitted as it is incompatible with the TM symmetry for $k_y\eq0$ and the coefficients vanish up to numerical precision.}
The results are represented in \figref{fig:Convergence} for both TM modes at the $3$ representative frequencies used in the main manuscript: \SI{400}{\tera\hertz} in the long wavelength limit, \SI{838}{\tera\hertz} at the dipole resonance, and \SI{900}{\tera\hertz} above the resonance.
Before discussing these results in detail, we note that the errors reported here, compare the numerical frequencies through back-feeding our results into a converged established numerical algorithm due to the lack of an established method that can produce $\kappa_z$ directly.
Particularly at and above the resonance, the obtained frequency $\omega_C$ varies very quickly with tiny changes in $\kappa_z$, so that the reported converged errors at and below $\SI{1}{\percent}$ underestimate the real numerical accuracy of our method by orders of magnitude, both in the eigenfields and the eigenvalues $\kappa_z$.

At \SI{400}{\tera\hertz}, the TM1 mode is already very well described with the MGA approximation, so that the convergence is purely limited by $N_G$ and no improvement can be found for polynomial degrees larger than (2,2).
As seen in \figref{fig:conv_TM1_400}, the predicted linear convergence behavior is seen for all degrees, except for pd\eq(2,2), which shows improved convergence for unknown reasons.
The situation is reversed for the TM2 mode, where the fields are underrepresented for polynomial degrees below (4,4).
As seen in \figref{fig:conv_TM2_400}, no convergence with $N_G$ is obeserved for pd\eq(4,2), where the error stays at \SI{10}{\percent}, while the linear convergence behavior is observed for pd\eq(4,4).
Similarly, both modes are evidently underrepresented for polynomial degree of $2$ in either $y$ or $z$ direction at \SI{838}{\tera\hertz}, with no convergence with increasing $N_G$ (\figref{fig:conv_TM1_838} and \ref{fig:conv_TM2_838}).
For pd\eq(4,3) or (4,4), however, both modes converge linearly with $N_G$, as expected, reaching errors below \SI{0.3}{\percent}.
At \SI{900}{\tera\hertz} above the resonance, as seen in \figref{fig:conv_TM2_900}, the TM2 mode is well resolved for any polynomial degree above (2,2).
The linear convergence with $N_G$, however, requires pd\eq(4,4).
The TM1 mode, on the other hand is slightly underrepresented even for pd\eq(4,4), leading to sub-linear convergence in $N_G$ (\figref{fig:conv_TM1_900}.
This can be understood from the complex field profile with intensity variations of more than an order of magnitude and strongly varying field lines (Figure 3(c) in the main text).

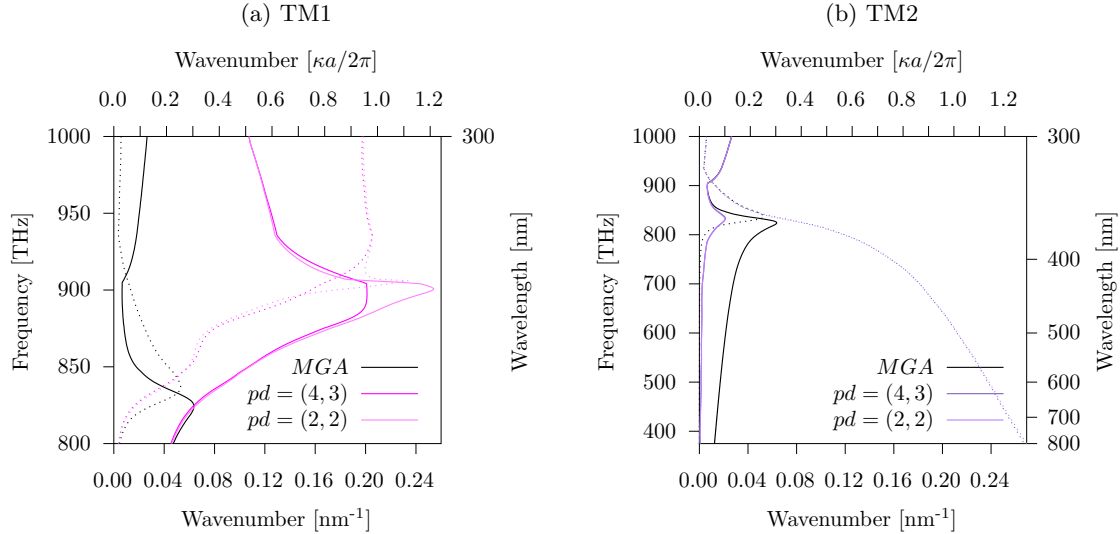
\begin{figure}[t]
    \centering
    \begin{subfigure}[b]{.48\textwidth}
    \centering
    \caption{TM1}\label{fig:pd_comp_TM1}
    \resizebox{.99\linewidth}{!}{\input{gnu/BS-pd22-43_TM1}}
    \end{subfigure} 
    \begin{subfigure}[b]{.48\textwidth}
    \centering
    \caption{TM2}\label{fig:pd_comp_TM2}
    \resizebox{.99\linewidth}{!}{\input{gnu/BS-pd22-43_TM2}}
    \end{subfigure} 
    \caption{Comparing $pd\eq(2,2)$, $pd\eq(4,3)$ and MGA [$\approx pd\eq(0,0)$] for the generic silver nanowire array with $R\eq\SI{10}{\nano\metre}$ and plane wave cut-off $N_G\eq800$.}
    \label{fig:polydegs_comp}
\end{figure}

We finally compare the full bandstructure obtained with pd\eq(2,2) with a well converged solution with pd\eq(4,3) and $N_G\eq800$ in \figref{fig:polydegs_comp}.
The TM2 solution is already well converged for pd\eq(2,2) over the whole spectrum, with a small deviation at longer wavelengths (\figref{fig:pd_comp_TM2}).
This can be seen as further evidence to our former comment that the solution is well converged, even where the numerical frequency error in \figref{fig:Convergence} suggests otherwise.
For the TM1 solution, we zoom in to the frequency region above \SI{800}{\tera\hertz} in \figref{fig:pd_comp_TM1}, below with the mode is well resolved with even the MGA model.
Close to \SI{900}{\tera\hertz}, above the MGA resonance, where the absolute value of $\kappa_z$ is largest, significant deviations between the pd\eq(2,2) and the converged pd\eq(4,3) solution are observed.
At these frequencies, the bandstructure is well outside the first Brillouin zone with boundary at $2\pi/(\sqrt{3}a)\,{\sim}\,\SI{0.12}{\per\nano\metre}$.
While the under-resolved solution reaches the extended $\Gamma$-point at approximately \SI{900}{\tera\hertz}, the real part of $\kappa_z$ plateaus at $\kappa_z'\,{\sim}\,\SI{0.2}{\per\nano\metre}$ in the converged solution.
Although not strictly applicable in the non-homogeneous multi-mode regime, the deviance of $\kappa_z''$ at and above \SI{900}{\tera\hertz} can be understood through application of the Kramers-Kronig relations using an effective refractive index $n_\text{eff}\deq\kappa_z/k_0$ \cite{sai2020designing}.
The kinks at approximately \SI{900}{\tera\hertz} and \SI{930}{\tera\hertz} in $\kappa_z'$ that are particularly prominent in the converged solution, can be traced back to the experimental permittivity data \cite{JCRI}, where a similar non-smooth behavior occurs at these frequencies.

\begin{figure}[t]
    \centering
    \begin{subfigure}[b]{.47\textwidth}
        \centering
        \caption{Real part $\kappa_z'$}\label{fig:Riemann_real}
        \resizebox{.99\linewidth}{!}{\includegraphics[bb=110 20 420 280,clip,width=\textwidth]{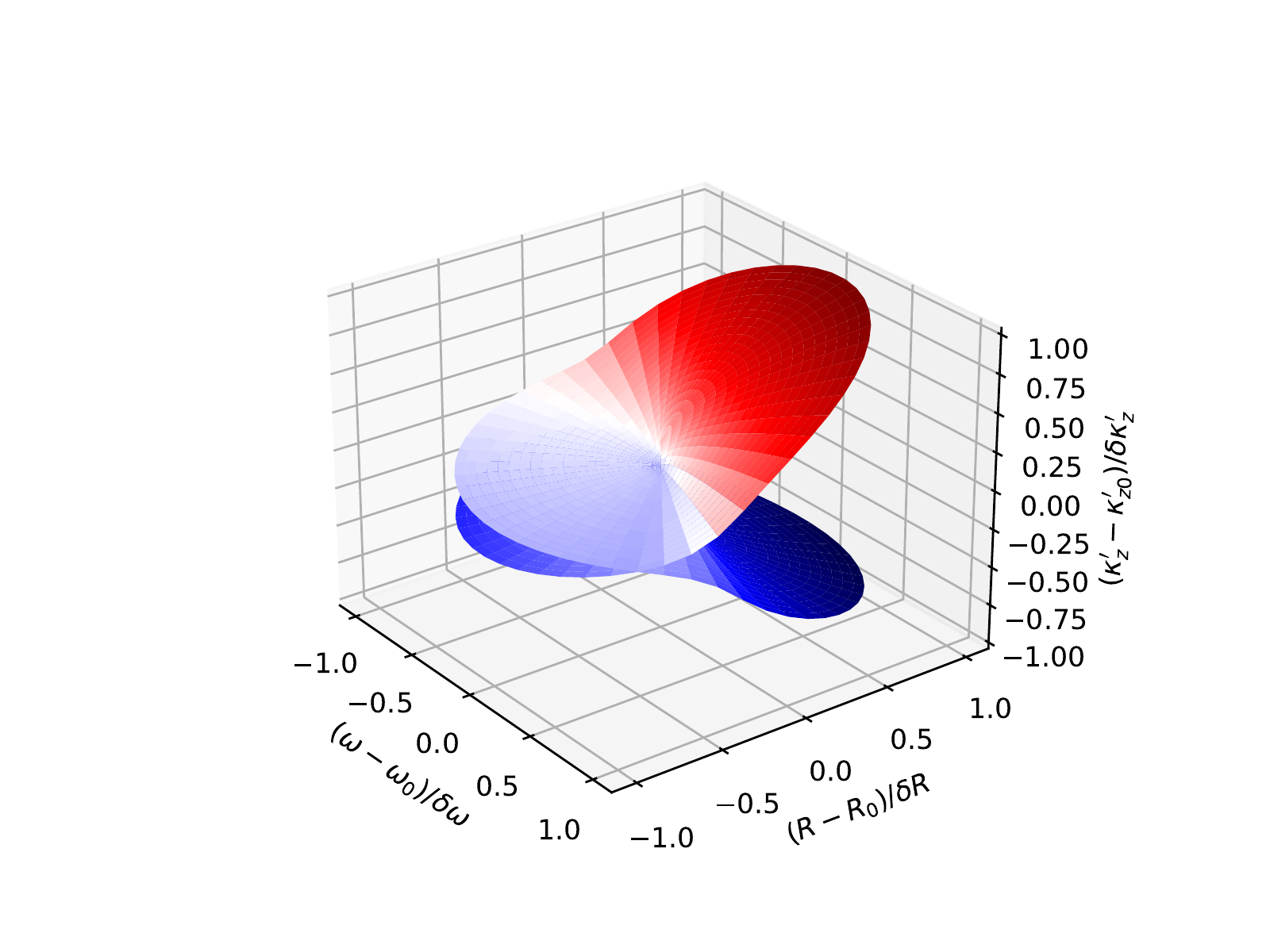}}
    \end{subfigure}
    \hfill
    \begin{subfigure}[b]{.49\textwidth}
        \centering
        \caption{Imaginary part $\kappa_z''$}\label{fig:Riemann_imag}
        \resizebox{.99\linewidth}{!}{\includegraphics[bb=110 20 420 280,clip,width=\textwidth]{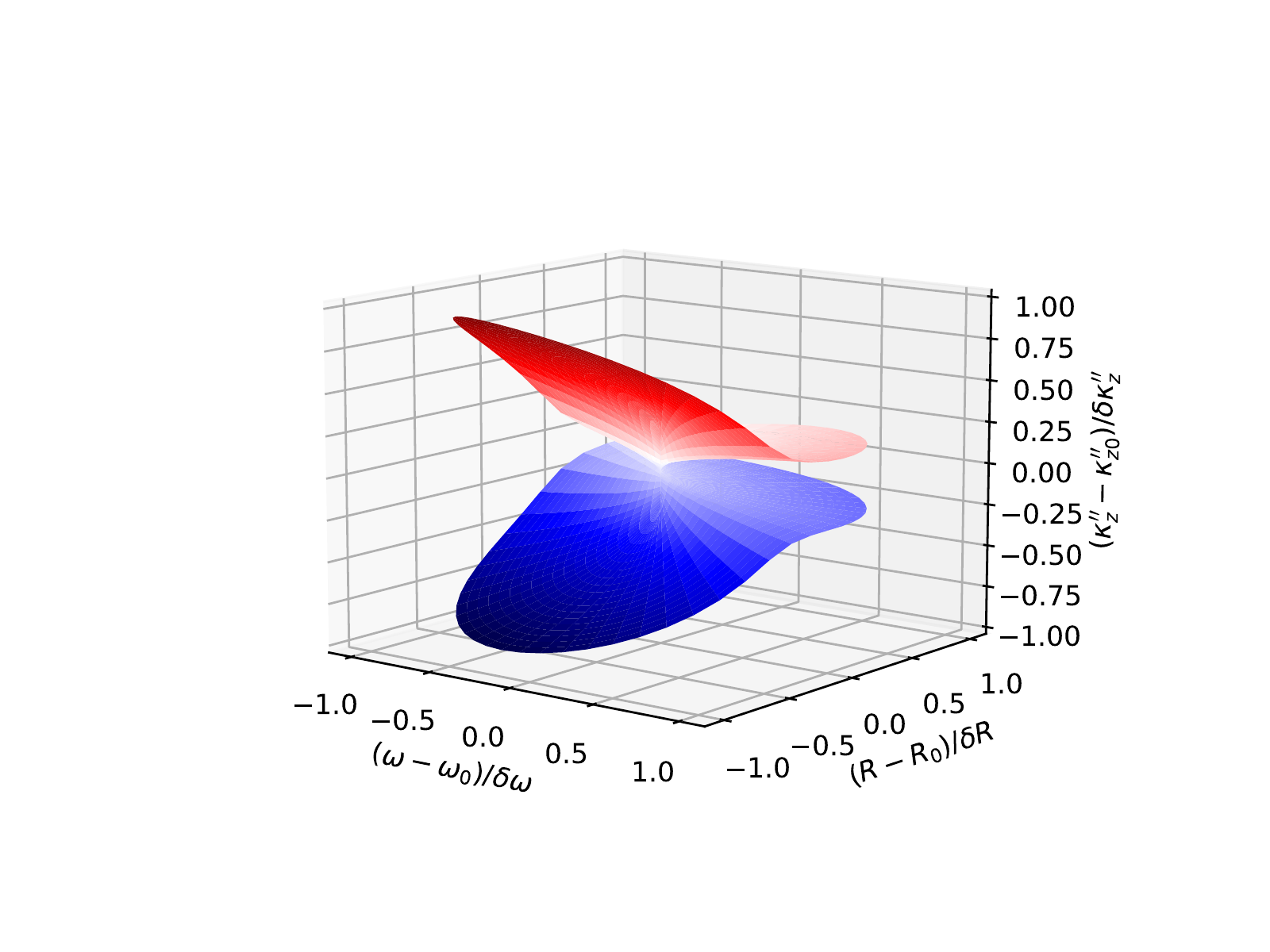}}
    \end{subfigure}
    \caption{
    Real part (a) and imaginary part (b) of the TM wave numbers $\kappa_z\eq\kappa_z'{+}\imath\kappa_z''$ as a function of cylinder radius $R$ and frequency $\nu\deq\omega/(2\pi)$ close to the band degeneracy at $R_0\,{\sim}\,\SI{8.775}{\nano\metre}$ and $\nu_0\,{\sim}\,\SI{851.25}{\tera\hertz}$ for a hexagonal silver nano-wire array with numerical parameters $N_G\eq50$ and pd\eq(2,2).
    The radii in parameter space are $\delta R\eq\SI{0.25}{\nano\metre}$ and $\delta\nu\eq\SI{1}{\tera\hertz}$.
    The wave number at the point of degeneracy is $k_{z0}\,{\sim}\,(0.0491{+}0.0545\imath)\si{\per\nano\metre}$, while its maximum extend is  $\delta k_{z}\,{\sim}\,(0.0272{+}0.0256\imath)\si{\per\nano\metre}$.}
    \label{fig:Riemann}
\end{figure}

\section{Non-trivial band topology from square-root-like Riemann surfaces} \label{sec:degen}

In \figref{fig:Au-Ag-radii}, we can see that the MGA formula predicts the dispersion relation of the TM1 mode well for a silver cylinder array with \SI{8}{\nano\metre} radius and a gold array with \SI{13}{\nano\metre} radius.
A second TM2 mode exists with large imaginary part of the eigenvalue $\kappa_z''$, but is physically less relevant.
Going to large radii, the behavior changes to the one reported in the main manuscript:
The TM1 mode is approximated well by the MGA formula for small frequencies, while the TM2 mode is approximated well for frequencies above the dipole resonance, while none is approximated well at frequencies close to the resonance.
A topological change must therefore occur in the $R$-$\omega$ parameter space in between, where the meaning of the somewhat arbitrary TM1/TM2 label changes at large frequencies.

To investigate the nature of the band topology, we first look at the nature of the eigenproblem at hand.
Since the linear operator in the NLEVP \eqref{eq:NLEVP} is analytical in $R$ and at least meromorphic in $\omega$,\footnote{It is analytical except where $k_0\eq|\kappa{+}\vec{G}|$, which can only happen at real valued $\kappa$ in our case.} we expect that same holds for the eigenvalues except at exceptional points, where two eigenvalues are degenerate \cite{kato1976perturbation}.
In order for the TM1/TM2 bands to interchange, we deduce that there must be an exceptional point associated with a non-trivial winding number in the parameter space.
This point can indeed be found slightly below $\SI{9}{\nano\metre}$ at approximately $\SI{750}{\tera\hertz}$.
Close to the exceptional point, the TM bandstructure topologically resembles the Riemann surface of the square-root function, that is both the real part and the imaginary part form self-intersection surfaces with a non-trivial winding number of $2$, as shown in \figref{fig:Riemann}.
Similar to the square-root function, the real part is doubly-degenerate along a straight line that emanates at the exceptional point in negative $R$ and positive $\omega$ direction, while the imaginary part is doubly-degenerate along a straight line in the opposite direction.
In contrast to the square-root function, the line of degeneracy does not define a mirror symmetry here.
At small radii, the imaginary parts of the two modes deviate substantially at all frequencies, deeming only one mode physically important.
The real parts on the other hand deviate strongly at large radii, but are confined to the inclination-normal Brillouin zone.
Both modes are physically relevant close to the resonance.

{\small
\bibliographystyle{unsrt}
\bibliography{bibliography}
}

%% file: gnu/BS-3modes_pd43.tex
\begingroup
  \makeatletter
  \providecommand\color[2][]{%
    \GenericError{(gnuplot) \space\space\space\@spaces}{%
      Package color not loaded in conjunction with
      terminal option `colourtext'%
    }{See the gnuplot documentation for explanation.%
    }{Either use 'blacktext' in gnuplot or load the package
      color.sty in LaTeX.}%
    \renewcommand\color[2][]{}%
  }%
  \providecommand\includegraphics[2][]{%
    \GenericError{(gnuplot) \space\space\space\@spaces}{%
      Package graphicx or graphics not loaded%
    }{See the gnuplot documentation for explanation.%
    }{The gnuplot epslatex terminal needs graphicx.sty or graphics.sty.}%
    \renewcommand\includegraphics[2][]{}%
  }%
  \providecommand\rotatebox[2]{#2}%
  \@ifundefined{ifGPcolor}{%
    \newif\ifGPcolor
    \GPcolortrue
  }{}%
  \@ifundefined{ifGPblacktext}{%
    \newif\ifGPblacktext
    \GPblacktexttrue
  }{}%
  \let\gplgaddtomacro\g@addto@macro
  \gdef\gplbacktext{}%
  \gdef\gplfronttext{}%
  \makeatother
  \ifGPblacktext
    \def\colorrgb#1{}%
    \def\colorgray#1{}%
  \else
    \ifGPcolor
      \def\colorrgb#1{\color[rgb]{#1}}%
      \def\colorgray#1{\color[gray]{#1}}%
      \expandafter\def\csname LTw\endcsname{\color{white}}%
      \expandafter\def\csname LTb\endcsname{\color{black}}%
      \expandafter\def\csname LTa\endcsname{\color{black}}%
      \expandafter\def\csname LT0\endcsname{\color[rgb]{1,0,0}}%
      \expandafter\def\csname LT1\endcsname{\color[rgb]{0,1,0}}%
      \expandafter\def\csname LT2\endcsname{\color[rgb]{0,0,1}}%
      \expandafter\def\csname LT3\endcsname{\color[rgb]{1,0,1}}%
      \expandafter\def\csname LT4\endcsname{\color[rgb]{0,1,1}}%
      \expandafter\def\csname LT5\endcsname{\color[rgb]{1,1,0}}%
      \expandafter\def\csname LT6\endcsname{\color[rgb]{0,0,0}}%
      \expandafter\def\csname LT7\endcsname{\color[rgb]{1,0.3,0}}%
      \expandafter\def\csname LT8\endcsname{\color[rgb]{0.5,0.5,0.5}}%
    \else
      \def\colorrgb#1{\color{black}}%
      \def\colorgray#1{\color[gray]{#1}}%
      \expandafter\def\csname LTw\endcsname{\color{white}}%
      \expandafter\def\csname LTb\endcsname{\color{black}}%
      \expandafter\def\csname LTa\endcsname{\color{black}}%
      \expandafter\def\csname LT0\endcsname{\color{black}}%
      \expandafter\def\csname LT1\endcsname{\color{black}}%
      \expandafter\def\csname LT2\endcsname{\color{black}}%
      \expandafter\def\csname LT3\endcsname{\color{black}}%
      \expandafter\def\csname LT4\endcsname{\color{black}}%
      \expandafter\def\csname LT5\endcsname{\color{black}}%
      \expandafter\def\csname LT6\endcsname{\color{black}}%
      \expandafter\def\csname LT7\endcsname{\color{black}}%
      \expandafter\def\csname LT8\endcsname{\color{black}}%
    \fi
  \fi
    \setlength{\unitlength}{0.0500bp}%
    \ifx\gptboxheight\undefined%
      \newlength{\gptboxheight}%
      \newlength{\gptboxwidth}%
      \newsavebox{\gptboxtext}%
    \fi%
    \setlength{\fboxrule}{0.5pt}%
    \setlength{\fboxsep}{1pt}%
    \definecolor{tbcol}{rgb}{1,1,1}%
\begin{picture}(7360.00,4520.00)%
    \gplgaddtomacro\gplbacktext{%
      \csname LTb\endcsname
      \put(816,972){\makebox(0,0)[r]{\strut{}400}}%
      \csname LTb\endcsname
      \put(816,1425){\makebox(0,0)[r]{\strut{}500}}%
      \csname LTb\endcsname
      \put(816,1877){\makebox(0,0)[r]{\strut{}600}}%
      \csname LTb\endcsname
      \put(816,2330){\makebox(0,0)[r]{\strut{}700}}%
      \csname LTb\endcsname
      \put(816,2782){\makebox(0,0)[r]{\strut{}800}}%
      \csname LTb\endcsname
      \put(816,3235){\makebox(0,0)[r]{\strut{}900}}%
      \csname LTb\endcsname
      \put(816,3687){\makebox(0,0)[r]{\strut{}1000}}%
      \csname LTb\endcsname
      \put(1016,525){\makebox(0,0){\strut{}0.00}}%
      \csname LTb\endcsname
      \put(1787,525){\makebox(0,0){\strut{}0.04}}%
      \csname LTb\endcsname
      \put(2557,525){\makebox(0,0){\strut{}0.08}}%
      \csname LTb\endcsname
      \put(3328,525){\makebox(0,0){\strut{}0.12}}%
      \csname LTb\endcsname
      \put(4099,525){\makebox(0,0){\strut{}0.16}}%
      \csname LTb\endcsname
      \put(4869,525){\makebox(0,0){\strut{}0.20}}%
      \csname LTb\endcsname
      \put(5640,525){\makebox(0,0){\strut{}0.24}}%
      \csname LTb\endcsname
      \put(6365,3687){\makebox(0,0)[l]{\strut{}300}}%
      \csname LTb\endcsname
      \put(6365,2556){\makebox(0,0)[l]{\strut{}400}}%
      \csname LTb\endcsname
      \put(6365,1877){\makebox(0,0)[l]{\strut{}500}}%
      \csname LTb\endcsname
      \put(6365,1425){\makebox(0,0)[l]{\strut{}600}}%
      \csname LTb\endcsname
      \put(6365,1101){\makebox(0,0)[l]{\strut{}700}}%
      \csname LTb\endcsname
      \put(6365,859){\makebox(0,0)[l]{\strut{}800}}%
      \csname LTb\endcsname
      \put(1016,4021){\makebox(0,0){\strut{}0.0}}%
      \csname LTb\endcsname
      \put(1823,4021){\makebox(0,0){\strut{}0.2}}%
      \csname LTb\endcsname
      \put(2630,4021){\makebox(0,0){\strut{}0.4}}%
      \csname LTb\endcsname
      \put(3437,4021){\makebox(0,0){\strut{}0.6}}%
      \csname LTb\endcsname
      \put(4244,4021){\makebox(0,0){\strut{}0.8}}%
      \csname LTb\endcsname
      \put(5051,4021){\makebox(0,0){\strut{}1.0}}%
      \csname LTb\endcsname
      \put(5858,4021){\makebox(0,0){\strut{}1.2}}%
    }%
    \gplgaddtomacro\gplfronttext{%
      \csname LTb\endcsname
      \put(198,2273){\rotatebox{-270}{\makebox(0,0){\strut{}Frequency [THz]}}}%
      \csname LTb\endcsname
      \put(6904,2273){\rotatebox{-270}{\makebox(0,0){\strut{}Wavelength [nm]}}}%
      \csname LTb\endcsname
      \put(3590,167){\makebox(0,0){\strut{}Wavenumber [nm\textsuperscript{-1}]}}%
      \csname LTb\endcsname
      \put(3590,4379){\makebox(0,0){\strut{}Wavenumber [$\kappa a / 2 \pi$]}}%
      \csname LTb\endcsname
      \put(3595,1790){\makebox(0,0)[r]{\strut{}$\kappa_{\rm{TE}}$}}%
      \csname LTb\endcsname
      \put(3595,1551){\makebox(0,0)[r]{\strut{}$\kappa_{\rm{TM}}^{\rm{MG}}$}}%
      \csname LTb\endcsname
      \put(3595,1312){\makebox(0,0)[r]{\strut{}$\kappa_{\rm{TM}1}$}}%
      \csname LTb\endcsname
      \put(3595,1073){\makebox(0,0)[r]{\strut{}$\kappa_{\rm{TM}2}$}}%
    }%
    \gplbacktext
    \put(0,0){\includegraphics[width={368.00bp},height={226.00bp}]{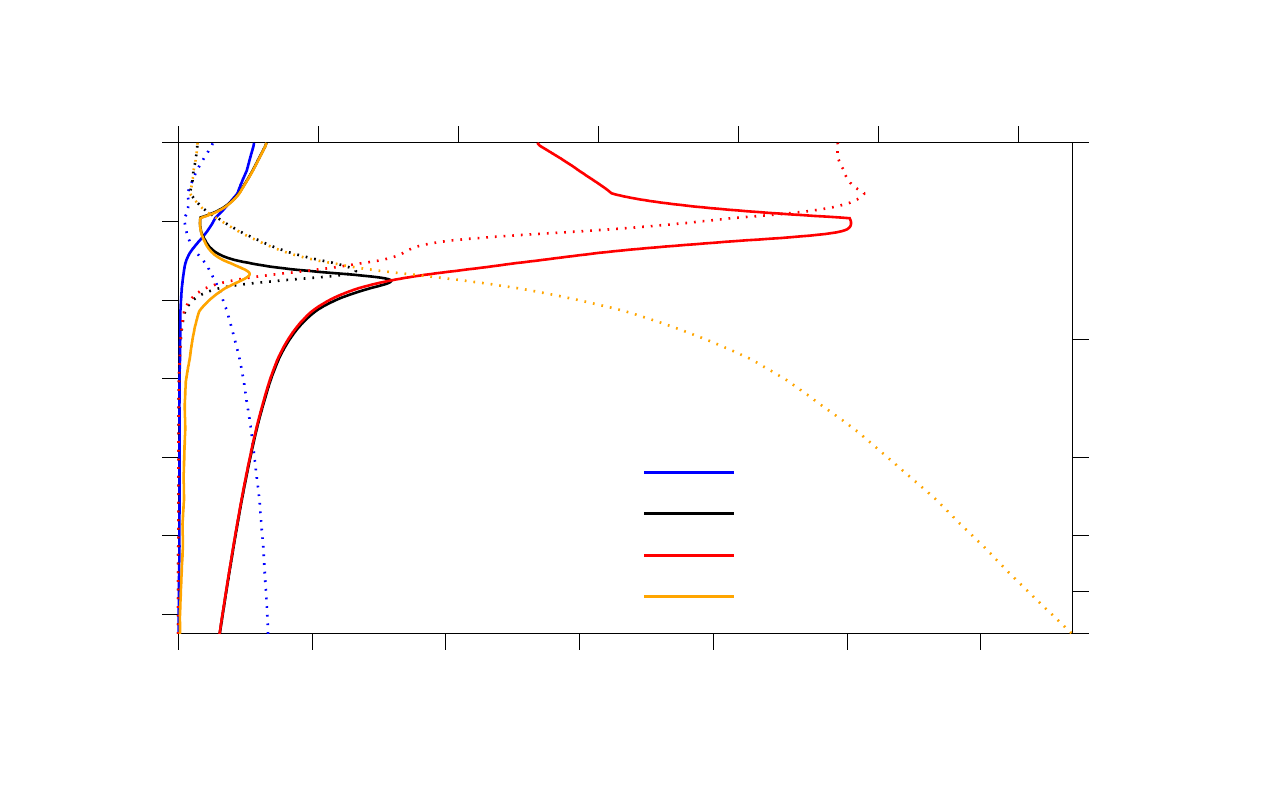}}%
    \gplfronttext
  \end{picture}%
\endgroup

%% file: gnu/BS-R13.tex
\begingroup
  \inputencoding{cp1252}%
  \makeatletter
  \providecommand\color[2][]{%
    \GenericError{(gnuplot) \space\space\space\@spaces}{%
      Package color not loaded in conjunction with
      terminal option `colourtext'%
    }{See the gnuplot documentation for explanation.%
    }{Either use 'blacktext' in gnuplot or load the package
      color.sty in LaTeX.}%
    \renewcommand\color[2][]{}%
  }%
  \providecommand\includegraphics[2][]{%
    \GenericError{(gnuplot) \space\space\space\@spaces}{%
      Package graphicx or graphics not loaded%
    }{See the gnuplot documentation for explanation.%
    }{The gnuplot epslatex terminal needs graphicx.sty or graphics.sty.}%
    \renewcommand\includegraphics[2][]{}%
  }%
  \providecommand\rotatebox[2]{#2}%
  \@ifundefined{ifGPcolor}{%
    \newif\ifGPcolor
    \GPcolortrue
  }{}%
  \@ifundefined{ifGPblacktext}{%
    \newif\ifGPblacktext
    \GPblacktexttrue
  }{}%
  \let\gplgaddtomacro\g@addto@macro
  \gdef\gplbacktext{}%
  \gdef\gplfronttext{}%
  \makeatother
  \ifGPblacktext
    \def\colorrgb#1{}%
    \def\colorgray#1{}%
  \else
    \ifGPcolor
      \def\colorrgb#1{\color[rgb]{#1}}%
      \def\colorgray#1{\color[gray]{#1}}%
      \expandafter\def\csname LTw\endcsname{\color{white}}%
      \expandafter\def\csname LTb\endcsname{\color{black}}%
      \expandafter\def\csname LTa\endcsname{\color{black}}%
      \expandafter\def\csname LT0\endcsname{\color[rgb]{1,0,0}}%
      \expandafter\def\csname LT1\endcsname{\color[rgb]{0,1,0}}%
      \expandafter\def\csname LT2\endcsname{\color[rgb]{0,0,1}}%
      \expandafter\def\csname LT3\endcsname{\color[rgb]{1,0,1}}%
      \expandafter\def\csname LT4\endcsname{\color[rgb]{0,1,1}}%
      \expandafter\def\csname LT5\endcsname{\color[rgb]{1,1,0}}%
      \expandafter\def\csname LT6\endcsname{\color[rgb]{0,0,0}}%
      \expandafter\def\csname LT7\endcsname{\color[rgb]{1,0.3,0}}%
      \expandafter\def\csname LT8\endcsname{\color[rgb]{0.5,0.5,0.5}}%
    \else
      \def\colorrgb#1{\color{black}}%
      \def\colorgray#1{\color[gray]{#1}}%
      \expandafter\def\csname LTw\endcsname{\color{white}}%
      \expandafter\def\csname LTb\endcsname{\color{black}}%
      \expandafter\def\csname LTa\endcsname{\color{black}}%
      \expandafter\def\csname LT0\endcsname{\color{black}}%
      \expandafter\def\csname LT1\endcsname{\color{black}}%
      \expandafter\def\csname LT2\endcsname{\color{black}}%
      \expandafter\def\csname LT3\endcsname{\color{black}}%
      \expandafter\def\csname LT4\endcsname{\color{black}}%
      \expandafter\def\csname LT5\endcsname{\color{black}}%
      \expandafter\def\csname LT6\endcsname{\color{black}}%
      \expandafter\def\csname LT7\endcsname{\color{black}}%
      \expandafter\def\csname LT8\endcsname{\color{black}}%
    \fi
  \fi
    \setlength{\unitlength}{0.0500bp}%
    \ifx\gptboxheight\undefined%
      \newlength{\gptboxheight}%
      \newlength{\gptboxwidth}%
      \newsavebox{\gptboxtext}%
    \fi%
    \setlength{\fboxrule}{0.5pt}%
    \setlength{\fboxsep}{1pt}%
    \definecolor{tbcol}{rgb}{1,1,1}%
\begin{picture}(5180.00,4480.00)%
    \gplgaddtomacro\gplbacktext{%
      \csname LTb\endcsname
      \put(816,971){\makebox(0,0)[r]{\strut{}400}}%
      \csname LTb\endcsname
      \put(816,1417){\makebox(0,0)[r]{\strut{}500}}%
      \csname LTb\endcsname
      \put(816,1863){\makebox(0,0)[r]{\strut{}600}}%
      \csname LTb\endcsname
      \put(816,2309){\makebox(0,0)[r]{\strut{}700}}%
      \csname LTb\endcsname
      \put(816,2755){\makebox(0,0)[r]{\strut{}800}}%
      \csname LTb\endcsname
      \put(816,3201){\makebox(0,0)[r]{\strut{}900}}%
      \csname LTb\endcsname
      \put(816,3647){\makebox(0,0)[r]{\strut{}1000}}%
      \csname LTb\endcsname
      \put(1016,525){\makebox(0,0){\strut{}0.00}}%
      \csname LTb\endcsname
      \put(1864,525){\makebox(0,0){\strut{}0.04}}%
      \csname LTb\endcsname
      \put(2713,525){\makebox(0,0){\strut{}0.08}}%
      \csname LTb\endcsname
      \put(3561,525){\makebox(0,0){\strut{}0.12}}%
      \csname LTb\endcsname
      \put(4185,3647){\makebox(0,0)[l]{\strut{}300}}%
      \csname LTb\endcsname
      \put(4185,2532){\makebox(0,0)[l]{\strut{}400}}%
      \csname LTb\endcsname
      \put(4185,1863){\makebox(0,0)[l]{\strut{}500}}%
      \csname LTb\endcsname
      \put(4185,1417){\makebox(0,0)[l]{\strut{}600}}%
      \csname LTb\endcsname
      \put(4185,1098){\makebox(0,0)[l]{\strut{}700}}%
      \csname LTb\endcsname
      \put(4185,859){\makebox(0,0)[l]{\strut{}800}}%
      \csname LTb\endcsname
      \put(1016,3981){\makebox(0,0){\strut{}0.0}}%
      \csname LTb\endcsname
      \put(1460,3981){\makebox(0,0){\strut{}0.1}}%
      \csname LTb\endcsname
      \put(1904,3981){\makebox(0,0){\strut{}0.2}}%
      \csname LTb\endcsname
      \put(2348,3981){\makebox(0,0){\strut{}0.3}}%
      \csname LTb\endcsname
      \put(2793,3981){\makebox(0,0){\strut{}0.4}}%
      \csname LTb\endcsname
      \put(3237,3981){\makebox(0,0){\strut{}0.5}}%
      \csname LTb\endcsname
      \put(3681,3981){\makebox(0,0){\strut{}0.6}}%
    }%
    \gplgaddtomacro\gplfronttext{%
      \csname LTb\endcsname
      \put(198,2253){\rotatebox{-270}{\makebox(0,0){\strut{}Frequency [THz]}}}%
      \csname LTb\endcsname
      \put(4724,2253){\rotatebox{-270}{\makebox(0,0){\strut{}Wavelength [nm]}}}%
      \csname LTb\endcsname
      \put(2500,167){\makebox(0,0){\strut{}Wavenumber [nm\textsuperscript{-1}]}}%
      \csname LTb\endcsname
      \put(2500,4339){\makebox(0,0){\strut{}Wavenumber [$\kappa a / 2 \pi$]}}%
      \csname LTb\endcsname
      \put(3155,1790){\makebox(0,0)[r]{\strut{}$\kappa_{TM}^{MG}$}}%
      \csname LTb\endcsname
      \put(3155,1551){\makebox(0,0)[r]{\strut{}$\kappa_{TE}$}}%
      \csname LTb\endcsname
      \put(3155,1312){\makebox(0,0)[r]{\strut{}$\kappa_{TM1}$}}%
      \csname LTb\endcsname
      \put(3155,1073){\makebox(0,0)[r]{\strut{}$\kappa_{TM2}$}}%
    }%
    \gplbacktext
    \put(0,0){\includegraphics[width={259.00bp},height={224.00bp}]{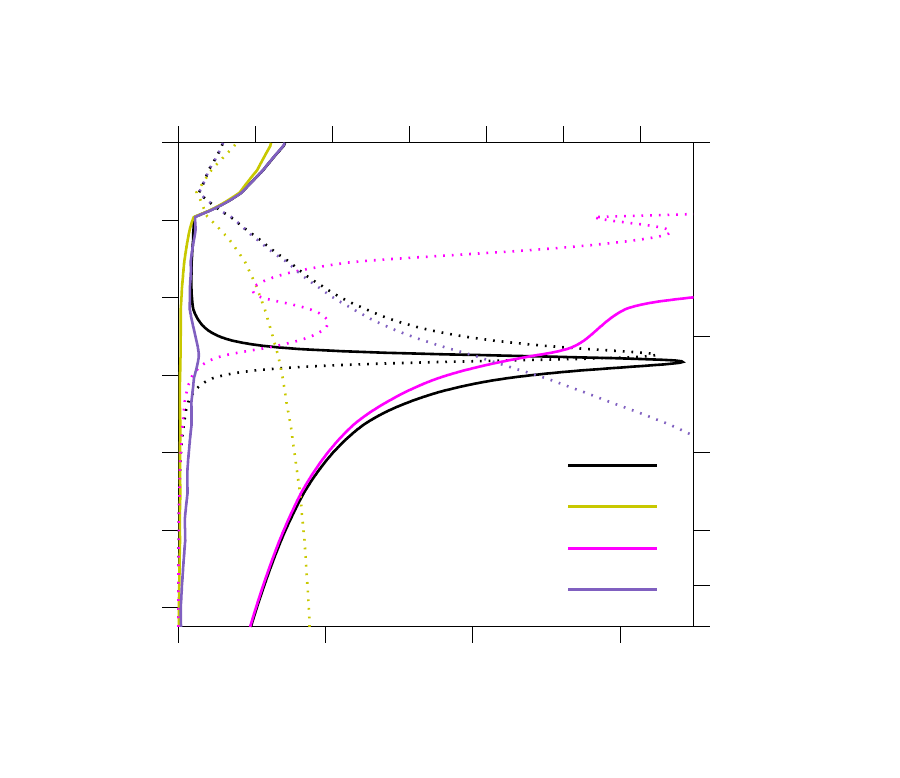}}%
    \gplfronttext
  \end{picture}%
\endgroup

%% file: gnu/BS-R8.tex
\begingroup
  \inputencoding{cp1252}%
  \makeatletter
  \providecommand\color[2][]{%
    \GenericError{(gnuplot) \space\space\space\@spaces}{%
      Package color not loaded in conjunction with
      terminal option `colourtext'%
    }{See the gnuplot documentation for explanation.%
    }{Either use 'blacktext' in gnuplot or load the package
      color.sty in LaTeX.}%
    \renewcommand\color[2][]{}%
  }%
  \providecommand\includegraphics[2][]{%
    \GenericError{(gnuplot) \space\space\space\@spaces}{%
      Package graphicx or graphics not loaded%
    }{See the gnuplot documentation for explanation.%
    }{The gnuplot epslatex terminal needs graphicx.sty or graphics.sty.}%
    \renewcommand\includegraphics[2][]{}%
  }%
  \providecommand\rotatebox[2]{#2}%
  \@ifundefined{ifGPcolor}{%
    \newif\ifGPcolor
    \GPcolortrue
  }{}%
  \@ifundefined{ifGPblacktext}{%
    \newif\ifGPblacktext
    \GPblacktexttrue
  }{}%
  \let\gplgaddtomacro\g@addto@macro
  \gdef\gplbacktext{}%
  \gdef\gplfronttext{}%
  \makeatother
  \ifGPblacktext
    \def\colorrgb#1{}%
    \def\colorgray#1{}%
  \else
    \ifGPcolor
      \def\colorrgb#1{\color[rgb]{#1}}%
      \def\colorgray#1{\color[gray]{#1}}%
      \expandafter\def\csname LTw\endcsname{\color{white}}%
      \expandafter\def\csname LTb\endcsname{\color{black}}%
      \expandafter\def\csname LTa\endcsname{\color{black}}%
      \expandafter\def\csname LT0\endcsname{\color[rgb]{1,0,0}}%
      \expandafter\def\csname LT1\endcsname{\color[rgb]{0,1,0}}%
      \expandafter\def\csname LT2\endcsname{\color[rgb]{0,0,1}}%
      \expandafter\def\csname LT3\endcsname{\color[rgb]{1,0,1}}%
      \expandafter\def\csname LT4\endcsname{\color[rgb]{0,1,1}}%
      \expandafter\def\csname LT5\endcsname{\color[rgb]{1,1,0}}%
      \expandafter\def\csname LT6\endcsname{\color[rgb]{0,0,0}}%
      \expandafter\def\csname LT7\endcsname{\color[rgb]{1,0.3,0}}%
      \expandafter\def\csname LT8\endcsname{\color[rgb]{0.5,0.5,0.5}}%
    \else
      \def\colorrgb#1{\color{black}}%
      \def\colorgray#1{\color[gray]{#1}}%
      \expandafter\def\csname LTw\endcsname{\color{white}}%
      \expandafter\def\csname LTb\endcsname{\color{black}}%
      \expandafter\def\csname LTa\endcsname{\color{black}}%
      \expandafter\def\csname LT0\endcsname{\color{black}}%
      \expandafter\def\csname LT1\endcsname{\color{black}}%
      \expandafter\def\csname LT2\endcsname{\color{black}}%
      \expandafter\def\csname LT3\endcsname{\color{black}}%
      \expandafter\def\csname LT4\endcsname{\color{black}}%
      \expandafter\def\csname LT5\endcsname{\color{black}}%
      \expandafter\def\csname LT6\endcsname{\color{black}}%
      \expandafter\def\csname LT7\endcsname{\color{black}}%
      \expandafter\def\csname LT8\endcsname{\color{black}}%
    \fi
  \fi
    \setlength{\unitlength}{0.0500bp}%
    \ifx\gptboxheight\undefined%
      \newlength{\gptboxheight}%
      \newlength{\gptboxwidth}%
      \newsavebox{\gptboxtext}%
    \fi%
    \setlength{\fboxrule}{0.5pt}%
    \setlength{\fboxsep}{1pt}%
    \definecolor{tbcol}{rgb}{1,1,1}%
\begin{picture}(5180.00,4480.00)%
    \gplgaddtomacro\gplbacktext{%
      \csname LTb\endcsname
      \put(816,971){\makebox(0,0)[r]{\strut{}400}}%
      \csname LTb\endcsname
      \put(816,1417){\makebox(0,0)[r]{\strut{}500}}%
      \csname LTb\endcsname
      \put(816,1863){\makebox(0,0)[r]{\strut{}600}}%
      \csname LTb\endcsname
      \put(816,2309){\makebox(0,0)[r]{\strut{}700}}%
      \csname LTb\endcsname
      \put(816,2755){\makebox(0,0)[r]{\strut{}800}}%
      \csname LTb\endcsname
      \put(816,3201){\makebox(0,0)[r]{\strut{}900}}%
      \csname LTb\endcsname
      \put(816,3647){\makebox(0,0)[r]{\strut{}1000}}%
      \csname LTb\endcsname
      \put(1016,525){\makebox(0,0){\strut{}0.00}}%
      \csname LTb\endcsname
      \put(1864,525){\makebox(0,0){\strut{}0.02}}%
      \csname LTb\endcsname
      \put(2713,525){\makebox(0,0){\strut{}0.04}}%
      \csname LTb\endcsname
      \put(3561,525){\makebox(0,0){\strut{}0.06}}%
      \csname LTb\endcsname
      \put(4185,3647){\makebox(0,0)[l]{\strut{}300}}%
      \csname LTb\endcsname
      \put(4185,2532){\makebox(0,0)[l]{\strut{}400}}%
      \csname LTb\endcsname
      \put(4185,1863){\makebox(0,0)[l]{\strut{}500}}%
      \csname LTb\endcsname
      \put(4185,1417){\makebox(0,0)[l]{\strut{}600}}%
      \csname LTb\endcsname
      \put(4185,1098){\makebox(0,0)[l]{\strut{}700}}%
      \csname LTb\endcsname
      \put(4185,859){\makebox(0,0)[l]{\strut{}800}}%
      \csname LTb\endcsname
      \put(1016,3981){\makebox(0,0){\strut{}0.0}}%
      \csname LTb\endcsname
      \put(1904,3981){\makebox(0,0){\strut{}0.1}}%
      \csname LTb\endcsname
      \put(2793,3981){\makebox(0,0){\strut{}0.2}}%
      \csname LTb\endcsname
      \put(3681,3981){\makebox(0,0){\strut{}0.3}}%
    }%
    \gplgaddtomacro\gplfronttext{%
      \csname LTb\endcsname
      \put(198,2253){\rotatebox{-270}{\makebox(0,0){\strut{}Frequency [THz]}}}%
      \csname LTb\endcsname
      \put(4724,2253){\rotatebox{-270}{\makebox(0,0){\strut{}Wavelength [nm]}}}%
      \csname LTb\endcsname
      \put(2500,167){\makebox(0,0){\strut{}Wavenumber [nm\textsuperscript{-1}]}}%
      \csname LTb\endcsname
      \put(2500,4339){\makebox(0,0){\strut{}Wavenumber [$\kappa a / 2 \pi$]}}%
      \csname LTb\endcsname
      \put(3155,1551){\makebox(0,0)[r]{\strut{}$\kappa_{TM}^{MG}$}}%
      \csname LTb\endcsname
      \put(3155,1312){\makebox(0,0)[r]{\strut{}$\kappa_{TM1}$}}%
      \csname LTb\endcsname
      \put(3155,1073){\makebox(0,0)[r]{\strut{}$\kappa_{TM2}$}}%
    }%
    \gplbacktext
    \put(0,0){\includegraphics[width={259.00bp},height={224.00bp}]{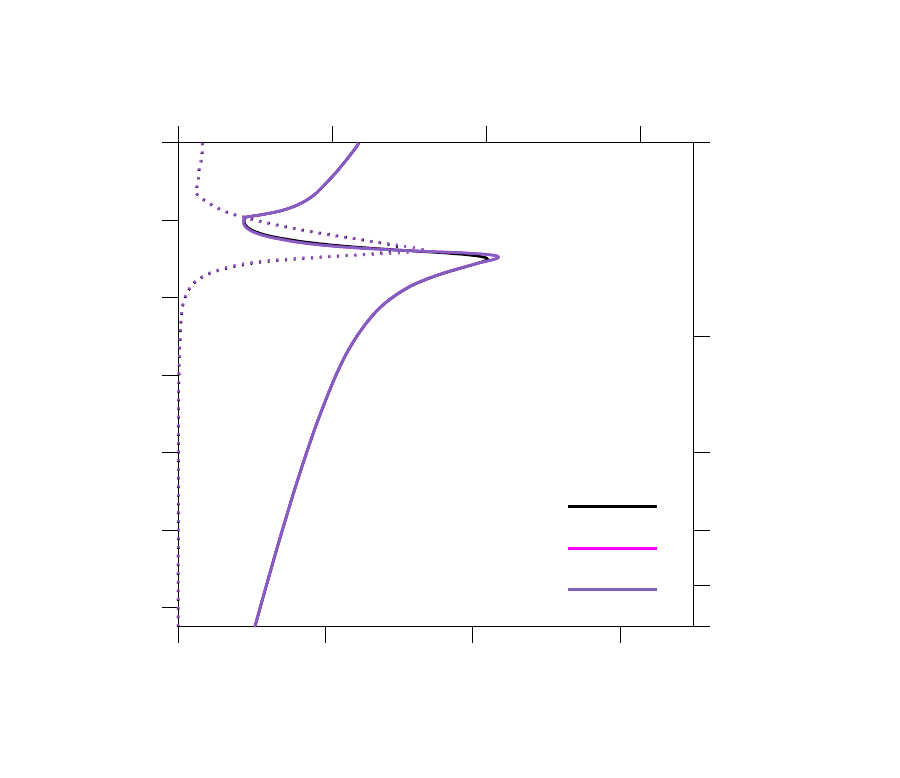}}%
    \gplfronttext
  \end{picture}%
\endgroup

%% file: gnu/BS-AuR13.tex
\begingroup
  \inputencoding{cp1252}%
  \makeatletter
  \providecommand\color[2][]{%
    \GenericError{(gnuplot) \space\space\space\@spaces}{%
      Package color not loaded in conjunction with
      terminal option `colourtext'%
    }{See the gnuplot documentation for explanation.%
    }{Either use 'blacktext' in gnuplot or load the package
      color.sty in LaTeX.}%
    \renewcommand\color[2][]{}%
  }%
  \providecommand\includegraphics[2][]{%
    \GenericError{(gnuplot) \space\space\space\@spaces}{%
      Package graphicx or graphics not loaded%
    }{See the gnuplot documentation for explanation.%
    }{The gnuplot epslatex terminal needs graphicx.sty or graphics.sty.}%
    \renewcommand\includegraphics[2][]{}%
  }%
  \providecommand\rotatebox[2]{#2}%
  \@ifundefined{ifGPcolor}{%
    \newif\ifGPcolor
    \GPcolortrue
  }{}%
  \@ifundefined{ifGPblacktext}{%
    \newif\ifGPblacktext
    \GPblacktexttrue
  }{}%
  \let\gplgaddtomacro\g@addto@macro
  \gdef\gplbacktext{}%
  \gdef\gplfronttext{}%
  \makeatother
  \ifGPblacktext
    \def\colorrgb#1{}%
    \def\colorgray#1{}%
  \else
    \ifGPcolor
      \def\colorrgb#1{\color[rgb]{#1}}%
      \def\colorgray#1{\color[gray]{#1}}%
      \expandafter\def\csname LTw\endcsname{\color{white}}%
      \expandafter\def\csname LTb\endcsname{\color{black}}%
      \expandafter\def\csname LTa\endcsname{\color{black}}%
      \expandafter\def\csname LT0\endcsname{\color[rgb]{1,0,0}}%
      \expandafter\def\csname LT1\endcsname{\color[rgb]{0,1,0}}%
      \expandafter\def\csname LT2\endcsname{\color[rgb]{0,0,1}}%
      \expandafter\def\csname LT3\endcsname{\color[rgb]{1,0,1}}%
      \expandafter\def\csname LT4\endcsname{\color[rgb]{0,1,1}}%
      \expandafter\def\csname LT5\endcsname{\color[rgb]{1,1,0}}%
      \expandafter\def\csname LT6\endcsname{\color[rgb]{0,0,0}}%
      \expandafter\def\csname LT7\endcsname{\color[rgb]{1,0.3,0}}%
      \expandafter\def\csname LT8\endcsname{\color[rgb]{0.5,0.5,0.5}}%
    \else
      \def\colorrgb#1{\color{black}}%
      \def\colorgray#1{\color[gray]{#1}}%
      \expandafter\def\csname LTw\endcsname{\color{white}}%
      \expandafter\def\csname LTb\endcsname{\color{black}}%
      \expandafter\def\csname LTa\endcsname{\color{black}}%
      \expandafter\def\csname LT0\endcsname{\color{black}}%
      \expandafter\def\csname LT1\endcsname{\color{black}}%
      \expandafter\def\csname LT2\endcsname{\color{black}}%
      \expandafter\def\csname LT3\endcsname{\color{black}}%
      \expandafter\def\csname LT4\endcsname{\color{black}}%
      \expandafter\def\csname LT5\endcsname{\color{black}}%
      \expandafter\def\csname LT6\endcsname{\color{black}}%
      \expandafter\def\csname LT7\endcsname{\color{black}}%
      \expandafter\def\csname LT8\endcsname{\color{black}}%
    \fi
  \fi
    \setlength{\unitlength}{0.0500bp}%
    \ifx\gptboxheight\undefined%
      \newlength{\gptboxheight}%
      \newlength{\gptboxwidth}%
      \newsavebox{\gptboxtext}%
    \fi%
    \setlength{\fboxrule}{0.5pt}%
    \setlength{\fboxsep}{1pt}%
    \definecolor{tbcol}{rgb}{1,1,1}%
\begin{picture}(5180.00,4480.00)%
    \gplgaddtomacro\gplbacktext{%
      \csname LTb\endcsname
      \put(816,971){\makebox(0,0)[r]{\strut{}400}}%
      \csname LTb\endcsname
      \put(816,1417){\makebox(0,0)[r]{\strut{}500}}%
      \csname LTb\endcsname
      \put(816,1863){\makebox(0,0)[r]{\strut{}600}}%
      \csname LTb\endcsname
      \put(816,2309){\makebox(0,0)[r]{\strut{}700}}%
      \csname LTb\endcsname
      \put(816,2755){\makebox(0,0)[r]{\strut{}800}}%
      \csname LTb\endcsname
      \put(816,3201){\makebox(0,0)[r]{\strut{}900}}%
      \csname LTb\endcsname
      \put(816,3647){\makebox(0,0)[r]{\strut{}1000}}%
      \csname LTb\endcsname
      \put(1016,525){\makebox(0,0){\strut{}0.00}}%
      \csname LTb\endcsname
      \put(1864,525){\makebox(0,0){\strut{}0.02}}%
      \csname LTb\endcsname
      \put(2713,525){\makebox(0,0){\strut{}0.04}}%
      \csname LTb\endcsname
      \put(3561,525){\makebox(0,0){\strut{}0.06}}%
      \csname LTb\endcsname
      \put(4185,3647){\makebox(0,0)[l]{\strut{}300}}%
      \csname LTb\endcsname
      \put(4185,2532){\makebox(0,0)[l]{\strut{}400}}%
      \csname LTb\endcsname
      \put(4185,1863){\makebox(0,0)[l]{\strut{}500}}%
      \csname LTb\endcsname
      \put(4185,1417){\makebox(0,0)[l]{\strut{}600}}%
      \csname LTb\endcsname
      \put(4185,1098){\makebox(0,0)[l]{\strut{}700}}%
      \csname LTb\endcsname
      \put(4185,859){\makebox(0,0)[l]{\strut{}800}}%
      \csname LTb\endcsname
      \put(1016,3981){\makebox(0,0){\strut{}0.0}}%
      \csname LTb\endcsname
      \put(1904,3981){\makebox(0,0){\strut{}0.1}}%
      \csname LTb\endcsname
      \put(2793,3981){\makebox(0,0){\strut{}0.2}}%
      \csname LTb\endcsname
      \put(3681,3981){\makebox(0,0){\strut{}0.3}}%
    }%
    \gplgaddtomacro\gplfronttext{%
      \csname LTb\endcsname
      \put(198,2253){\rotatebox{-270}{\makebox(0,0){\strut{}Frequency [THz]}}}%
      \csname LTb\endcsname
      \put(4724,2253){\rotatebox{-270}{\makebox(0,0){\strut{}Wavelength [nm]}}}%
      \csname LTb\endcsname
      \put(2500,167){\makebox(0,0){\strut{}Wavenumber [nm\textsuperscript{-1}]}}%
      \csname LTb\endcsname
      \put(2500,4339){\makebox(0,0){\strut{}Wavenumber [$\kappa a / 2 \pi$]}}%
      \csname LTb\endcsname
      \put(3155,1551){\makebox(0,0)[r]{\strut{}$\kappa_{TM}^{MG}$}}%
      \csname LTb\endcsname
      \put(3155,1312){\makebox(0,0)[r]{\strut{}$\kappa_{TM1}$}}%
      \csname LTb\endcsname
      \put(3155,1073){\makebox(0,0)[r]{\strut{}$\kappa_{TM2}$}}%
    }%
    \gplbacktext
    \put(0,0){\includegraphics[width={259.00bp},height={224.00bp}]{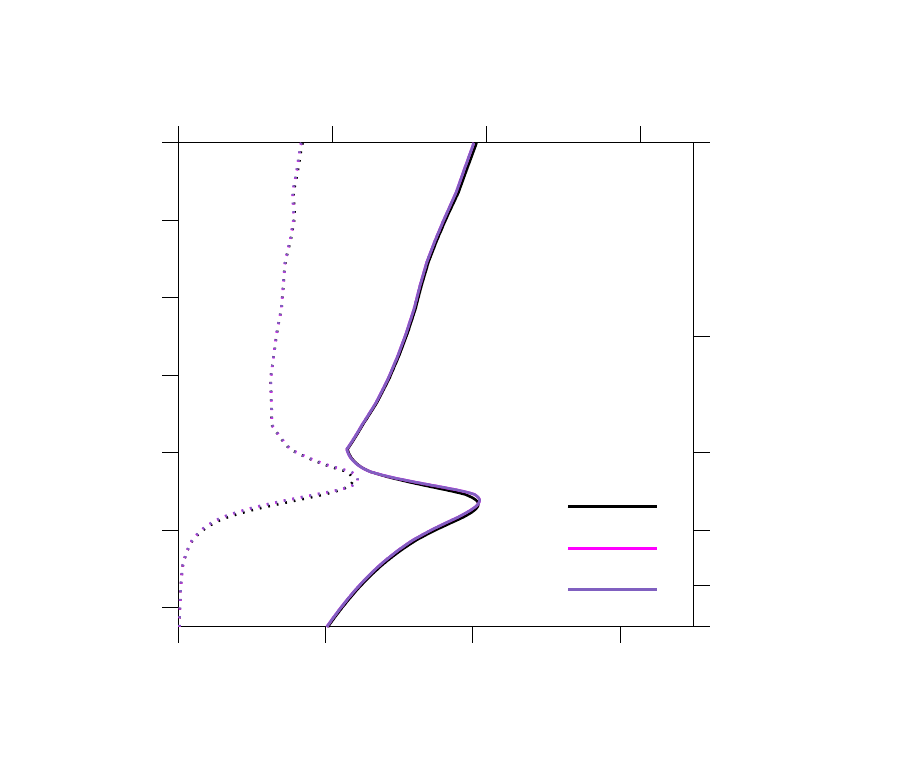}}%
    \gplfronttext
  \end{picture}%
\endgroup

%% file: gnu/BS-AuR14.tex
\begingroup
  \inputencoding{cp1252}%
  \makeatletter
  \providecommand\color[2][]{%
    \GenericError{(gnuplot) \space\space\space\@spaces}{%
      Package color not loaded in conjunction with
      terminal option `colourtext'%
    }{See the gnuplot documentation for explanation.%
    }{Either use 'blacktext' in gnuplot or load the package
      color.sty in LaTeX.}%
    \renewcommand\color[2][]{}%
  }%
  \providecommand\includegraphics[2][]{%
    \GenericError{(gnuplot) \space\space\space\@spaces}{%
      Package graphicx or graphics not loaded%
    }{See the gnuplot documentation for explanation.%
    }{The gnuplot epslatex terminal needs graphicx.sty or graphics.sty.}%
    \renewcommand\includegraphics[2][]{}%
  }%
  \providecommand\rotatebox[2]{#2}%
  \@ifundefined{ifGPcolor}{%
    \newif\ifGPcolor
    \GPcolortrue
  }{}%
  \@ifundefined{ifGPblacktext}{%
    \newif\ifGPblacktext
    \GPblacktexttrue
  }{}%
  \let\gplgaddtomacro\g@addto@macro
  \gdef\gplbacktext{}%
  \gdef\gplfronttext{}%
  \makeatother
  \ifGPblacktext
    \def\colorrgb#1{}%
    \def\colorgray#1{}%
  \else
    \ifGPcolor
      \def\colorrgb#1{\color[rgb]{#1}}%
      \def\colorgray#1{\color[gray]{#1}}%
      \expandafter\def\csname LTw\endcsname{\color{white}}%
      \expandafter\def\csname LTb\endcsname{\color{black}}%
      \expandafter\def\csname LTa\endcsname{\color{black}}%
      \expandafter\def\csname LT0\endcsname{\color[rgb]{1,0,0}}%
      \expandafter\def\csname LT1\endcsname{\color[rgb]{0,1,0}}%
      \expandafter\def\csname LT2\endcsname{\color[rgb]{0,0,1}}%
      \expandafter\def\csname LT3\endcsname{\color[rgb]{1,0,1}}%
      \expandafter\def\csname LT4\endcsname{\color[rgb]{0,1,1}}%
      \expandafter\def\csname LT5\endcsname{\color[rgb]{1,1,0}}%
      \expandafter\def\csname LT6\endcsname{\color[rgb]{0,0,0}}%
      \expandafter\def\csname LT7\endcsname{\color[rgb]{1,0.3,0}}%
      \expandafter\def\csname LT8\endcsname{\color[rgb]{0.5,0.5,0.5}}%
    \else
      \def\colorrgb#1{\color{black}}%
      \def\colorgray#1{\color[gray]{#1}}%
      \expandafter\def\csname LTw\endcsname{\color{white}}%
      \expandafter\def\csname LTb\endcsname{\color{black}}%
      \expandafter\def\csname LTa\endcsname{\color{black}}%
      \expandafter\def\csname LT0\endcsname{\color{black}}%
      \expandafter\def\csname LT1\endcsname{\color{black}}%
      \expandafter\def\csname LT2\endcsname{\color{black}}%
      \expandafter\def\csname LT3\endcsname{\color{black}}%
      \expandafter\def\csname LT4\endcsname{\color{black}}%
      \expandafter\def\csname LT5\endcsname{\color{black}}%
      \expandafter\def\csname LT6\endcsname{\color{black}}%
      \expandafter\def\csname LT7\endcsname{\color{black}}%
      \expandafter\def\csname LT8\endcsname{\color{black}}%
    \fi
  \fi
    \setlength{\unitlength}{0.0500bp}%
    \ifx\gptboxheight\undefined%
      \newlength{\gptboxheight}%
      \newlength{\gptboxwidth}%
      \newsavebox{\gptboxtext}%
    \fi%
    \setlength{\fboxrule}{0.5pt}%
    \setlength{\fboxsep}{1pt}%
    \definecolor{tbcol}{rgb}{1,1,1}%
\begin{picture}(5180.00,4480.00)%
    \gplgaddtomacro\gplbacktext{%
      \csname LTb\endcsname
      \put(816,971){\makebox(0,0)[r]{\strut{}400}}%
      \csname LTb\endcsname
      \put(816,1417){\makebox(0,0)[r]{\strut{}500}}%
      \csname LTb\endcsname
      \put(816,1863){\makebox(0,0)[r]{\strut{}600}}%
      \csname LTb\endcsname
      \put(816,2309){\makebox(0,0)[r]{\strut{}700}}%
      \csname LTb\endcsname
      \put(816,2755){\makebox(0,0)[r]{\strut{}800}}%
      \csname LTb\endcsname
      \put(816,3201){\makebox(0,0)[r]{\strut{}900}}%
      \csname LTb\endcsname
      \put(816,3647){\makebox(0,0)[r]{\strut{}1000}}%
      \csname LTb\endcsname
      \put(1016,525){\makebox(0,0){\strut{}0.00}}%
      \csname LTb\endcsname
      \put(1691,525){\makebox(0,0){\strut{}0.05}}%
      \csname LTb\endcsname
      \put(2366,525){\makebox(0,0){\strut{}0.10}}%
      \csname LTb\endcsname
      \put(3040,525){\makebox(0,0){\strut{}0.15}}%
      \csname LTb\endcsname
      \put(3715,525){\makebox(0,0){\strut{}0.20}}%
      \csname LTb\endcsname
      \put(4185,3647){\makebox(0,0)[l]{\strut{}300}}%
      \csname LTb\endcsname
      \put(4185,2532){\makebox(0,0)[l]{\strut{}400}}%
      \csname LTb\endcsname
      \put(4185,1863){\makebox(0,0)[l]{\strut{}500}}%
      \csname LTb\endcsname
      \put(4185,1417){\makebox(0,0)[l]{\strut{}600}}%
      \csname LTb\endcsname
      \put(4185,1098){\makebox(0,0)[l]{\strut{}700}}%
      \csname LTb\endcsname
      \put(4185,859){\makebox(0,0)[l]{\strut{}800}}%
      \csname LTb\endcsname
      \put(1016,3981){\makebox(0,0){\strut{}0.0}}%
      \csname LTb\endcsname
      \put(1581,3981){\makebox(0,0){\strut{}0.2}}%
      \csname LTb\endcsname
      \put(2147,3981){\makebox(0,0){\strut{}0.4}}%
      \csname LTb\endcsname
      \put(2712,3981){\makebox(0,0){\strut{}0.6}}%
      \csname LTb\endcsname
      \put(3277,3981){\makebox(0,0){\strut{}0.8}}%
      \csname LTb\endcsname
      \put(3842,3981){\makebox(0,0){\strut{}1.0}}%
    }%
    \gplgaddtomacro\gplfronttext{%
      \csname LTb\endcsname
      \put(198,2253){\rotatebox{-270}{\makebox(0,0){\strut{}Frequency [THz]}}}%
      \csname LTb\endcsname
      \put(4724,2253){\rotatebox{-270}{\makebox(0,0){\strut{}Wavelength [nm]}}}%
      \csname LTb\endcsname
      \put(2500,167){\makebox(0,0){\strut{}Wavenumber [nm\textsuperscript{-1}]}}%
      \csname LTb\endcsname
      \put(2500,4339){\makebox(0,0){\strut{}Wavenumber [$\kappa a / 2 \pi$]}}%
      \csname LTb\endcsname
      \put(3155,1551){\makebox(0,0)[r]{\strut{}$\kappa_{TM}^{MG}$}}%
      \csname LTb\endcsname
      \put(3155,1312){\makebox(0,0)[r]{\strut{}$\kappa_{TM1}$}}%
      \csname LTb\endcsname
      \put(3155,1073){\makebox(0,0)[r]{\strut{}$\kappa_{TM2}$}}%
    }%
    \gplbacktext
    \put(0,0){\includegraphics[width={259.00bp},height={224.00bp}]{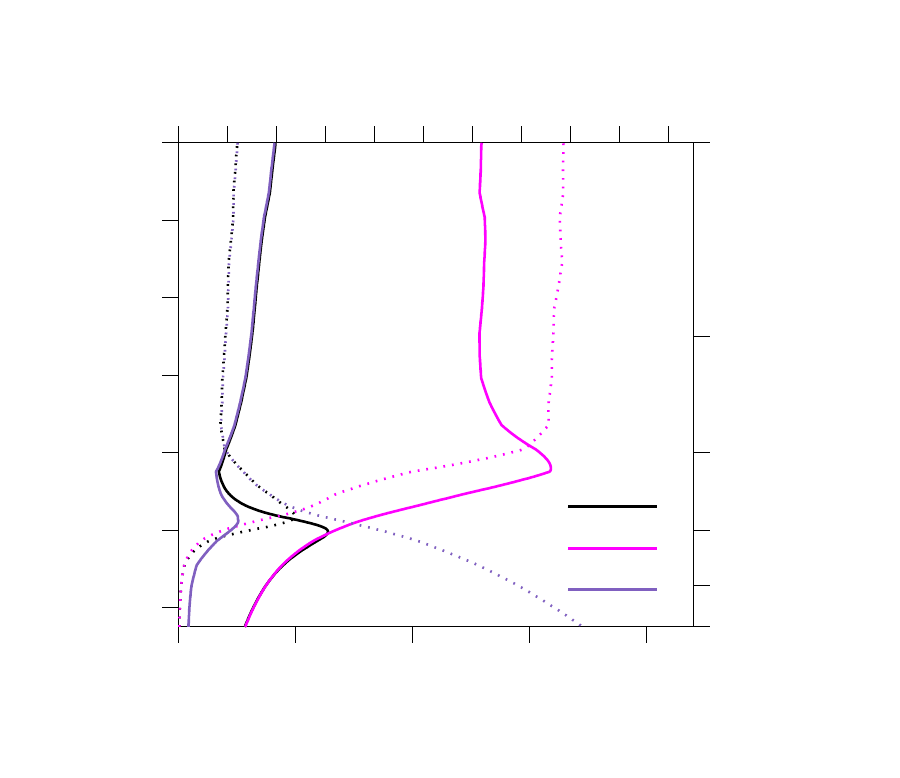}}%
    \gplfronttext
  \end{picture}%
\endgroup

%% file: gnu/TM1_COMSOL_CHECK_nearest.tex
\begingroup
  \inputencoding{cp1252}%
  \makeatletter
  \providecommand\color[2][]{%
    \GenericError{(gnuplot) \space\space\space\@spaces}{%
      Package color not loaded in conjunction with
      terminal option `colourtext'%
    }{See the gnuplot documentation for explanation.%
    }{Either use 'blacktext' in gnuplot or load the package
      color.sty in LaTeX.}%
    \renewcommand\color[2][]{}%
  }%
  \providecommand\includegraphics[2][]{%
    \GenericError{(gnuplot) \space\space\space\@spaces}{%
      Package graphicx or graphics not loaded%
    }{See the gnuplot documentation for explanation.%
    }{The gnuplot epslatex terminal needs graphicx.sty or graphics.sty.}%
    \renewcommand\includegraphics[2][]{}%
  }%
  \providecommand\rotatebox[2]{#2}%
  \@ifundefined{ifGPcolor}{%
    \newif\ifGPcolor
    \GPcolortrue
  }{}%
  \@ifundefined{ifGPblacktext}{%
    \newif\ifGPblacktext
    \GPblacktexttrue
  }{}%
  \let\gplgaddtomacro\g@addto@macro
  \gdef\gplbacktext{}%
  \gdef\gplfronttext{}%
  \makeatother
  \ifGPblacktext
    \def\colorrgb#1{}%
    \def\colorgray#1{}%
  \else
    \ifGPcolor
      \def\colorrgb#1{\color[rgb]{#1}}%
      \def\colorgray#1{\color[gray]{#1}}%
      \expandafter\def\csname LTw\endcsname{\color{white}}%
      \expandafter\def\csname LTb\endcsname{\color{black}}%
      \expandafter\def\csname LTa\endcsname{\color{black}}%
      \expandafter\def\csname LT0\endcsname{\color[rgb]{1,0,0}}%
      \expandafter\def\csname LT1\endcsname{\color[rgb]{0,1,0}}%
      \expandafter\def\csname LT2\endcsname{\color[rgb]{0,0,1}}%
      \expandafter\def\csname LT3\endcsname{\color[rgb]{1,0,1}}%
      \expandafter\def\csname LT4\endcsname{\color[rgb]{0,1,1}}%
      \expandafter\def\csname LT5\endcsname{\color[rgb]{1,1,0}}%
      \expandafter\def\csname LT6\endcsname{\color[rgb]{0,0,0}}%
      \expandafter\def\csname LT7\endcsname{\color[rgb]{1,0.3,0}}%
      \expandafter\def\csname LT8\endcsname{\color[rgb]{0.5,0.5,0.5}}%
    \else
      \def\colorrgb#1{\color{black}}%
      \def\colorgray#1{\color[gray]{#1}}%
      \expandafter\def\csname LTw\endcsname{\color{white}}%
      \expandafter\def\csname LTb\endcsname{\color{black}}%
      \expandafter\def\csname LTa\endcsname{\color{black}}%
      \expandafter\def\csname LT0\endcsname{\color{black}}%
      \expandafter\def\csname LT1\endcsname{\color{black}}%
      \expandafter\def\csname LT2\endcsname{\color{black}}%
      \expandafter\def\csname LT3\endcsname{\color{black}}%
      \expandafter\def\csname LT4\endcsname{\color{black}}%
      \expandafter\def\csname LT5\endcsname{\color{black}}%
      \expandafter\def\csname LT6\endcsname{\color{black}}%
      \expandafter\def\csname LT7\endcsname{\color{black}}%
      \expandafter\def\csname LT8\endcsname{\color{black}}%
    \fi
  \fi
    \setlength{\unitlength}{0.0500bp}%
    \ifx\gptboxheight\undefined%
      \newlength{\gptboxheight}%
      \newlength{\gptboxwidth}%
      \newsavebox{\gptboxtext}%
    \fi%
    \setlength{\fboxrule}{0.5pt}%
    \setlength{\fboxsep}{1pt}%
    \definecolor{tbcol}{rgb}{1,1,1}%
\begin{picture}(6800.00,3400.00)%
    \gplgaddtomacro\gplbacktext{%
      \csname LTb\endcsname
      \put(816,961){\makebox(0,0)[r]{\strut{}0}}%
      \csname LTb\endcsname
      \put(816,1301){\makebox(0,0)[r]{\strut{}200}}%
      \csname LTb\endcsname
      \put(816,1641){\makebox(0,0)[r]{\strut{}400}}%
      \csname LTb\endcsname
      \put(816,1982){\makebox(0,0)[r]{\strut{}600}}%
      \csname LTb\endcsname
      \put(816,2322){\makebox(0,0)[r]{\strut{}800}}%
      \csname LTb\endcsname
      \put(816,2662){\makebox(0,0)[r]{\strut{}1000}}%
      \csname LTb\endcsname
      \put(1234,525){\makebox(0,0){\strut{}400}}%
      \csname LTb\endcsname
      \put(2106,525){\makebox(0,0){\strut{}500}}%
      \csname LTb\endcsname
      \put(2977,525){\makebox(0,0){\strut{}600}}%
      \csname LTb\endcsname
      \put(3849,525){\makebox(0,0){\strut{}700}}%
      \csname LTb\endcsname
      \put(4721,525){\makebox(0,0){\strut{}800}}%
      \csname LTb\endcsname
      \put(5592,525){\makebox(0,0){\strut{}900}}%
      \csname LTb\endcsname
      \put(6464,525){\makebox(0,0){\strut{}1000}}%
      \csname LTb\endcsname
      \put(6464,2901){\makebox(0,0){\strut{}300}}%
      \csname LTb\endcsname
      \put(4285,2901){\makebox(0,0){\strut{}400}}%
      \csname LTb\endcsname
      \put(2977,2901){\makebox(0,0){\strut{}500}}%
      \csname LTb\endcsname
      \put(2106,2901){\makebox(0,0){\strut{}600}}%
      \csname LTb\endcsname
      \put(1483,2901){\makebox(0,0){\strut{}700}}%
      \csname LTb\endcsname
      \put(1016,2901){\makebox(0,0){\strut{}800}}%
    }%
    \gplgaddtomacro\gplfronttext{%
      \csname LTb\endcsname
      \put(198,1760){\rotatebox{-270}{\makebox(0,0){\strut{}Frequency [THz]}}}%
      \csname LTb\endcsname
      \put(3740,167){\makebox(0,0){\strut{}Frequency $\nu_0$ [THz]}}%
      \csname LTb\endcsname
      \put(3740,3259){\makebox(0,0){\strut{}Wavelength in [nm]}}%
      \csname LTb\endcsname
      \put(5634,1999){\makebox(0,0)[r]{\strut{}$\Re(\nu_C)$}}%
      \csname LTb\endcsname
      \put(5634,1760){\makebox(0,0)[r]{\strut{}$\Im(\nu_C)$}}%
      \csname LTb\endcsname
      \put(5634,1521){\makebox(0,0)[r]{\strut{}$\nu_0$}}%
    }%
    \gplbacktext
    \put(0,0){\includegraphics[width={340.00bp},height={170.00bp}]{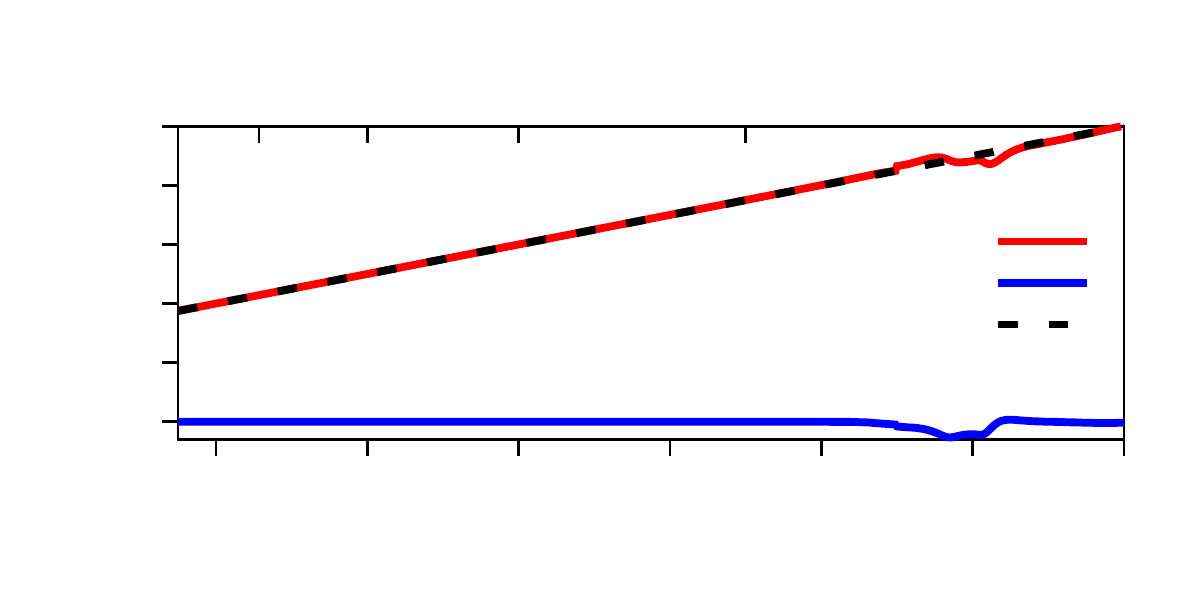}}%
    \gplfronttext
  \end{picture}%
\endgroup

%% file: gnu/TM2_COMSOL_CHECK_nearest.tex
\begingroup
  \inputencoding{cp1252}%
  \makeatletter
  \providecommand\color[2][]{%
    \GenericError{(gnuplot) \space\space\space\@spaces}{%
      Package color not loaded in conjunction with
      terminal option `colourtext'%
    }{See the gnuplot documentation for explanation.%
    }{Either use 'blacktext' in gnuplot or load the package
      color.sty in LaTeX.}%
    \renewcommand\color[2][]{}%
  }%
  \providecommand\includegraphics[2][]{%
    \GenericError{(gnuplot) \space\space\space\@spaces}{%
      Package graphicx or graphics not loaded%
    }{See the gnuplot documentation for explanation.%
    }{The gnuplot epslatex terminal needs graphicx.sty or graphics.sty.}%
    \renewcommand\includegraphics[2][]{}%
  }%
  \providecommand\rotatebox[2]{#2}%
  \@ifundefined{ifGPcolor}{%
    \newif\ifGPcolor
    \GPcolortrue
  }{}%
  \@ifundefined{ifGPblacktext}{%
    \newif\ifGPblacktext
    \GPblacktexttrue
  }{}%
  \let\gplgaddtomacro\g@addto@macro
  \gdef\gplbacktext{}%
  \gdef\gplfronttext{}%
  \makeatother
  \ifGPblacktext
    \def\colorrgb#1{}%
    \def\colorgray#1{}%
  \else
    \ifGPcolor
      \def\colorrgb#1{\color[rgb]{#1}}%
      \def\colorgray#1{\color[gray]{#1}}%
      \expandafter\def\csname LTw\endcsname{\color{white}}%
      \expandafter\def\csname LTb\endcsname{\color{black}}%
      \expandafter\def\csname LTa\endcsname{\color{black}}%
      \expandafter\def\csname LT0\endcsname{\color[rgb]{1,0,0}}%
      \expandafter\def\csname LT1\endcsname{\color[rgb]{0,1,0}}%
      \expandafter\def\csname LT2\endcsname{\color[rgb]{0,0,1}}%
      \expandafter\def\csname LT3\endcsname{\color[rgb]{1,0,1}}%
      \expandafter\def\csname LT4\endcsname{\color[rgb]{0,1,1}}%
      \expandafter\def\csname LT5\endcsname{\color[rgb]{1,1,0}}%
      \expandafter\def\csname LT6\endcsname{\color[rgb]{0,0,0}}%
      \expandafter\def\csname LT7\endcsname{\color[rgb]{1,0.3,0}}%
      \expandafter\def\csname LT8\endcsname{\color[rgb]{0.5,0.5,0.5}}%
    \else
      \def\colorrgb#1{\color{black}}%
      \def\colorgray#1{\color[gray]{#1}}%
      \expandafter\def\csname LTw\endcsname{\color{white}}%
      \expandafter\def\csname LTb\endcsname{\color{black}}%
      \expandafter\def\csname LTa\endcsname{\color{black}}%
      \expandafter\def\csname LT0\endcsname{\color{black}}%
      \expandafter\def\csname LT1\endcsname{\color{black}}%
      \expandafter\def\csname LT2\endcsname{\color{black}}%
      \expandafter\def\csname LT3\endcsname{\color{black}}%
      \expandafter\def\csname LT4\endcsname{\color{black}}%
      \expandafter\def\csname LT5\endcsname{\color{black}}%
      \expandafter\def\csname LT6\endcsname{\color{black}}%
      \expandafter\def\csname LT7\endcsname{\color{black}}%
      \expandafter\def\csname LT8\endcsname{\color{black}}%
    \fi
  \fi
    \setlength{\unitlength}{0.0500bp}%
    \ifx\gptboxheight\undefined%
      \newlength{\gptboxheight}%
      \newlength{\gptboxwidth}%
      \newsavebox{\gptboxtext}%
    \fi%
    \setlength{\fboxrule}{0.5pt}%
    \setlength{\fboxsep}{1pt}%
    \definecolor{tbcol}{rgb}{1,1,1}%
\begin{picture}(6800.00,3400.00)%
    \gplgaddtomacro\gplbacktext{%
      \csname LTb\endcsname
      \put(816,868){\makebox(0,0)[r]{\strut{}0}}%
      \csname LTb\endcsname
      \put(816,1227){\makebox(0,0)[r]{\strut{}200}}%
      \csname LTb\endcsname
      \put(816,1586){\makebox(0,0)[r]{\strut{}400}}%
      \csname LTb\endcsname
      \put(816,1944){\makebox(0,0)[r]{\strut{}600}}%
      \csname LTb\endcsname
      \put(816,2303){\makebox(0,0)[r]{\strut{}800}}%
      \csname LTb\endcsname
      \put(816,2662){\makebox(0,0)[r]{\strut{}1000}}%
      \csname LTb\endcsname
      \put(1234,525){\makebox(0,0){\strut{}400}}%
      \csname LTb\endcsname
      \put(2106,525){\makebox(0,0){\strut{}500}}%
      \csname LTb\endcsname
      \put(2977,525){\makebox(0,0){\strut{}600}}%
      \csname LTb\endcsname
      \put(3849,525){\makebox(0,0){\strut{}700}}%
      \csname LTb\endcsname
      \put(4721,525){\makebox(0,0){\strut{}800}}%
      \csname LTb\endcsname
      \put(5592,525){\makebox(0,0){\strut{}900}}%
      \csname LTb\endcsname
      \put(6464,525){\makebox(0,0){\strut{}1000}}%
      \csname LTb\endcsname
      \put(6464,2901){\makebox(0,0){\strut{}300}}%
      \csname LTb\endcsname
      \put(4285,2901){\makebox(0,0){\strut{}400}}%
      \csname LTb\endcsname
      \put(2977,2901){\makebox(0,0){\strut{}500}}%
      \csname LTb\endcsname
      \put(2106,2901){\makebox(0,0){\strut{}600}}%
      \csname LTb\endcsname
      \put(1483,2901){\makebox(0,0){\strut{}700}}%
      \csname LTb\endcsname
      \put(1016,2901){\makebox(0,0){\strut{}800}}%
    }%
    \gplgaddtomacro\gplfronttext{%
      \csname LTb\endcsname
      \put(198,1760){\rotatebox{-270}{\makebox(0,0){\strut{}Frequency [THz]}}}%
      \csname LTb\endcsname
      \put(3740,167){\makebox(0,0){\strut{}Frequency $\nu_0$ [THz]}}%
      \csname LTb\endcsname
      \put(3740,3259){\makebox(0,0){\strut{}Wavelength in [nm]}}%
      \csname LTb\endcsname
      \put(5634,1999){\makebox(0,0)[r]{\strut{}$\Re(\nu_C)$}}%
      \csname LTb\endcsname
      \put(5634,1760){\makebox(0,0)[r]{\strut{}$\Im(\nu_C)$}}%
      \csname LTb\endcsname
      \put(5634,1521){\makebox(0,0)[r]{\strut{}$\nu_0$}}%
    }%
    \gplbacktext
    \put(0,0){\includegraphics[width={340.00bp},height={170.00bp}]{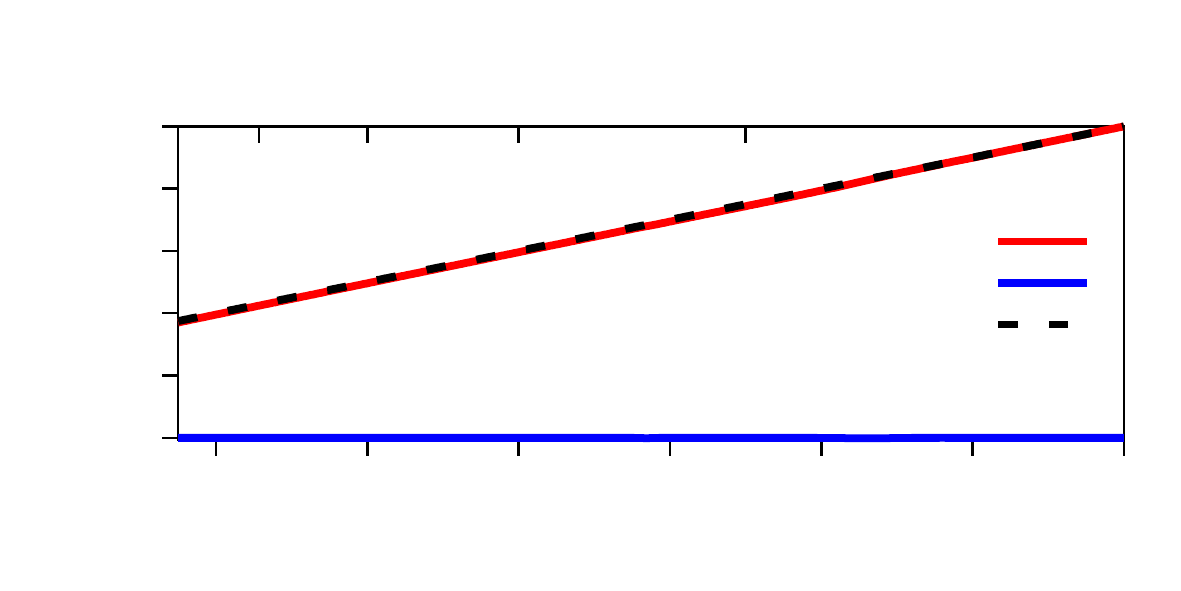}}%
    \gplfronttext
  \end{picture}%
\endgroup

%% file: gnu/Conv_TM1_400THz.tex
\begingroup
  \inputencoding{cp1252}%
  \makeatletter
  \providecommand\color[2][]{%
    \GenericError{(gnuplot) \space\space\space\@spaces}{%
      Package color not loaded in conjunction with
      terminal option `colourtext'%
    }{See the gnuplot documentation for explanation.%
    }{Either use 'blacktext' in gnuplot or load the package
      color.sty in LaTeX.}%
    \renewcommand\color[2][]{}%
  }%
  \providecommand\includegraphics[2][]{%
    \GenericError{(gnuplot) \space\space\space\@spaces}{%
      Package graphicx or graphics not loaded%
    }{See the gnuplot documentation for explanation.%
    }{The gnuplot epslatex terminal needs graphicx.sty or graphics.sty.}%
    \renewcommand\includegraphics[2][]{}%
  }%
  \providecommand\rotatebox[2]{#2}%
  \@ifundefined{ifGPcolor}{%
    \newif\ifGPcolor
    \GPcolortrue
  }{}%
  \@ifundefined{ifGPblacktext}{%
    \newif\ifGPblacktext
    \GPblacktexttrue
  }{}%
  \let\gplgaddtomacro\g@addto@macro
  \gdef\gplbacktext{}%
  \gdef\gplfronttext{}%
  \makeatother
  \ifGPblacktext
    \def\colorrgb#1{}%
    \def\colorgray#1{}%
  \else
    \ifGPcolor
      \def\colorrgb#1{\color[rgb]{#1}}%
      \def\colorgray#1{\color[gray]{#1}}%
      \expandafter\def\csname LTw\endcsname{\color{white}}%
      \expandafter\def\csname LTb\endcsname{\color{black}}%
      \expandafter\def\csname LTa\endcsname{\color{black}}%
      \expandafter\def\csname LT0\endcsname{\color[rgb]{1,0,0}}%
      \expandafter\def\csname LT1\endcsname{\color[rgb]{0,1,0}}%
      \expandafter\def\csname LT2\endcsname{\color[rgb]{0,0,1}}%
      \expandafter\def\csname LT3\endcsname{\color[rgb]{1,0,1}}%
      \expandafter\def\csname LT4\endcsname{\color[rgb]{0,1,1}}%
      \expandafter\def\csname LT5\endcsname{\color[rgb]{1,1,0}}%
      \expandafter\def\csname LT6\endcsname{\color[rgb]{0,0,0}}%
      \expandafter\def\csname LT7\endcsname{\color[rgb]{1,0.3,0}}%
      \expandafter\def\csname LT8\endcsname{\color[rgb]{0.5,0.5,0.5}}%
    \else
      \def\colorrgb#1{\color{black}}%
      \def\colorgray#1{\color[gray]{#1}}%
      \expandafter\def\csname LTw\endcsname{\color{white}}%
      \expandafter\def\csname LTb\endcsname{\color{black}}%
      \expandafter\def\csname LTa\endcsname{\color{black}}%
      \expandafter\def\csname LT0\endcsname{\color{black}}%
      \expandafter\def\csname LT1\endcsname{\color{black}}%
      \expandafter\def\csname LT2\endcsname{\color{black}}%
      \expandafter\def\csname LT3\endcsname{\color{black}}%
      \expandafter\def\csname LT4\endcsname{\color{black}}%
      \expandafter\def\csname LT5\endcsname{\color{black}}%
      \expandafter\def\csname LT6\endcsname{\color{black}}%
      \expandafter\def\csname LT7\endcsname{\color{black}}%
      \expandafter\def\csname LT8\endcsname{\color{black}}%
    \fi
  \fi
    \setlength{\unitlength}{0.0500bp}%
    \ifx\gptboxheight\undefined%
      \newlength{\gptboxheight}%
      \newlength{\gptboxwidth}%
      \newsavebox{\gptboxtext}%
    \fi%
    \setlength{\fboxrule}{0.5pt}%
    \setlength{\fboxsep}{1pt}%
    \definecolor{tbcol}{rgb}{1,1,1}%
\begin{picture}(5760.00,5040.00)%
    \gplgaddtomacro\gplbacktext{%
      \csname LTb\endcsname
      \put(816,764){\makebox(0,0)[r]{\strut{}$-5.5$}}%
      \csname LTb\endcsname
      \put(816,1567){\makebox(0,0)[r]{\strut{}$-5$}}%
      \csname LTb\endcsname
      \put(816,2370){\makebox(0,0)[r]{\strut{}$-4.5$}}%
      \csname LTb\endcsname
      \put(816,3174){\makebox(0,0)[r]{\strut{}$-4$}}%
      \csname LTb\endcsname
      \put(816,3977){\makebox(0,0)[r]{\strut{}$-3.5$}}%
      \csname LTb\endcsname
      \put(816,4780){\makebox(0,0)[r]{\strut{}$-3$}}%
      \csname LTb\endcsname
      \put(1016,525){\makebox(0,0){\strut{}$2.3$}}%
      \csname LTb\endcsname
      \put(1457,525){\makebox(0,0){\strut{}$2.4$}}%
      \csname LTb\endcsname
      \put(1898,525){\makebox(0,0){\strut{}$2.5$}}%
      \csname LTb\endcsname
      \put(2338,525){\makebox(0,0){\strut{}$2.6$}}%
      \csname LTb\endcsname
      \put(2779,525){\makebox(0,0){\strut{}$2.7$}}%
      \csname LTb\endcsname
      \put(3220,525){\makebox(0,0){\strut{}$2.8$}}%
      \csname LTb\endcsname
      \put(3661,525){\makebox(0,0){\strut{}$2.9$}}%
      \csname LTb\endcsname
      \put(4102,525){\makebox(0,0){\strut{}$3$}}%
      \csname LTb\endcsname
      \put(4542,525){\makebox(0,0){\strut{}$3.1$}}%
      \csname LTb\endcsname
      \put(4983,525){\makebox(0,0){\strut{}$3.2$}}%
      \csname LTb\endcsname
      \put(5424,525){\makebox(0,0){\strut{}$3.3$}}%
    }%
    \gplgaddtomacro\gplfronttext{%
      \csname LTb\endcsname
      \put(198,2772){\rotatebox{-270}{\makebox(0,0){\strut{}Relative Error}}}%
      \csname LTb\endcsname
      \put(3220,167){\makebox(0,0){\strut{}$\log(NG)/\log(10)$}}%
      \csname LTb\endcsname
      \put(1961,2173){\makebox(0,0)[r]{\strut{}$pd=(4,2)$}}%
      \csname LTb\endcsname
      \put(1961,1934){\makebox(0,0)[r]{\strut{}$pd=(4,3)$}}%
      \csname LTb\endcsname
      \put(1961,1695){\makebox(0,0)[r]{\strut{}$pd=(4,4)$}}%
      \csname LTb\endcsname
      \put(1961,1456){\makebox(0,0)[r]{\strut{}$pd=(2,2)$}}%
      \csname LTb\endcsname
      \put(1961,1217){\makebox(0,0)[r]{\strut{}$pd=(2,3)$}}%
      \csname LTb\endcsname
      \put(1961,978){\makebox(0,0)[r]{\strut{}$pd=(2,4)$}}%
    }%
    \gplbacktext
    \put(0,0){\includegraphics[width={288.00bp},height={252.00bp}]{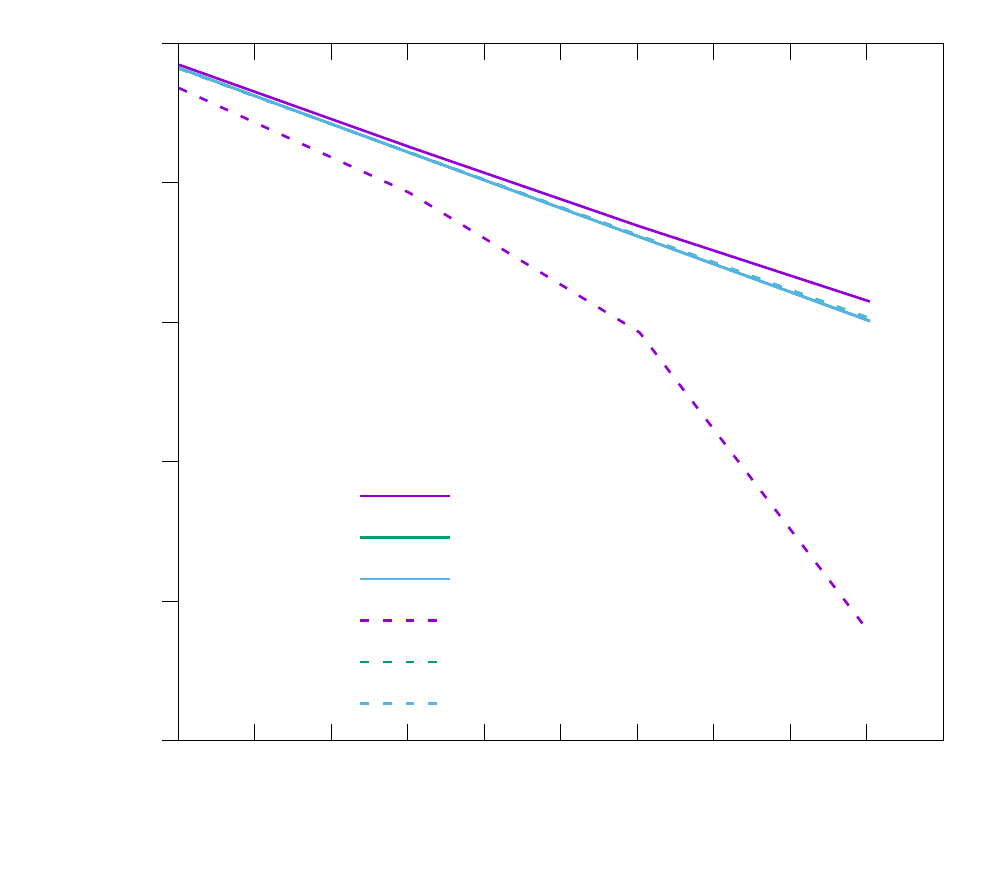}}%
    \gplfronttext
  \end{picture}%
\endgroup

%% file: gnu/Conv_TM2_400THz.tex
\begingroup
  \inputencoding{cp1252}%
  \makeatletter
  \providecommand\color[2][]{%
    \GenericError{(gnuplot) \space\space\space\@spaces}{%
      Package color not loaded in conjunction with
      terminal option `colourtext'%
    }{See the gnuplot documentation for explanation.%
    }{Either use 'blacktext' in gnuplot or load the package
      color.sty in LaTeX.}%
    \renewcommand\color[2][]{}%
  }%
  \providecommand\includegraphics[2][]{%
    \GenericError{(gnuplot) \space\space\space\@spaces}{%
      Package graphicx or graphics not loaded%
    }{See the gnuplot documentation for explanation.%
    }{The gnuplot epslatex terminal needs graphicx.sty or graphics.sty.}%
    \renewcommand\includegraphics[2][]{}%
  }%
  \providecommand\rotatebox[2]{#2}%
  \@ifundefined{ifGPcolor}{%
    \newif\ifGPcolor
    \GPcolortrue
  }{}%
  \@ifundefined{ifGPblacktext}{%
    \newif\ifGPblacktext
    \GPblacktexttrue
  }{}%
  \let\gplgaddtomacro\g@addto@macro
  \gdef\gplbacktext{}%
  \gdef\gplfronttext{}%
  \makeatother
  \ifGPblacktext
    \def\colorrgb#1{}%
    \def\colorgray#1{}%
  \else
    \ifGPcolor
      \def\colorrgb#1{\color[rgb]{#1}}%
      \def\colorgray#1{\color[gray]{#1}}%
      \expandafter\def\csname LTw\endcsname{\color{white}}%
      \expandafter\def\csname LTb\endcsname{\color{black}}%
      \expandafter\def\csname LTa\endcsname{\color{black}}%
      \expandafter\def\csname LT0\endcsname{\color[rgb]{1,0,0}}%
      \expandafter\def\csname LT1\endcsname{\color[rgb]{0,1,0}}%
      \expandafter\def\csname LT2\endcsname{\color[rgb]{0,0,1}}%
      \expandafter\def\csname LT3\endcsname{\color[rgb]{1,0,1}}%
      \expandafter\def\csname LT4\endcsname{\color[rgb]{0,1,1}}%
      \expandafter\def\csname LT5\endcsname{\color[rgb]{1,1,0}}%
      \expandafter\def\csname LT6\endcsname{\color[rgb]{0,0,0}}%
      \expandafter\def\csname LT7\endcsname{\color[rgb]{1,0.3,0}}%
      \expandafter\def\csname LT8\endcsname{\color[rgb]{0.5,0.5,0.5}}%
    \else
      \def\colorrgb#1{\color{black}}%
      \def\colorgray#1{\color[gray]{#1}}%
      \expandafter\def\csname LTw\endcsname{\color{white}}%
      \expandafter\def\csname LTb\endcsname{\color{black}}%
      \expandafter\def\csname LTa\endcsname{\color{black}}%
      \expandafter\def\csname LT0\endcsname{\color{black}}%
      \expandafter\def\csname LT1\endcsname{\color{black}}%
      \expandafter\def\csname LT2\endcsname{\color{black}}%
      \expandafter\def\csname LT3\endcsname{\color{black}}%
      \expandafter\def\csname LT4\endcsname{\color{black}}%
      \expandafter\def\csname LT5\endcsname{\color{black}}%
      \expandafter\def\csname LT6\endcsname{\color{black}}%
      \expandafter\def\csname LT7\endcsname{\color{black}}%
      \expandafter\def\csname LT8\endcsname{\color{black}}%
    \fi
  \fi
    \setlength{\unitlength}{0.0500bp}%
    \ifx\gptboxheight\undefined%
      \newlength{\gptboxheight}%
      \newlength{\gptboxwidth}%
      \newsavebox{\gptboxtext}%
    \fi%
    \setlength{\fboxrule}{0.5pt}%
    \setlength{\fboxsep}{1pt}%
    \definecolor{tbcol}{rgb}{1,1,1}%
\begin{picture}(5760.00,5040.00)%
    \gplgaddtomacro\gplbacktext{%
      \csname LTb\endcsname
      \put(816,764){\makebox(0,0)[r]{\strut{}$-3$}}%
      \csname LTb\endcsname
      \put(816,1129){\makebox(0,0)[r]{\strut{}$-2.8$}}%
      \csname LTb\endcsname
      \put(816,1494){\makebox(0,0)[r]{\strut{}$-2.6$}}%
      \csname LTb\endcsname
      \put(816,1859){\makebox(0,0)[r]{\strut{}$-2.4$}}%
      \csname LTb\endcsname
      \put(816,2224){\makebox(0,0)[r]{\strut{}$-2.2$}}%
      \csname LTb\endcsname
      \put(816,2589){\makebox(0,0)[r]{\strut{}$-2$}}%
      \csname LTb\endcsname
      \put(816,2955){\makebox(0,0)[r]{\strut{}$-1.8$}}%
      \csname LTb\endcsname
      \put(816,3320){\makebox(0,0)[r]{\strut{}$-1.6$}}%
      \csname LTb\endcsname
      \put(816,3685){\makebox(0,0)[r]{\strut{}$-1.4$}}%
      \csname LTb\endcsname
      \put(816,4050){\makebox(0,0)[r]{\strut{}$-1.2$}}%
      \csname LTb\endcsname
      \put(816,4415){\makebox(0,0)[r]{\strut{}$-1$}}%
      \csname LTb\endcsname
      \put(816,4780){\makebox(0,0)[r]{\strut{}$-0.8$}}%
      \csname LTb\endcsname
      \put(1016,525){\makebox(0,0){\strut{}$2.3$}}%
      \csname LTb\endcsname
      \put(1457,525){\makebox(0,0){\strut{}$2.4$}}%
      \csname LTb\endcsname
      \put(1898,525){\makebox(0,0){\strut{}$2.5$}}%
      \csname LTb\endcsname
      \put(2338,525){\makebox(0,0){\strut{}$2.6$}}%
      \csname LTb\endcsname
      \put(2779,525){\makebox(0,0){\strut{}$2.7$}}%
      \csname LTb\endcsname
      \put(3220,525){\makebox(0,0){\strut{}$2.8$}}%
      \csname LTb\endcsname
      \put(3661,525){\makebox(0,0){\strut{}$2.9$}}%
      \csname LTb\endcsname
      \put(4102,525){\makebox(0,0){\strut{}$3$}}%
      \csname LTb\endcsname
      \put(4542,525){\makebox(0,0){\strut{}$3.1$}}%
      \csname LTb\endcsname
      \put(4983,525){\makebox(0,0){\strut{}$3.2$}}%
      \csname LTb\endcsname
      \put(5424,525){\makebox(0,0){\strut{}$3.3$}}%
    }%
    \gplgaddtomacro\gplfronttext{%
      \csname LTb\endcsname
      \put(198,2772){\rotatebox{-270}{\makebox(0,0){\strut{}Relative Error}}}%
      \csname LTb\endcsname
      \put(3220,167){\makebox(0,0){\strut{}$\log(NG)/\log(10)$}}%
      \csname LTb\endcsname
      \put(1961,1456){\makebox(0,0)[r]{\strut{}$pd=(4,2)$}}%
      \csname LTb\endcsname
      \put(1961,1217){\makebox(0,0)[r]{\strut{}$pd=(4,3)$}}%
      \csname LTb\endcsname
      \put(1961,978){\makebox(0,0)[r]{\strut{}$pd=(4,4)$}}%
    }%
    \gplbacktext
    \put(0,0){\includegraphics[width={288.00bp},height={252.00bp}]{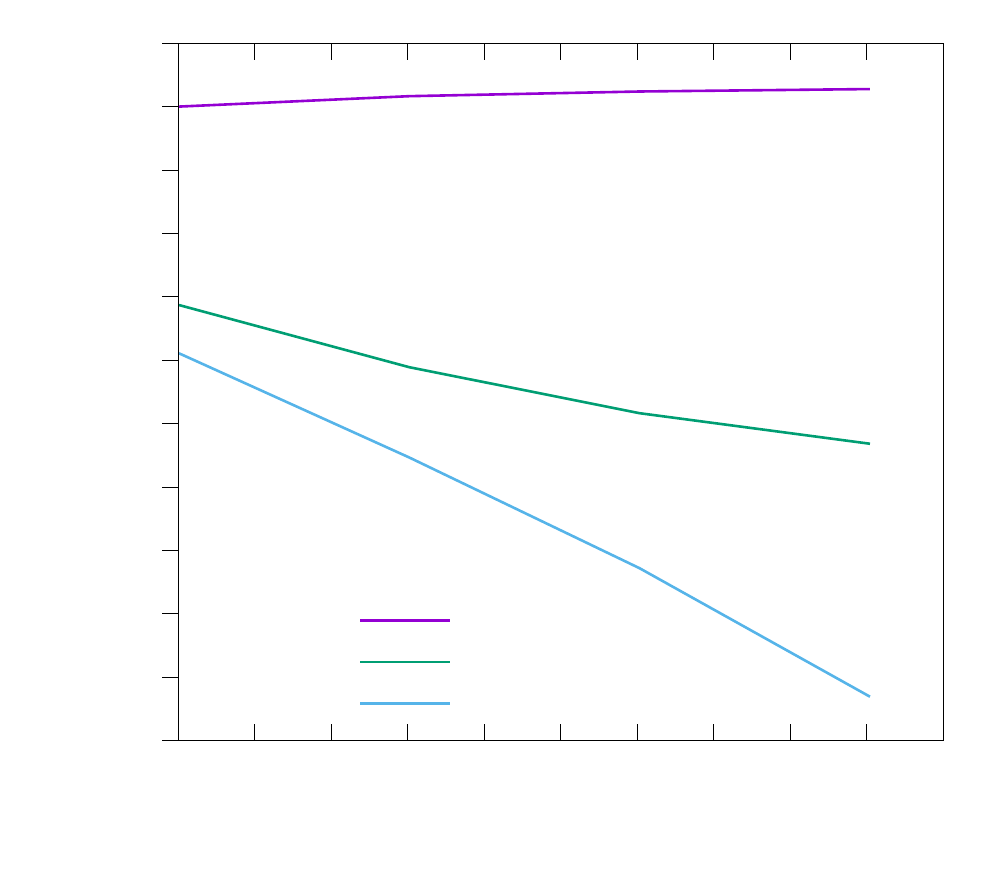}}%
    \gplfronttext
  \end{picture}%
\endgroup

%% file: gnu/Conv_TM1_838THz.tex
\begingroup
  \inputencoding{cp1252}%
  \makeatletter
  \providecommand\color[2][]{%
    \GenericError{(gnuplot) \space\space\space\@spaces}{%
      Package color not loaded in conjunction with
      terminal option `colourtext'%
    }{See the gnuplot documentation for explanation.%
    }{Either use 'blacktext' in gnuplot or load the package
      color.sty in LaTeX.}%
    \renewcommand\color[2][]{}%
  }%
  \providecommand\includegraphics[2][]{%
    \GenericError{(gnuplot) \space\space\space\@spaces}{%
      Package graphicx or graphics not loaded%
    }{See the gnuplot documentation for explanation.%
    }{The gnuplot epslatex terminal needs graphicx.sty or graphics.sty.}%
    \renewcommand\includegraphics[2][]{}%
  }%
  \providecommand\rotatebox[2]{#2}%
  \@ifundefined{ifGPcolor}{%
    \newif\ifGPcolor
    \GPcolortrue
  }{}%
  \@ifundefined{ifGPblacktext}{%
    \newif\ifGPblacktext
    \GPblacktexttrue
  }{}%
  \let\gplgaddtomacro\g@addto@macro
  \gdef\gplbacktext{}%
  \gdef\gplfronttext{}%
  \makeatother
  \ifGPblacktext
    \def\colorrgb#1{}%
    \def\colorgray#1{}%
  \else
    \ifGPcolor
      \def\colorrgb#1{\color[rgb]{#1}}%
      \def\colorgray#1{\color[gray]{#1}}%
      \expandafter\def\csname LTw\endcsname{\color{white}}%
      \expandafter\def\csname LTb\endcsname{\color{black}}%
      \expandafter\def\csname LTa\endcsname{\color{black}}%
      \expandafter\def\csname LT0\endcsname{\color[rgb]{1,0,0}}%
      \expandafter\def\csname LT1\endcsname{\color[rgb]{0,1,0}}%
      \expandafter\def\csname LT2\endcsname{\color[rgb]{0,0,1}}%
      \expandafter\def\csname LT3\endcsname{\color[rgb]{1,0,1}}%
      \expandafter\def\csname LT4\endcsname{\color[rgb]{0,1,1}}%
      \expandafter\def\csname LT5\endcsname{\color[rgb]{1,1,0}}%
      \expandafter\def\csname LT6\endcsname{\color[rgb]{0,0,0}}%
      \expandafter\def\csname LT7\endcsname{\color[rgb]{1,0.3,0}}%
      \expandafter\def\csname LT8\endcsname{\color[rgb]{0.5,0.5,0.5}}%
    \else
      \def\colorrgb#1{\color{black}}%
      \def\colorgray#1{\color[gray]{#1}}%
      \expandafter\def\csname LTw\endcsname{\color{white}}%
      \expandafter\def\csname LTb\endcsname{\color{black}}%
      \expandafter\def\csname LTa\endcsname{\color{black}}%
      \expandafter\def\csname LT0\endcsname{\color{black}}%
      \expandafter\def\csname LT1\endcsname{\color{black}}%
      \expandafter\def\csname LT2\endcsname{\color{black}}%
      \expandafter\def\csname LT3\endcsname{\color{black}}%
      \expandafter\def\csname LT4\endcsname{\color{black}}%
      \expandafter\def\csname LT5\endcsname{\color{black}}%
      \expandafter\def\csname LT6\endcsname{\color{black}}%
      \expandafter\def\csname LT7\endcsname{\color{black}}%
      \expandafter\def\csname LT8\endcsname{\color{black}}%
    \fi
  \fi
    \setlength{\unitlength}{0.0500bp}%
    \ifx\gptboxheight\undefined%
      \newlength{\gptboxheight}%
      \newlength{\gptboxwidth}%
      \newsavebox{\gptboxtext}%
    \fi%
    \setlength{\fboxrule}{0.5pt}%
    \setlength{\fboxsep}{1pt}%
    \definecolor{tbcol}{rgb}{1,1,1}%
\begin{picture}(5760.00,5040.00)%
    \gplgaddtomacro\gplbacktext{%
      \csname LTb\endcsname
      \put(816,764){\makebox(0,0)[r]{\strut{}$-2.5$}}%
      \csname LTb\endcsname
      \put(816,1129){\makebox(0,0)[r]{\strut{}$-2.4$}}%
      \csname LTb\endcsname
      \put(816,1494){\makebox(0,0)[r]{\strut{}$-2.3$}}%
      \csname LTb\endcsname
      \put(816,1859){\makebox(0,0)[r]{\strut{}$-2.2$}}%
      \csname LTb\endcsname
      \put(816,2224){\makebox(0,0)[r]{\strut{}$-2.1$}}%
      \csname LTb\endcsname
      \put(816,2589){\makebox(0,0)[r]{\strut{}$-2$}}%
      \csname LTb\endcsname
      \put(816,2955){\makebox(0,0)[r]{\strut{}$-1.9$}}%
      \csname LTb\endcsname
      \put(816,3320){\makebox(0,0)[r]{\strut{}$-1.8$}}%
      \csname LTb\endcsname
      \put(816,3685){\makebox(0,0)[r]{\strut{}$-1.7$}}%
      \csname LTb\endcsname
      \put(816,4050){\makebox(0,0)[r]{\strut{}$-1.6$}}%
      \csname LTb\endcsname
      \put(816,4415){\makebox(0,0)[r]{\strut{}$-1.5$}}%
      \csname LTb\endcsname
      \put(816,4780){\makebox(0,0)[r]{\strut{}$-1.4$}}%
      \csname LTb\endcsname
      \put(1016,525){\makebox(0,0){\strut{}$2.3$}}%
      \csname LTb\endcsname
      \put(1457,525){\makebox(0,0){\strut{}$2.4$}}%
      \csname LTb\endcsname
      \put(1898,525){\makebox(0,0){\strut{}$2.5$}}%
      \csname LTb\endcsname
      \put(2338,525){\makebox(0,0){\strut{}$2.6$}}%
      \csname LTb\endcsname
      \put(2779,525){\makebox(0,0){\strut{}$2.7$}}%
      \csname LTb\endcsname
      \put(3220,525){\makebox(0,0){\strut{}$2.8$}}%
      \csname LTb\endcsname
      \put(3661,525){\makebox(0,0){\strut{}$2.9$}}%
      \csname LTb\endcsname
      \put(4102,525){\makebox(0,0){\strut{}$3$}}%
      \csname LTb\endcsname
      \put(4542,525){\makebox(0,0){\strut{}$3.1$}}%
      \csname LTb\endcsname
      \put(4983,525){\makebox(0,0){\strut{}$3.2$}}%
      \csname LTb\endcsname
      \put(5424,525){\makebox(0,0){\strut{}$3.3$}}%
    }%
    \gplgaddtomacro\gplfronttext{%
      \csname LTb\endcsname
      \put(198,2772){\rotatebox{-270}{\makebox(0,0){\strut{}Relative Error}}}%
      \csname LTb\endcsname
      \put(3220,167){\makebox(0,0){\strut{}$\log(NG)/\log(10)$}}%
      \csname LTb\endcsname
      \put(1961,2173){\makebox(0,0)[r]{\strut{}$pd=(4,2)$}}%
      \csname LTb\endcsname
      \put(1961,1934){\makebox(0,0)[r]{\strut{}$pd=(4,3)$}}%
      \csname LTb\endcsname
      \put(1961,1695){\makebox(0,0)[r]{\strut{}$pd=(4,4)$}}%
      \csname LTb\endcsname
      \put(1961,1456){\makebox(0,0)[r]{\strut{}$pd=(2,2)$}}%
      \csname LTb\endcsname
      \put(1961,1217){\makebox(0,0)[r]{\strut{}$pd=(2,3)$}}%
      \csname LTb\endcsname
      \put(1961,978){\makebox(0,0)[r]{\strut{}$pd=(2,4)$}}%
    }%
    \gplbacktext
    \put(0,0){\includegraphics[width={288.00bp},height={252.00bp}]{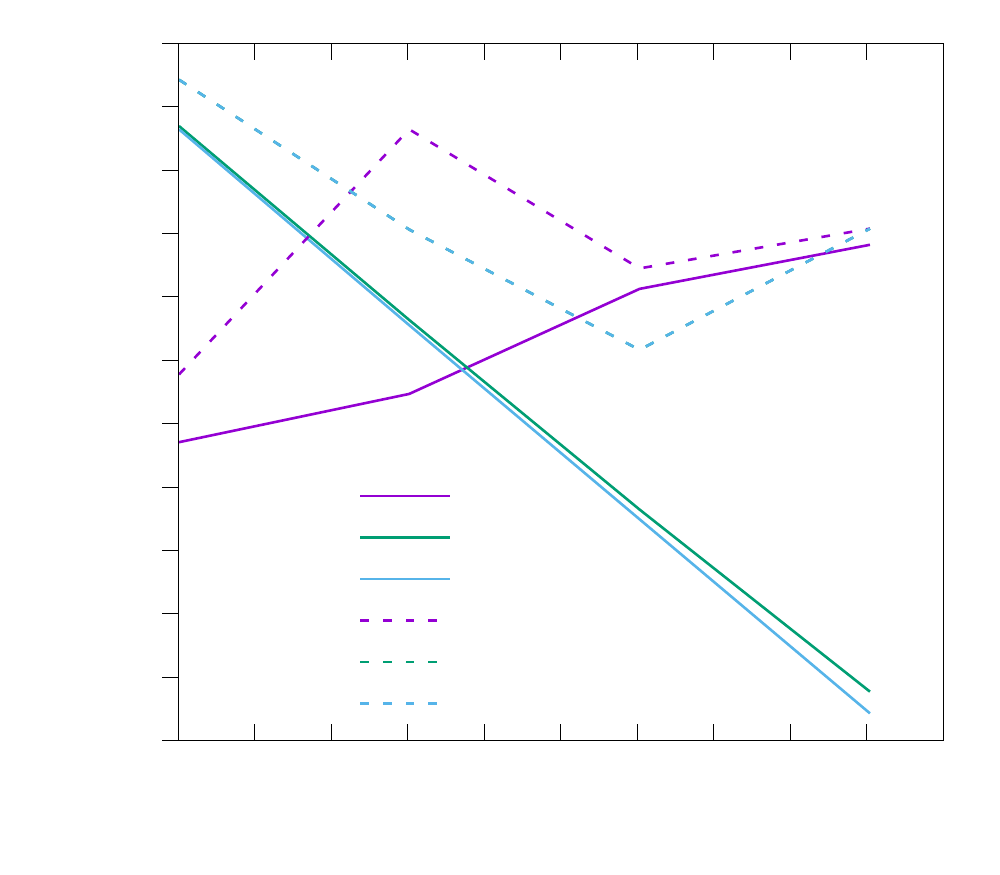}}%
    \gplfronttext
  \end{picture}%
\endgroup

%% file: gnu/Conv_TM2_838THz.tex
\begingroup
  \inputencoding{cp1252}%
  \makeatletter
  \providecommand\color[2][]{%
    \GenericError{(gnuplot) \space\space\space\@spaces}{%
      Package color not loaded in conjunction with
      terminal option `colourtext'%
    }{See the gnuplot documentation for explanation.%
    }{Either use 'blacktext' in gnuplot or load the package
      color.sty in LaTeX.}%
    \renewcommand\color[2][]{}%
  }%
  \providecommand\includegraphics[2][]{%
    \GenericError{(gnuplot) \space\space\space\@spaces}{%
      Package graphicx or graphics not loaded%
    }{See the gnuplot documentation for explanation.%
    }{The gnuplot epslatex terminal needs graphicx.sty or graphics.sty.}%
    \renewcommand\includegraphics[2][]{}%
  }%
  \providecommand\rotatebox[2]{#2}%
  \@ifundefined{ifGPcolor}{%
    \newif\ifGPcolor
    \GPcolortrue
  }{}%
  \@ifundefined{ifGPblacktext}{%
    \newif\ifGPblacktext
    \GPblacktexttrue
  }{}%
  \let\gplgaddtomacro\g@addto@macro
  \gdef\gplbacktext{}%
  \gdef\gplfronttext{}%
  \makeatother
  \ifGPblacktext
    \def\colorrgb#1{}%
    \def\colorgray#1{}%
  \else
    \ifGPcolor
      \def\colorrgb#1{\color[rgb]{#1}}%
      \def\colorgray#1{\color[gray]{#1}}%
      \expandafter\def\csname LTw\endcsname{\color{white}}%
      \expandafter\def\csname LTb\endcsname{\color{black}}%
      \expandafter\def\csname LTa\endcsname{\color{black}}%
      \expandafter\def\csname LT0\endcsname{\color[rgb]{1,0,0}}%
      \expandafter\def\csname LT1\endcsname{\color[rgb]{0,1,0}}%
      \expandafter\def\csname LT2\endcsname{\color[rgb]{0,0,1}}%
      \expandafter\def\csname LT3\endcsname{\color[rgb]{1,0,1}}%
      \expandafter\def\csname LT4\endcsname{\color[rgb]{0,1,1}}%
      \expandafter\def\csname LT5\endcsname{\color[rgb]{1,1,0}}%
      \expandafter\def\csname LT6\endcsname{\color[rgb]{0,0,0}}%
      \expandafter\def\csname LT7\endcsname{\color[rgb]{1,0.3,0}}%
      \expandafter\def\csname LT8\endcsname{\color[rgb]{0.5,0.5,0.5}}%
    \else
      \def\colorrgb#1{\color{black}}%
      \def\colorgray#1{\color[gray]{#1}}%
      \expandafter\def\csname LTw\endcsname{\color{white}}%
      \expandafter\def\csname LTb\endcsname{\color{black}}%
      \expandafter\def\csname LTa\endcsname{\color{black}}%
      \expandafter\def\csname LT0\endcsname{\color{black}}%
      \expandafter\def\csname LT1\endcsname{\color{black}}%
      \expandafter\def\csname LT2\endcsname{\color{black}}%
      \expandafter\def\csname LT3\endcsname{\color{black}}%
      \expandafter\def\csname LT4\endcsname{\color{black}}%
      \expandafter\def\csname LT5\endcsname{\color{black}}%
      \expandafter\def\csname LT6\endcsname{\color{black}}%
      \expandafter\def\csname LT7\endcsname{\color{black}}%
      \expandafter\def\csname LT8\endcsname{\color{black}}%
    \fi
  \fi
    \setlength{\unitlength}{0.0500bp}%
    \ifx\gptboxheight\undefined%
      \newlength{\gptboxheight}%
      \newlength{\gptboxwidth}%
      \newsavebox{\gptboxtext}%
    \fi%
    \setlength{\fboxrule}{0.5pt}%
    \setlength{\fboxsep}{1pt}%
    \definecolor{tbcol}{rgb}{1,1,1}%
\begin{picture}(5760.00,5040.00)%
    \gplgaddtomacro\gplbacktext{%
      \csname LTb\endcsname
      \put(816,764){\makebox(0,0)[r]{\strut{}$-3.6$}}%
      \csname LTb\endcsname
      \put(816,1210){\makebox(0,0)[r]{\strut{}$-3.4$}}%
      \csname LTb\endcsname
      \put(816,1656){\makebox(0,0)[r]{\strut{}$-3.2$}}%
      \csname LTb\endcsname
      \put(816,2103){\makebox(0,0)[r]{\strut{}$-3$}}%
      \csname LTb\endcsname
      \put(816,2549){\makebox(0,0)[r]{\strut{}$-2.8$}}%
      \csname LTb\endcsname
      \put(816,2995){\makebox(0,0)[r]{\strut{}$-2.6$}}%
      \csname LTb\endcsname
      \put(816,3441){\makebox(0,0)[r]{\strut{}$-2.4$}}%
      \csname LTb\endcsname
      \put(816,3888){\makebox(0,0)[r]{\strut{}$-2.2$}}%
      \csname LTb\endcsname
      \put(816,4334){\makebox(0,0)[r]{\strut{}$-2$}}%
      \csname LTb\endcsname
      \put(816,4780){\makebox(0,0)[r]{\strut{}$-1.8$}}%
      \csname LTb\endcsname
      \put(1016,525){\makebox(0,0){\strut{}$2.3$}}%
      \csname LTb\endcsname
      \put(1457,525){\makebox(0,0){\strut{}$2.4$}}%
      \csname LTb\endcsname
      \put(1898,525){\makebox(0,0){\strut{}$2.5$}}%
      \csname LTb\endcsname
      \put(2338,525){\makebox(0,0){\strut{}$2.6$}}%
      \csname LTb\endcsname
      \put(2779,525){\makebox(0,0){\strut{}$2.7$}}%
      \csname LTb\endcsname
      \put(3220,525){\makebox(0,0){\strut{}$2.8$}}%
      \csname LTb\endcsname
      \put(3661,525){\makebox(0,0){\strut{}$2.9$}}%
      \csname LTb\endcsname
      \put(4102,525){\makebox(0,0){\strut{}$3$}}%
      \csname LTb\endcsname
      \put(4542,525){\makebox(0,0){\strut{}$3.1$}}%
      \csname LTb\endcsname
      \put(4983,525){\makebox(0,0){\strut{}$3.2$}}%
      \csname LTb\endcsname
      \put(5424,525){\makebox(0,0){\strut{}$3.3$}}%
    }%
    \gplgaddtomacro\gplfronttext{%
      \csname LTb\endcsname
      \put(198,2772){\rotatebox{-270}{\makebox(0,0){\strut{}Relative Error}}}%
      \csname LTb\endcsname
      \put(3220,167){\makebox(0,0){\strut{}$\log(NG)/\log(10)$}}%
      \csname LTb\endcsname
      \put(1961,2173){\makebox(0,0)[r]{\strut{}$pd=(4,2)$}}%
      \csname LTb\endcsname
      \put(1961,1934){\makebox(0,0)[r]{\strut{}$pd=(4,3)$}}%
      \csname LTb\endcsname
      \put(1961,1695){\makebox(0,0)[r]{\strut{}$pd=(4,4)$}}%
      \csname LTb\endcsname
      \put(1961,1456){\makebox(0,0)[r]{\strut{}$pd=(2,2)$}}%
      \csname LTb\endcsname
      \put(1961,1217){\makebox(0,0)[r]{\strut{}$pd=(2,3)$}}%
      \csname LTb\endcsname
      \put(1961,978){\makebox(0,0)[r]{\strut{}$pd=(2,4)$}}%
    }%
    \gplbacktext
    \put(0,0){\includegraphics[width={288.00bp},height={252.00bp}]{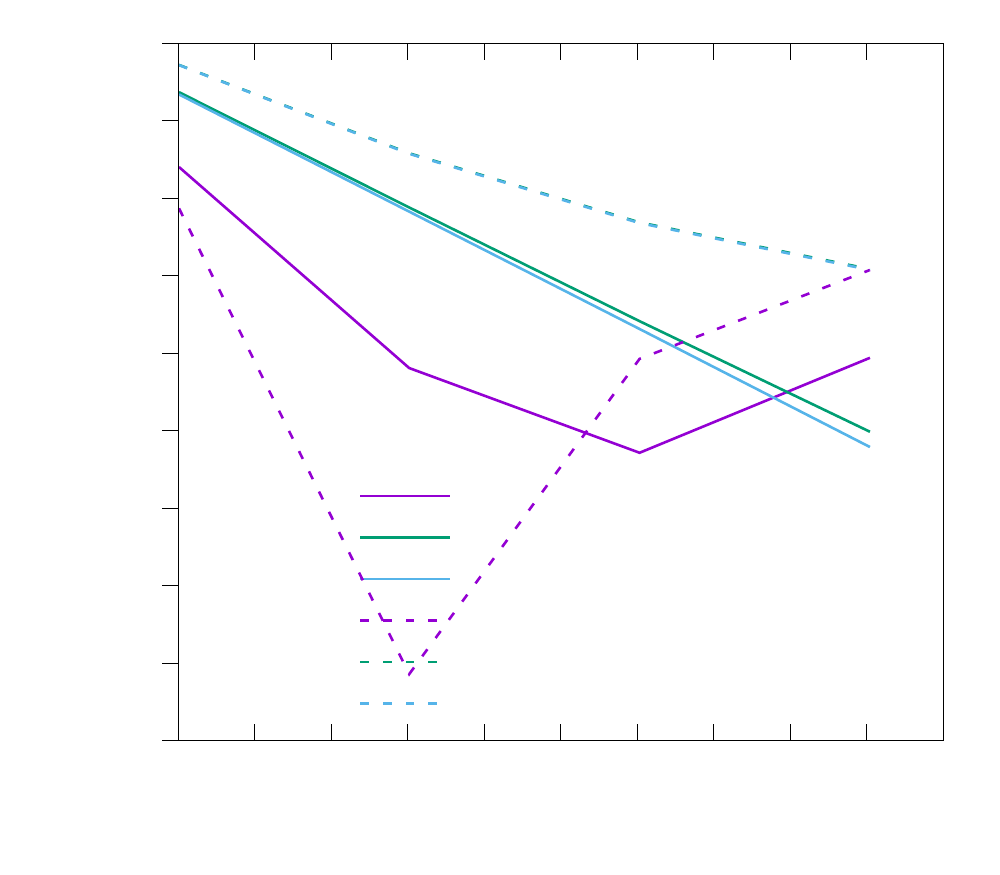}}%
    \gplfronttext
  \end{picture}%
\endgroup

%% file: gnu/Conv_TM1_900THz.tex
\begingroup
  \inputencoding{cp1252}%
  \makeatletter
  \providecommand\color[2][]{%
    \GenericError{(gnuplot) \space\space\space\@spaces}{%
      Package color not loaded in conjunction with
      terminal option `colourtext'%
    }{See the gnuplot documentation for explanation.%
    }{Either use 'blacktext' in gnuplot or load the package
      color.sty in LaTeX.}%
    \renewcommand\color[2][]{}%
  }%
  \providecommand\includegraphics[2][]{%
    \GenericError{(gnuplot) \space\space\space\@spaces}{%
      Package graphicx or graphics not loaded%
    }{See the gnuplot documentation for explanation.%
    }{The gnuplot epslatex terminal needs graphicx.sty or graphics.sty.}%
    \renewcommand\includegraphics[2][]{}%
  }%
  \providecommand\rotatebox[2]{#2}%
  \@ifundefined{ifGPcolor}{%
    \newif\ifGPcolor
    \GPcolortrue
  }{}%
  \@ifundefined{ifGPblacktext}{%
    \newif\ifGPblacktext
    \GPblacktexttrue
  }{}%
  \let\gplgaddtomacro\g@addto@macro
  \gdef\gplbacktext{}%
  \gdef\gplfronttext{}%
  \makeatother
  \ifGPblacktext
    \def\colorrgb#1{}%
    \def\colorgray#1{}%
  \else
    \ifGPcolor
      \def\colorrgb#1{\color[rgb]{#1}}%
      \def\colorgray#1{\color[gray]{#1}}%
      \expandafter\def\csname LTw\endcsname{\color{white}}%
      \expandafter\def\csname LTb\endcsname{\color{black}}%
      \expandafter\def\csname LTa\endcsname{\color{black}}%
      \expandafter\def\csname LT0\endcsname{\color[rgb]{1,0,0}}%
      \expandafter\def\csname LT1\endcsname{\color[rgb]{0,1,0}}%
      \expandafter\def\csname LT2\endcsname{\color[rgb]{0,0,1}}%
      \expandafter\def\csname LT3\endcsname{\color[rgb]{1,0,1}}%
      \expandafter\def\csname LT4\endcsname{\color[rgb]{0,1,1}}%
      \expandafter\def\csname LT5\endcsname{\color[rgb]{1,1,0}}%
      \expandafter\def\csname LT6\endcsname{\color[rgb]{0,0,0}}%
      \expandafter\def\csname LT7\endcsname{\color[rgb]{1,0.3,0}}%
      \expandafter\def\csname LT8\endcsname{\color[rgb]{0.5,0.5,0.5}}%
    \else
      \def\colorrgb#1{\color{black}}%
      \def\colorgray#1{\color[gray]{#1}}%
      \expandafter\def\csname LTw\endcsname{\color{white}}%
      \expandafter\def\csname LTb\endcsname{\color{black}}%
      \expandafter\def\csname LTa\endcsname{\color{black}}%
      \expandafter\def\csname LT0\endcsname{\color{black}}%
      \expandafter\def\csname LT1\endcsname{\color{black}}%
      \expandafter\def\csname LT2\endcsname{\color{black}}%
      \expandafter\def\csname LT3\endcsname{\color{black}}%
      \expandafter\def\csname LT4\endcsname{\color{black}}%
      \expandafter\def\csname LT5\endcsname{\color{black}}%
      \expandafter\def\csname LT6\endcsname{\color{black}}%
      \expandafter\def\csname LT7\endcsname{\color{black}}%
      \expandafter\def\csname LT8\endcsname{\color{black}}%
    \fi
  \fi
    \setlength{\unitlength}{0.0500bp}%
    \ifx\gptboxheight\undefined%
      \newlength{\gptboxheight}%
      \newlength{\gptboxwidth}%
      \newsavebox{\gptboxtext}%
    \fi%
    \setlength{\fboxrule}{0.5pt}%
    \setlength{\fboxsep}{1pt}%
    \definecolor{tbcol}{rgb}{1,1,1}%
\begin{picture}(5760.00,5040.00)%
    \gplgaddtomacro\gplbacktext{%
      \csname LTb\endcsname
      \put(816,764){\makebox(0,0)[r]{\strut{}$-1.8$}}%
      \csname LTb\endcsname
      \put(816,1338){\makebox(0,0)[r]{\strut{}$-1.6$}}%
      \csname LTb\endcsname
      \put(816,1911){\makebox(0,0)[r]{\strut{}$-1.4$}}%
      \csname LTb\endcsname
      \put(816,2485){\makebox(0,0)[r]{\strut{}$-1.2$}}%
      \csname LTb\endcsname
      \put(816,3059){\makebox(0,0)[r]{\strut{}$-1$}}%
      \csname LTb\endcsname
      \put(816,3633){\makebox(0,0)[r]{\strut{}$-0.8$}}%
      \csname LTb\endcsname
      \put(816,4206){\makebox(0,0)[r]{\strut{}$-0.6$}}%
      \csname LTb\endcsname
      \put(816,4780){\makebox(0,0)[r]{\strut{}$-0.4$}}%
      \csname LTb\endcsname
      \put(1016,525){\makebox(0,0){\strut{}$2.3$}}%
      \csname LTb\endcsname
      \put(1457,525){\makebox(0,0){\strut{}$2.4$}}%
      \csname LTb\endcsname
      \put(1898,525){\makebox(0,0){\strut{}$2.5$}}%
      \csname LTb\endcsname
      \put(2338,525){\makebox(0,0){\strut{}$2.6$}}%
      \csname LTb\endcsname
      \put(2779,525){\makebox(0,0){\strut{}$2.7$}}%
      \csname LTb\endcsname
      \put(3220,525){\makebox(0,0){\strut{}$2.8$}}%
      \csname LTb\endcsname
      \put(3661,525){\makebox(0,0){\strut{}$2.9$}}%
      \csname LTb\endcsname
      \put(4102,525){\makebox(0,0){\strut{}$3$}}%
      \csname LTb\endcsname
      \put(4542,525){\makebox(0,0){\strut{}$3.1$}}%
      \csname LTb\endcsname
      \put(4983,525){\makebox(0,0){\strut{}$3.2$}}%
      \csname LTb\endcsname
      \put(5424,525){\makebox(0,0){\strut{}$3.3$}}%
    }%
    \gplgaddtomacro\gplfronttext{%
      \csname LTb\endcsname
      \put(198,2772){\rotatebox{-270}{\makebox(0,0){\strut{}Relative Error}}}%
      \csname LTb\endcsname
      \put(3220,167){\makebox(0,0){\strut{}$\log(NG)/\log(10)$}}%
      \csname LTb\endcsname
      \put(1961,1456){\makebox(0,0)[r]{\strut{}$pd=(4,2)$}}%
      \csname LTb\endcsname
      \put(1961,1217){\makebox(0,0)[r]{\strut{}$pd=(4,3)$}}%
      \csname LTb\endcsname
      \put(1961,978){\makebox(0,0)[r]{\strut{}$pd=(4,4)$}}%
    }%
    \gplbacktext
    \put(0,0){\includegraphics[width={288.00bp},height={252.00bp}]{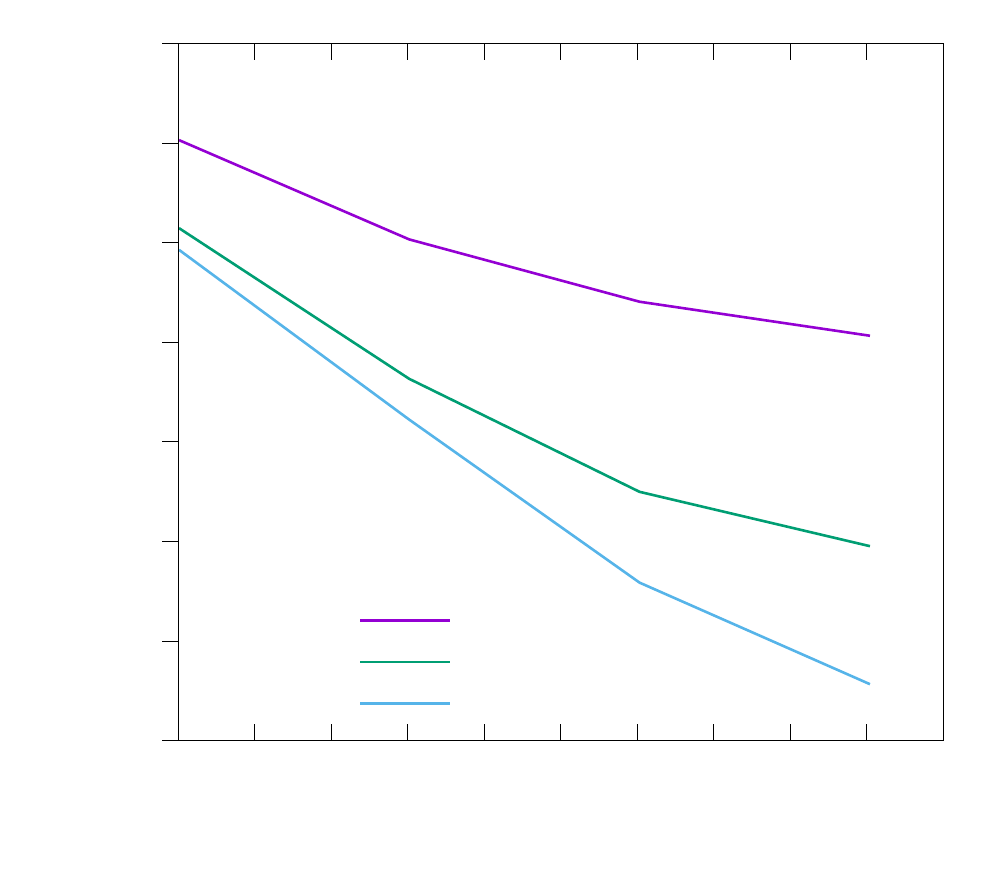}}%
    \gplfronttext
  \end{picture}%
\endgroup

%% file: gnu/Conv_TM2_900THz.tex
\begingroup
  \inputencoding{cp1252}%
  \makeatletter
  \providecommand\color[2][]{%
    \GenericError{(gnuplot) \space\space\space\@spaces}{%
      Package color not loaded in conjunction with
      terminal option `colourtext'%
    }{See the gnuplot documentation for explanation.%
    }{Either use 'blacktext' in gnuplot or load the package
      color.sty in LaTeX.}%
    \renewcommand\color[2][]{}%
  }%
  \providecommand\includegraphics[2][]{%
    \GenericError{(gnuplot) \space\space\space\@spaces}{%
      Package graphicx or graphics not loaded%
    }{See the gnuplot documentation for explanation.%
    }{The gnuplot epslatex terminal needs graphicx.sty or graphics.sty.}%
    \renewcommand\includegraphics[2][]{}%
  }%
  \providecommand\rotatebox[2]{#2}%
  \@ifundefined{ifGPcolor}{%
    \newif\ifGPcolor
    \GPcolortrue
  }{}%
  \@ifundefined{ifGPblacktext}{%
    \newif\ifGPblacktext
    \GPblacktexttrue
  }{}%
  \let\gplgaddtomacro\g@addto@macro
  \gdef\gplbacktext{}%
  \gdef\gplfronttext{}%
  \makeatother
  \ifGPblacktext
    \def\colorrgb#1{}%
    \def\colorgray#1{}%
  \else
    \ifGPcolor
      \def\colorrgb#1{\color[rgb]{#1}}%
      \def\colorgray#1{\color[gray]{#1}}%
      \expandafter\def\csname LTw\endcsname{\color{white}}%
      \expandafter\def\csname LTb\endcsname{\color{black}}%
      \expandafter\def\csname LTa\endcsname{\color{black}}%
      \expandafter\def\csname LT0\endcsname{\color[rgb]{1,0,0}}%
      \expandafter\def\csname LT1\endcsname{\color[rgb]{0,1,0}}%
      \expandafter\def\csname LT2\endcsname{\color[rgb]{0,0,1}}%
      \expandafter\def\csname LT3\endcsname{\color[rgb]{1,0,1}}%
      \expandafter\def\csname LT4\endcsname{\color[rgb]{0,1,1}}%
      \expandafter\def\csname LT5\endcsname{\color[rgb]{1,1,0}}%
      \expandafter\def\csname LT6\endcsname{\color[rgb]{0,0,0}}%
      \expandafter\def\csname LT7\endcsname{\color[rgb]{1,0.3,0}}%
      \expandafter\def\csname LT8\endcsname{\color[rgb]{0.5,0.5,0.5}}%
    \else
      \def\colorrgb#1{\color{black}}%
      \def\colorgray#1{\color[gray]{#1}}%
      \expandafter\def\csname LTw\endcsname{\color{white}}%
      \expandafter\def\csname LTb\endcsname{\color{black}}%
      \expandafter\def\csname LTa\endcsname{\color{black}}%
      \expandafter\def\csname LT0\endcsname{\color{black}}%
      \expandafter\def\csname LT1\endcsname{\color{black}}%
      \expandafter\def\csname LT2\endcsname{\color{black}}%
      \expandafter\def\csname LT3\endcsname{\color{black}}%
      \expandafter\def\csname LT4\endcsname{\color{black}}%
      \expandafter\def\csname LT5\endcsname{\color{black}}%
      \expandafter\def\csname LT6\endcsname{\color{black}}%
      \expandafter\def\csname LT7\endcsname{\color{black}}%
      \expandafter\def\csname LT8\endcsname{\color{black}}%
    \fi
  \fi
    \setlength{\unitlength}{0.0500bp}%
    \ifx\gptboxheight\undefined%
      \newlength{\gptboxheight}%
      \newlength{\gptboxwidth}%
      \newsavebox{\gptboxtext}%
    \fi%
    \setlength{\fboxrule}{0.5pt}%
    \setlength{\fboxsep}{1pt}%
    \definecolor{tbcol}{rgb}{1,1,1}%
\begin{picture}(5760.00,5040.00)%
    \gplgaddtomacro\gplbacktext{%
      \csname LTb\endcsname
      \put(816,764){\makebox(0,0)[r]{\strut{}$-3.6$}}%
      \csname LTb\endcsname
      \put(816,1338){\makebox(0,0)[r]{\strut{}$-3.4$}}%
      \csname LTb\endcsname
      \put(816,1911){\makebox(0,0)[r]{\strut{}$-3.2$}}%
      \csname LTb\endcsname
      \put(816,2485){\makebox(0,0)[r]{\strut{}$-3$}}%
      \csname LTb\endcsname
      \put(816,3059){\makebox(0,0)[r]{\strut{}$-2.8$}}%
      \csname LTb\endcsname
      \put(816,3633){\makebox(0,0)[r]{\strut{}$-2.6$}}%
      \csname LTb\endcsname
      \put(816,4206){\makebox(0,0)[r]{\strut{}$-2.4$}}%
      \csname LTb\endcsname
      \put(816,4780){\makebox(0,0)[r]{\strut{}$-2.2$}}%
      \csname LTb\endcsname
      \put(1016,525){\makebox(0,0){\strut{}$2.3$}}%
      \csname LTb\endcsname
      \put(1457,525){\makebox(0,0){\strut{}$2.4$}}%
      \csname LTb\endcsname
      \put(1898,525){\makebox(0,0){\strut{}$2.5$}}%
      \csname LTb\endcsname
      \put(2338,525){\makebox(0,0){\strut{}$2.6$}}%
      \csname LTb\endcsname
      \put(2779,525){\makebox(0,0){\strut{}$2.7$}}%
      \csname LTb\endcsname
      \put(3220,525){\makebox(0,0){\strut{}$2.8$}}%
      \csname LTb\endcsname
      \put(3661,525){\makebox(0,0){\strut{}$2.9$}}%
      \csname LTb\endcsname
      \put(4102,525){\makebox(0,0){\strut{}$3$}}%
      \csname LTb\endcsname
      \put(4542,525){\makebox(0,0){\strut{}$3.1$}}%
      \csname LTb\endcsname
      \put(4983,525){\makebox(0,0){\strut{}$3.2$}}%
      \csname LTb\endcsname
      \put(5424,525){\makebox(0,0){\strut{}$3.3$}}%
    }%
    \gplgaddtomacro\gplfronttext{%
      \csname LTb\endcsname
      \put(198,2772){\rotatebox{-270}{\makebox(0,0){\strut{}Relative Error}}}%
      \csname LTb\endcsname
      \put(3220,167){\makebox(0,0){\strut{}$\log(NG)/\log(10)$}}%
      \csname LTb\endcsname
      \put(1961,2173){\makebox(0,0)[r]{\strut{}$pd=(4,2)$}}%
      \csname LTb\endcsname
      \put(1961,1934){\makebox(0,0)[r]{\strut{}$pd=(4,3)$}}%
      \csname LTb\endcsname
      \put(1961,1695){\makebox(0,0)[r]{\strut{}$pd=(4,4)$}}%
      \csname LTb\endcsname
      \put(1961,1456){\makebox(0,0)[r]{\strut{}$pd=(2,2)$}}%
      \csname LTb\endcsname
      \put(1961,1217){\makebox(0,0)[r]{\strut{}$pd=(2,3)$}}%
      \csname LTb\endcsname
      \put(1961,978){\makebox(0,0)[r]{\strut{}$pd=(2,4)$}}%
    }%
    \gplbacktext
    \put(0,0){\includegraphics[width={288.00bp},height={252.00bp}]{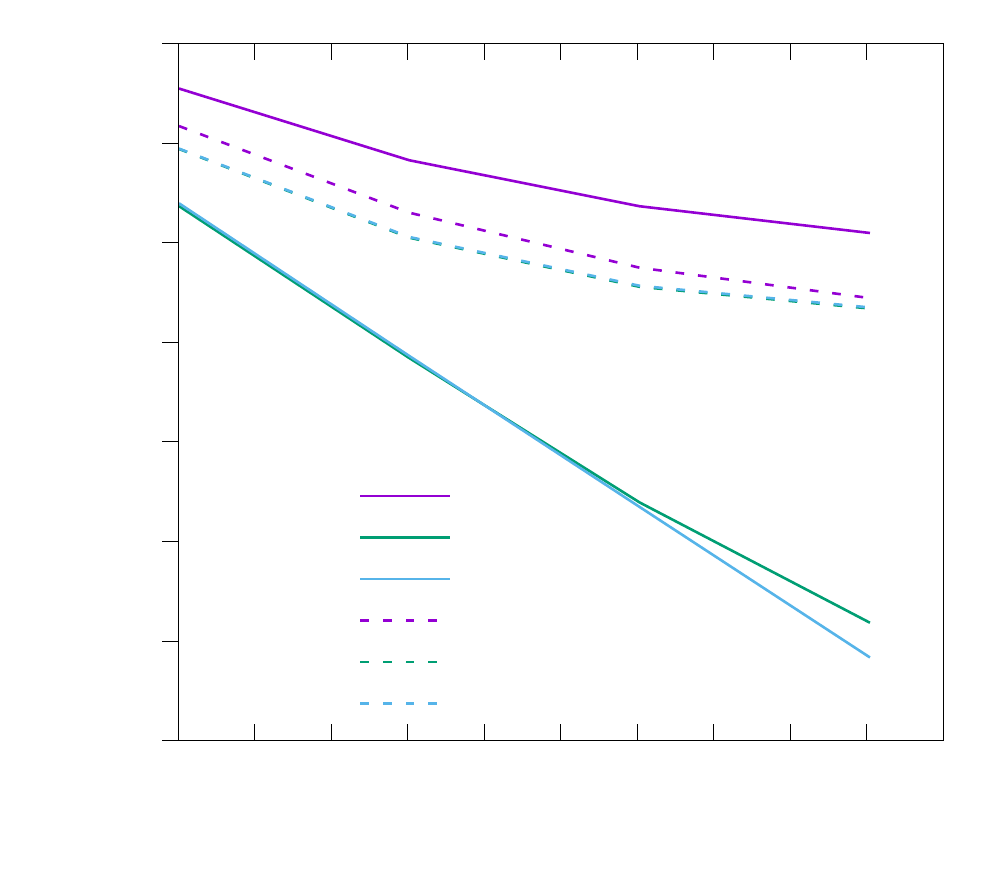}}%
    \gplfronttext
  \end{picture}%
\endgroup

%% file: gnu/BS-pd22-43_TM1.tex
\begingroup
  \inputencoding{cp1252}%
  \makeatletter
  \providecommand\color[2][]{%
    \GenericError{(gnuplot) \space\space\space\@spaces}{%
      Package color not loaded in conjunction with
      terminal option `colourtext'%
    }{See the gnuplot documentation for explanation.%
    }{Either use 'blacktext' in gnuplot or load the package
      color.sty in LaTeX.}%
    \renewcommand\color[2][]{}%
  }%
  \providecommand\includegraphics[2][]{%
    \GenericError{(gnuplot) \space\space\space\@spaces}{%
      Package graphicx or graphics not loaded%
    }{See the gnuplot documentation for explanation.%
    }{The gnuplot epslatex terminal needs graphicx.sty or graphics.sty.}%
    \renewcommand\includegraphics[2][]{}%
  }%
  \providecommand\rotatebox[2]{#2}%
  \@ifundefined{ifGPcolor}{%
    \newif\ifGPcolor
    \GPcolortrue
  }{}%
  \@ifundefined{ifGPblacktext}{%
    \newif\ifGPblacktext
    \GPblacktexttrue
  }{}%
  \let\gplgaddtomacro\g@addto@macro
  \gdef\gplbacktext{}%
  \gdef\gplfronttext{}%
  \makeatother
  \ifGPblacktext
    \def\colorrgb#1{}%
    \def\colorgray#1{}%
  \else
    \ifGPcolor
      \def\colorrgb#1{\color[rgb]{#1}}%
      \def\colorgray#1{\color[gray]{#1}}%
      \expandafter\def\csname LTw\endcsname{\color{white}}%
      \expandafter\def\csname LTb\endcsname{\color{black}}%
      \expandafter\def\csname LTa\endcsname{\color{black}}%
      \expandafter\def\csname LT0\endcsname{\color[rgb]{1,0,0}}%
      \expandafter\def\csname LT1\endcsname{\color[rgb]{0,1,0}}%
      \expandafter\def\csname LT2\endcsname{\color[rgb]{0,0,1}}%
      \expandafter\def\csname LT3\endcsname{\color[rgb]{1,0,1}}%
      \expandafter\def\csname LT4\endcsname{\color[rgb]{0,1,1}}%
      \expandafter\def\csname LT5\endcsname{\color[rgb]{1,1,0}}%
      \expandafter\def\csname LT6\endcsname{\color[rgb]{0,0,0}}%
      \expandafter\def\csname LT7\endcsname{\color[rgb]{1,0.3,0}}%
      \expandafter\def\csname LT8\endcsname{\color[rgb]{0.5,0.5,0.5}}%
    \else
      \def\colorrgb#1{\color{black}}%
      \def\colorgray#1{\color[gray]{#1}}%
      \expandafter\def\csname LTw\endcsname{\color{white}}%
      \expandafter\def\csname LTb\endcsname{\color{black}}%
      \expandafter\def\csname LTa\endcsname{\color{black}}%
      \expandafter\def\csname LT0\endcsname{\color{black}}%
      \expandafter\def\csname LT1\endcsname{\color{black}}%
      \expandafter\def\csname LT2\endcsname{\color{black}}%
      \expandafter\def\csname LT3\endcsname{\color{black}}%
      \expandafter\def\csname LT4\endcsname{\color{black}}%
      \expandafter\def\csname LT5\endcsname{\color{black}}%
      \expandafter\def\csname LT6\endcsname{\color{black}}%
      \expandafter\def\csname LT7\endcsname{\color{black}}%
      \expandafter\def\csname LT8\endcsname{\color{black}}%
    \fi
  \fi
    \setlength{\unitlength}{0.0500bp}%
    \ifx\gptboxheight\undefined%
      \newlength{\gptboxheight}%
      \newlength{\gptboxwidth}%
      \newsavebox{\gptboxtext}%
    \fi%
    \setlength{\fboxrule}{0.5pt}%
    \setlength{\fboxsep}{1pt}%
    \definecolor{tbcol}{rgb}{1,1,1}%
\begin{picture}(5180.00,4480.00)%
    \gplgaddtomacro\gplbacktext{%
      \csname LTb\endcsname
      \put(816,859){\makebox(0,0)[r]{\strut{}800}}%
      \csname LTb\endcsname
      \put(816,1556){\makebox(0,0)[r]{\strut{}850}}%
      \csname LTb\endcsname
      \put(816,2253){\makebox(0,0)[r]{\strut{}900}}%
      \csname LTb\endcsname
      \put(816,2950){\makebox(0,0)[r]{\strut{}950}}%
      \csname LTb\endcsname
      \put(816,3647){\makebox(0,0)[r]{\strut{}1000}}%
      \csname LTb\endcsname
      \put(1016,525){\makebox(0,0){\strut{}0.00}}%
      \csname LTb\endcsname
      \put(1473,525){\makebox(0,0){\strut{}0.04}}%
      \csname LTb\endcsname
      \put(1930,525){\makebox(0,0){\strut{}0.08}}%
      \csname LTb\endcsname
      \put(2386,525){\makebox(0,0){\strut{}0.12}}%
      \csname LTb\endcsname
      \put(2843,525){\makebox(0,0){\strut{}0.16}}%
      \csname LTb\endcsname
      \put(3300,525){\makebox(0,0){\strut{}0.20}}%
      \csname LTb\endcsname
      \put(3757,525){\makebox(0,0){\strut{}0.24}}%
      \csname LTb\endcsname
      \put(4185,3647){\makebox(0,0)[l]{\strut{}300}}%
      \csname LTb\endcsname
      \put(1016,3981){\makebox(0,0){\strut{}0.0}}%
      \csname LTb\endcsname
      \put(1494,3981){\makebox(0,0){\strut{}0.2}}%
      \csname LTb\endcsname
      \put(1973,3981){\makebox(0,0){\strut{}0.4}}%
      \csname LTb\endcsname
      \put(2451,3981){\makebox(0,0){\strut{}0.6}}%
      \csname LTb\endcsname
      \put(2929,3981){\makebox(0,0){\strut{}0.8}}%
      \csname LTb\endcsname
      \put(3408,3981){\makebox(0,0){\strut{}1.0}}%
      \csname LTb\endcsname
      \put(3886,3981){\makebox(0,0){\strut{}1.2}}%
    }%
    \gplgaddtomacro\gplfronttext{%
      \csname LTb\endcsname
      \put(198,2253){\rotatebox{-270}{\makebox(0,0){\strut{}Frequency [THz]}}}%
      \csname LTb\endcsname
      \put(4724,2253){\rotatebox{-270}{\makebox(0,0){\strut{}Wavelength [nm]}}}%
      \csname LTb\endcsname
      \put(2500,167){\makebox(0,0){\strut{}Wavenumber [nm\textsuperscript{-1}]}}%
      \csname LTb\endcsname
      \put(2500,4339){\makebox(0,0){\strut{}Wavenumber [$\kappa a / 2 \pi$]}}%
      \csname LTb\endcsname
      \put(3155,1551){\makebox(0,0)[r]{\strut{}$MGA$}}%
      \csname LTb\endcsname
      \put(3155,1312){\makebox(0,0)[r]{\strut{}$pd=(4,3)$}}%
      \csname LTb\endcsname
      \put(3155,1073){\makebox(0,0)[r]{\strut{}$pd=(2,2)$}}%
    }%
    \gplbacktext
    \put(0,0){\includegraphics[width={259.00bp},height={224.00bp}]{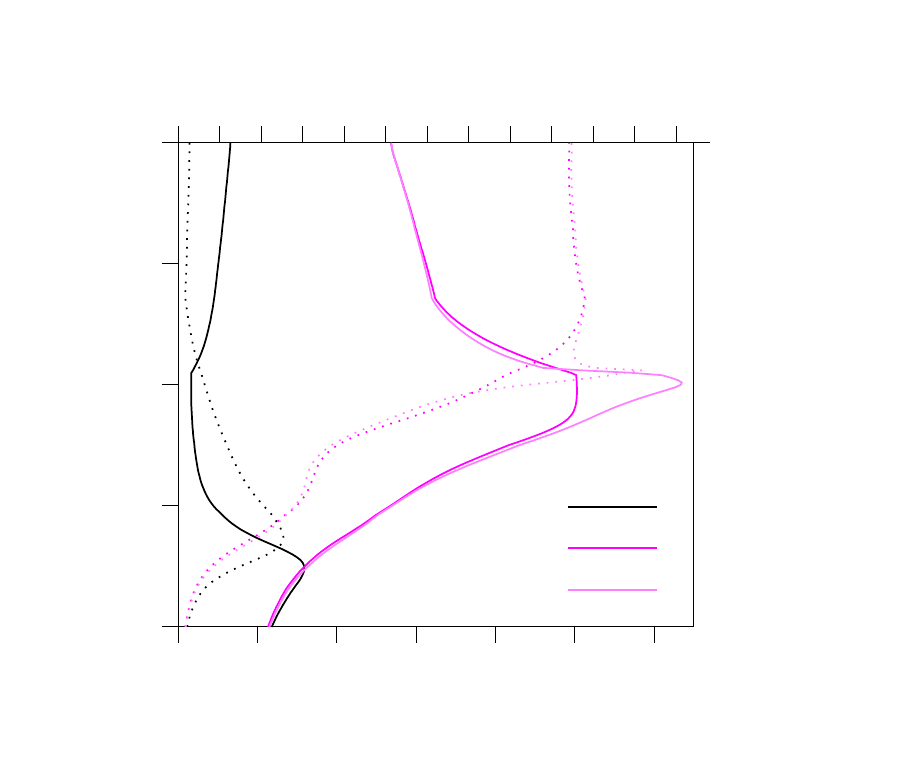}}%
    \gplfronttext
  \end{picture}%
\endgroup

%% file: gnu/BS-pd22-43_TM2.tex
\begingroup
  \inputencoding{cp1252}%
  \makeatletter
  \providecommand\color[2][]{%
    \GenericError{(gnuplot) \space\space\space\@spaces}{%
      Package color not loaded in conjunction with
      terminal option `colourtext'%
    }{See the gnuplot documentation for explanation.%
    }{Either use 'blacktext' in gnuplot or load the package
      color.sty in LaTeX.}%
    \renewcommand\color[2][]{}%
  }%
  \providecommand\includegraphics[2][]{%
    \GenericError{(gnuplot) \space\space\space\@spaces}{%
      Package graphicx or graphics not loaded%
    }{See the gnuplot documentation for explanation.%
    }{The gnuplot epslatex terminal needs graphicx.sty or graphics.sty.}%
    \renewcommand\includegraphics[2][]{}%
  }%
  \providecommand\rotatebox[2]{#2}%
  \@ifundefined{ifGPcolor}{%
    \newif\ifGPcolor
    \GPcolortrue
  }{}%
  \@ifundefined{ifGPblacktext}{%
    \newif\ifGPblacktext
    \GPblacktexttrue
  }{}%
  \let\gplgaddtomacro\g@addto@macro
  \gdef\gplbacktext{}%
  \gdef\gplfronttext{}%
  \makeatother
  \ifGPblacktext
    \def\colorrgb#1{}%
    \def\colorgray#1{}%
  \else
    \ifGPcolor
      \def\colorrgb#1{\color[rgb]{#1}}%
      \def\colorgray#1{\color[gray]{#1}}%
      \expandafter\def\csname LTw\endcsname{\color{white}}%
      \expandafter\def\csname LTb\endcsname{\color{black}}%
      \expandafter\def\csname LTa\endcsname{\color{black}}%
      \expandafter\def\csname LT0\endcsname{\color[rgb]{1,0,0}}%
      \expandafter\def\csname LT1\endcsname{\color[rgb]{0,1,0}}%
      \expandafter\def\csname LT2\endcsname{\color[rgb]{0,0,1}}%
      \expandafter\def\csname LT3\endcsname{\color[rgb]{1,0,1}}%
      \expandafter\def\csname LT4\endcsname{\color[rgb]{0,1,1}}%
      \expandafter\def\csname LT5\endcsname{\color[rgb]{1,1,0}}%
      \expandafter\def\csname LT6\endcsname{\color[rgb]{0,0,0}}%
      \expandafter\def\csname LT7\endcsname{\color[rgb]{1,0.3,0}}%
      \expandafter\def\csname LT8\endcsname{\color[rgb]{0.5,0.5,0.5}}%
    \else
      \def\colorrgb#1{\color{black}}%
      \def\colorgray#1{\color[gray]{#1}}%
      \expandafter\def\csname LTw\endcsname{\color{white}}%
      \expandafter\def\csname LTb\endcsname{\color{black}}%
      \expandafter\def\csname LTa\endcsname{\color{black}}%
      \expandafter\def\csname LT0\endcsname{\color{black}}%
      \expandafter\def\csname LT1\endcsname{\color{black}}%
      \expandafter\def\csname LT2\endcsname{\color{black}}%
      \expandafter\def\csname LT3\endcsname{\color{black}}%
      \expandafter\def\csname LT4\endcsname{\color{black}}%
      \expandafter\def\csname LT5\endcsname{\color{black}}%
      \expandafter\def\csname LT6\endcsname{\color{black}}%
      \expandafter\def\csname LT7\endcsname{\color{black}}%
      \expandafter\def\csname LT8\endcsname{\color{black}}%
    \fi
  \fi
    \setlength{\unitlength}{0.0500bp}%
    \ifx\gptboxheight\undefined%
      \newlength{\gptboxheight}%
      \newlength{\gptboxwidth}%
      \newsavebox{\gptboxtext}%
    \fi%
    \setlength{\fboxrule}{0.5pt}%
    \setlength{\fboxsep}{1pt}%
    \definecolor{tbcol}{rgb}{1,1,1}%
\begin{picture}(5180.00,4480.00)%
    \gplgaddtomacro\gplbacktext{%
      \csname LTb\endcsname
      \put(816,971){\makebox(0,0)[r]{\strut{}400}}%
      \csname LTb\endcsname
      \put(816,1417){\makebox(0,0)[r]{\strut{}500}}%
      \csname LTb\endcsname
      \put(816,1863){\makebox(0,0)[r]{\strut{}600}}%
      \csname LTb\endcsname
      \put(816,2309){\makebox(0,0)[r]{\strut{}700}}%
      \csname LTb\endcsname
      \put(816,2755){\makebox(0,0)[r]{\strut{}800}}%
      \csname LTb\endcsname
      \put(816,3201){\makebox(0,0)[r]{\strut{}900}}%
      \csname LTb\endcsname
      \put(816,3647){\makebox(0,0)[r]{\strut{}1000}}%
      \csname LTb\endcsname
      \put(1016,525){\makebox(0,0){\strut{}0.00}}%
      \csname LTb\endcsname
      \put(1457,525){\makebox(0,0){\strut{}0.04}}%
      \csname LTb\endcsname
      \put(1898,525){\makebox(0,0){\strut{}0.08}}%
      \csname LTb\endcsname
      \put(2339,525){\makebox(0,0){\strut{}0.12}}%
      \csname LTb\endcsname
      \put(2780,525){\makebox(0,0){\strut{}0.16}}%
      \csname LTb\endcsname
      \put(3221,525){\makebox(0,0){\strut{}0.20}}%
      \csname LTb\endcsname
      \put(3662,525){\makebox(0,0){\strut{}0.24}}%
      \csname LTb\endcsname
      \put(4185,3647){\makebox(0,0)[l]{\strut{}300}}%
      \csname LTb\endcsname
      \put(4185,2532){\makebox(0,0)[l]{\strut{}400}}%
      \csname LTb\endcsname
      \put(4185,1863){\makebox(0,0)[l]{\strut{}500}}%
      \csname LTb\endcsname
      \put(4185,1417){\makebox(0,0)[l]{\strut{}600}}%
      \csname LTb\endcsname
      \put(4185,1098){\makebox(0,0)[l]{\strut{}700}}%
      \csname LTb\endcsname
      \put(4185,859){\makebox(0,0)[l]{\strut{}800}}%
      \csname LTb\endcsname
      \put(1016,3981){\makebox(0,0){\strut{}0.0}}%
      \csname LTb\endcsname
      \put(1478,3981){\makebox(0,0){\strut{}0.2}}%
      \csname LTb\endcsname
      \put(1940,3981){\makebox(0,0){\strut{}0.4}}%
      \csname LTb\endcsname
      \put(2402,3981){\makebox(0,0){\strut{}0.6}}%
      \csname LTb\endcsname
      \put(2864,3981){\makebox(0,0){\strut{}0.8}}%
      \csname LTb\endcsname
      \put(3325,3981){\makebox(0,0){\strut{}1.0}}%
      \csname LTb\endcsname
      \put(3787,3981){\makebox(0,0){\strut{}1.2}}%
    }%
    \gplgaddtomacro\gplfronttext{%
      \csname LTb\endcsname
      \put(198,2253){\rotatebox{-270}{\makebox(0,0){\strut{}Frequency [THz]}}}%
      \csname LTb\endcsname
      \put(4724,2253){\rotatebox{-270}{\makebox(0,0){\strut{}Wavelength [nm]}}}%
      \csname LTb\endcsname
      \put(2500,167){\makebox(0,0){\strut{}Wavenumber [nm\textsuperscript{-1}]}}%
      \csname LTb\endcsname
      \put(2500,4339){\makebox(0,0){\strut{}Wavenumber [$\kappa a / 2 \pi$]}}%
      \csname LTb\endcsname
      \put(3155,1551){\makebox(0,0)[r]{\strut{}$MGA$}}%
      \csname LTb\endcsname
      \put(3155,1312){\makebox(0,0)[r]{\strut{}$pd=(4,3)$}}%
      \csname LTb\endcsname
      \put(3155,1073){\makebox(0,0)[r]{\strut{}$pd=(2,2)$}}%
    }%
    \gplbacktext
    \put(0,0){\includegraphics[width={259.00bp},height={224.00bp}]{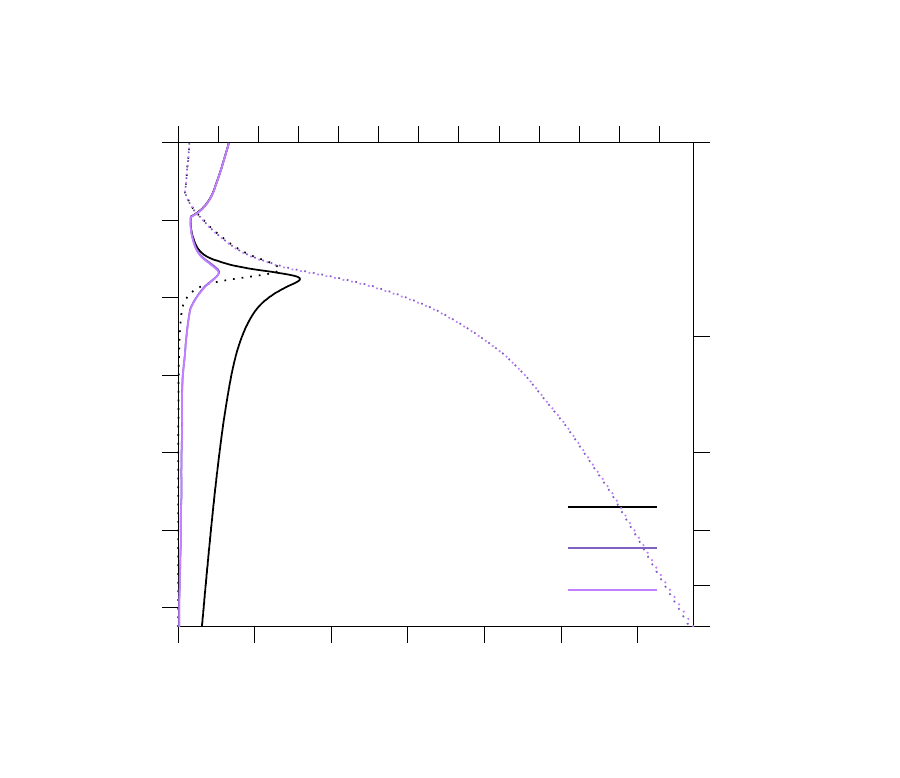}}%
    \gplfronttext
  \end{picture}%
\endgroup